\documentclass[a4paper,fleqn,usenatbib]{mnras}

\usepackage{newtxtext,newtxmath}
\usepackage[T1]{fontenc}
\usepackage{ae,aecompl}

\usepackage{xcolor}

\usepackage{subfig}
\usepackage[export]{adjustbox}
\usepackage{ragged2e}

\usepackage{graphicx}	% Including figure files
\usepackage{amsmath}	% Advanced maths commands
\usepackage{amssymb}	% Extra maths symbols
\usepackage{verbatim}

\graphicspath{{./plots-reduced/}}

\newcommand{\vzrms}{$v_{\mathrm{z,rms}}$}	
\newcommand{\vsq}{$v^2_{\mathrm{z,rms}}$}
                     
\newcommand{\logg}{$\log g$ }                         
\newcommand{\teff}{$T_{\mathrm{eff}}$ }
\newcommand{\tautop}{$\tau_{\mathrm{R,top}}$ }
\newcommand{\taubot}{$\tau_{\mathrm{R,bot}}$ }
\newcommand{\tauR}{$\tau_{\mathrm{R}}$ }

\newcommand{\zrms}{$z_{\mathrm{rms}}$}
\newcommand{\zstarms}{$z^{*}_{\mathrm{rms}}$}
\newcommand{\thalf}{$\sqrt{t}$ }

\newcommand{\cobold}{CO$^5$BOLD}

\newcommand{\T}[1]{$T_{\mathrm{eff}} = #100\,{\mathrm{K}}$}

\newcommand{\tdiff}{$t_{\mathrm{diff}}$}
\newcommand{\zst}{$z_{\mathrm{S,top}}$}

\title[Convective Overshoot in DA White Dwarfs]{Convective Overshoot and Macroscopic Diffusion in Pure-Hydrogen Atmosphere White Dwarfs}

\author[T.~Cunningham et. al]{Tim Cunningham,$^{1}$\thanks{E-mail: t.cunningham@warwick.ac.uk}
Pier-Emmanuel Tremblay,$^{1}$ Bernd Freytag,$^{2}$ 
\newauthor{Hans-G\"unter Ludwig,$^{3}$ and Detlev Koester$^{4}$}
\\
$^{1}$Department of Physics, University of Warwick, Coventry, CV4 7AL, UK\\
$^{2}$Department of Physics and Astronomy, Uppsala University, Box 516, 751 20 Uppsala, Sweden \\
$^{3}$Zentrum f\"ur Astronomie der Universit\"at Heidelberg, Landessternwarte, K\"onigstuhl 12, 69117 Heidelberg, Germany \\
$^{4}$Institut f\"ur Theoretische Physik und Astrophysik, University of Kiel, 24098 Kiel, Germany
}

\date{Accepted XXX. Received YYY; in original form ZZZ}

\pubyear{2019}

\begin{document}
\label{firstpage}
\pagerange{\pageref{firstpage}--\pageref{lastpage}}
\maketitle

\begin{abstract}
We present a theoretical description of macroscopic diffusion caused by convective overshoot in pure-hydrogen DA white dwarfs using three-dimensional (3D), closed-bottom, radiation hydrodynamics CO$^5$BOLD simulations. We rely on a new grid of deep 3D white dwarf models in the temperature range 11\,400~K $\leq T_{\mathrm{eff}} \leq 18\,000$~K where tracer particles and a tracer density are used to derive macroscopic diffusion coefficients driven by convective overshoot. These diffusion coefficients are compared to microscopic diffusion coefficients from one-dimensional structures. We find that the mass of the fully mixed region is likely to increase by up to 2.5 orders of magnitude while inferred accretion rates increase by a more moderate order of magnitude. We present evidence that an increase in settling time of up to 2 orders of magnitude is to be expected which is of significance for time-variability studies of polluted white dwarfs. Our grid also provides the most robust constraint on the onset of convective instabilities in DA white dwarfs to be in the effective temperature range from 18\,000 to 18\,250~K.
\end{abstract}

\begin{keywords}
white dwarfs -- convection -- hydrodynamics
\end{keywords}

%%%%%%%%%%%%%%%%%%%%%%%%%%%%%%%%%%%%%%%%%%%%%%%%%%

%%%%%%%%%%%%%%%%% BODY OF PAPER %%%%%%%%%%%%%%%%%%

\section{Introduction}

For decades, understanding the diffusion of heavy elements through stellar plasmas has been central to research in solar and stellar evolution \citep{michaud70,thoul94}, including the cooling of white dwarfs \citep{schatzman45}. In the latter case, large surface gravities ($7.0\lesssim\log g\lesssim9.0$) imply the rapid settling of the initial composition into a structure stratified according to atomic weight with an outer shell made of hydrogen and helium. 

The related scenario of metal accretion onto white dwarfs has now matured into an extremely active field of research around evolved planetary systems \citep{veras16}. For a small fraction of these systems it is possible to study the debris disk or gas around the white dwarf \citep{jura03,farihi09,manser16,xu18,manser19,wilson19} but in most cases the primary evidence is metal pollution from tidally disrupted asteroids in the stellar atmosphere \citep{vanMaanen17,zuckerman03,gaensicke12,vanderburg16}. To transform photospheric metal abundances into the parent body properties requires an accurate model describing the volume of the stellar envelope over which this abundance is prevalent. The accretion-diffusion model is usually employed, where microscopic diffusion removes the heavy elements out of the observable layers \citep{paquette86a,paquette86b,pelletier86,koester09}.

Constraining the mass and composition of the accreted material has wide-reaching implications for the chemical compositions of other rocky worlds and planetary evolutionary models \citep[see, e.g.,][]{zuckerman07,dufour12,farihi13,farihi16,wilson16,xu17,melis17}. Since the emergence of space-based UV spectroscopy the ability to detect metal pollution has significantly increased, putting the fraction of degenerate stars with evolved planetary systems at $\approx 50\%$ \citep{zuckerman10,koester14}. These advances have allowed to describe the frequency and properties of accretion events across white dwarf cooling age, mass, and spectral type, yet the possibly observed correlations are not fully understood \citep{farihi12,koester14,hollands18}, and systematic model effects have been debated \citep{wachlin17,kupka18,bauer18,bauer19}. In this work we concentrate on theoretical aspects of the accretion-diffusion scenario in the context of hydrogen-atmosphere DA white dwarfs.

Young DA white dwarfs (cooling ages $t_{\mathrm{cool}} < 10^8$\,yr corresponding to effective temperatures $T_{\rm eff} \gtrapprox 20\,000$\,K) have an atmosphere dominated by radiative energy transport. In this temperature regime, radiative levitation can still maintain some heavier chemical elements in the photosphere \citep{chayer95a,chayer95b}, although ongoing accretion from planetesimals is invoked in a majority of cases \citep{koester14}. Once DA white dwarfs have cooled sufficiently to develop a convection zone, small characteristic convective turnover timescales imply that metals become fully mixed within these turbulent layers, and that microscopic diffusion only takes place at the base of the convection zone. As this boundary layer is denser than the surface layers, this implies longer settling times of the metals of up to 10\,000\,yrs \citep{koester09,bauer19}.

Going back to the work of \citet{prandtl25}, convection has often been modeled using 1D mixing-length theory (MLT). This model requires that the Schwarzschild criterion for convective instability be satisfied and where it is not, the models show no convective motions \citep{bohm-vitense58,tassoul90}. In the 1D MLT picture the base of the convection zone is at the Schwarzschild boundary. Beneath this, the dominant process for mass transport is that of microscopic diffusion. This microscopic diffusion comprises contributions from gravitational settling, thermal diffusion, {radiative diffusion} and diffusion driven by concentration gradients, unless the diffusion concerns tracer particles, in which case the latter term is neglected \citep{paquette86b, koester09}. The characteristic velocities of microscopic diffusion near the Schwarzschild boundary are $v_{\mathrm{diff}}\sim 10^{-7}$\,km\,s$^{-1}$. This can be up to seven orders of magnitude less than the characteristic convective velocities of $\sim 1\, \mathrm{km\,s^{-1}}$ in the adjacent layer above. 

{The discontinuity between convective and non-convective layers in the 1D model is unphysical \citep{spiegel63,roxburgh78,zahn91} and multi-dimensional numerical simulations have also confirmed this for the case of white dwarfs \citep{freytag96,tremblay15,kupka18}.} Whilst the layers beneath the convectively unstable region are not able to accelerate material to greater depths, convective cells accelerated at the base of the unstable region can have sufficient momentum to penetrate the deeper layers before dissipating their kinetic energy. This process is known as convective overshoot and it has been shown from earlier 2D simulations to be capable of mixing material \citep{freytag96}. Constraining precisely the mass fraction of mixed material is crucial for our understanding of accretion onto white dwarfs.

We present the first direct tests of mixing due to convective overshoot using tracer particles and trace density in 3D radiation-hydrodynamics (RHD) CO$^5$BOLD simulations in the context of DA white dwarfs with shallow surface convection zones. Building upon the work of \citet{freytag96,freytag12}, we estimate macroscopic diffusion coefficients below the Schwarzschild unstable region from direct 3D experiments (Sections~\ref{sec2}-\ref{sec:Dover}). We employ the dependence of macroscopic diffusion on mean convective velocities from these direct experiments to infer the increase in convectively mixed mass for a wide range of atmospheric parameters, using a grid of CO$^5$BOLD simulations covering 11\,400\,K $< T_{\rm eff} <$ 18\,000\,K at $\log g = 8.0$ (Section~\ref{sec:results}). We discuss the implications on accretion rates and diffusion timescales (Section~\ref{discussion}) and conclude in Section \ref{conclusions}. 

\section{Numerical Setup}
\label{sec2}

The simulations presented in this work have been run using the 3D RHD code CO$^5$BOLD as described in \citet{freytag12}. On a Cartesian grid, CO$^5$BOLD solves the coupled equations of compressible hydrodynamics and non-local radiation transport, implicitly respecting conservation laws of energy, mass and momentum. The vertical grid spacing is depth dependent, allowing for better resolution of radiative transfer in upper layers, whilst the grid spacing does not vary in the horizontal plane. For simulations presented throughout this study radiation is handled with opacity and EOS tables from \citet{tremblay13a,tremblay13c}, using both grey (Table~\ref{tb:main}) and non-grey schemes (Table~\ref{tb:deep}). The upper (surface) boundary is open to mass flows and radiation (see \citealt{freytag17}). The lower boundary is closed, requiring velocities to be zero, and with a fixed radiative flux for a given effective temperature. The four horizontal boundaries have periodic boundary conditions.

To increase stability around discontinuities a numerical reconstruction scheme is required by the hydrodynamics solver. For all simulations presented in this study the chosen scheme is designated as FRweno using a reconstruction by 2nd order polynomials \citep{freytag13}. For the handling of trace density arrays with CO$^5$BOLD, which will be discussed in detail in the following section, {a piecewise-parabolic reconstruction scheme is used \citep{colella84}.} 

The effective temperature is an input parameter for our simulations, as its value determines the radiative flux put into the simulation lower boundary. Following relaxation, the effective temperature is confirmed by the spatially and temporally averaged emergent radiative flux. For brevity the effective temperatures in Tables~\ref{tb:main}-\ref{tb:deep} will be referred to by rounding to the nearest 100 throughout this study, though any calculations involving the effective temperature will use the more precise values shown in the tables.

\subsection{Diffusion Coefficient Experiments}
Table \ref{tb:main} shows the numerical setups for the simulations analysed in the direct study of macroscopic diffusion in Section~\ref{sec:Dover}. All simulations have been built from previously relaxed 3D simulations and given sufficient time to relax after any parameter change.  

\begin{table*}
	\centering
	\caption{Numerical setup for grey opacity simulations analysed directly for macroscopic diffusion with methods of \cobold\, tracer density and path integration. All simulations have a pure-hydrogen composition, closed bottom boundary and \logg = 8.0.}
	\label{tab:grid}
	\begin{tabular}{lcccccccccccc}
		\hline
		Sim. ID & \teff & z & x,y & log \tautop & log \taubot & $\Delta H_{p}$ & $z_{\mathrm{S,top}}$ & $H_{p, \mathrm{S,bot}}$ & $v_{\mathrm{z,S,bot}}$ & $[x,y,z]$  & $t_{\mathrm{diff~exp }} $ & $N_{\mathrm{tr~dens}}$\\
		& [K] & [km] & [km] &  &         &     & [km] & [km] & [kms$^{-1}$] & (grid points) &[s]   & \\		
		\hline
		A1 & 11\,992   & 4.7 & 7.5 & 	$-$4.28  & 3.63 & 5.19  & 1.06 & 0.53 &  4.48 & 250, 250, 250  & 6.02  & 10 \\ \\
		
		B1& 13\,000   & 4.5 &   7.5 & $-$3.35	  & 3.01 &  5.06 &  0.66 & 0.40 &   & 150, 150, 150  & 9.05 & 20 \\	
		B2 & 12\,999   & 4.5 &  7.5  & $-$3.33	  & 3.01 &  5.02 &  0.43 & 0.38 & 2.78  & 250, 250, 250  & 3.16 & 10 \\ \\
				
		C1 & 13\,498   & 3.6  &  7.5 & $-$3.19 & 2.69 	& 4.37 &  0.52 & 0.37 &   & 150, 150, 150  & 6.20 & 20\\
		C2 & 13\,498   & 3.6  &  7.5 & $-$3.18 & 2.69 	& 4.37 &  0.51  & 0.38 &   & 250, 250, 150  & 7.91 & 20\\	
		C3 & 13\,498   & 3.6  &  7.5 & $-$3.20 & 2.69 	& 4.39 &  0.50 & 0.35 & 1.52  & 250, 250, 250  & 8.41 & 10\\ \\
		
		C1-2 & 13\,499   & 3.6  &  15.0 & $-$3.19 & 2.69	& 4.39 &  0.50 & 0.37   &   & 150, 150, 150  & 8.14 & 20\\
		\hline
	\end{tabular}
	\label{tb:main}\\
	\justify
	{Notes: {The vertical (horizontal) extent of the simulation box is indicated by $z$ ($x,y$).}  The number of pressure scale heights from the horizontally averaged layer $\tau_{\mathrm{R}} = 1$ and the simulation base is given by $\Delta H_{p}$. All simulations have been fully relaxed prior to the addition of tracer density at which point the diffusion experiment begins and its duration is denoted by $t_{\mathrm{diff~exp}} $. The number of unique depths at which tracer densities are added is given by $N_{\mathrm{tr~dens}}$. We also include the geometric position of the upper Schwarzschild boundary, $z_{\mathrm{S,top}}$, relative to the lower Schwarzschild boundary which is set to be at the origin of the depth axis, i.e., where $z=0$~km. Denoted by $v_{\mathrm{z,S,bot}}$ is the time-averaged RMS vertical velocity at the lower Schwarzschild boundary. $T_{\rm eff}$ is derived from the time and spatially averaged outgoing radiative flux. \tautop and \taubot are geometrical and time averages of the Rosseland optical depth at the boundary layers {of the simulation box}. Characteristic granule sizes for all simulations range from 1.0--1.2~km and data sampled from simulations with period $\Delta t = 0.01$~s.}
\end{table*}

\subsection{Extended Grid of Closed 3D Simulations}
\label{sec:deep-grid}

Diffusion experiments using the direct methods laid out in Section~\ref{sec:Dover} are computationally expensive, in terms of both random-access memory during the simulation run proper (Section~\ref{sec:dust}) and data storage after the fact (Section~\ref{sec:pathint}). Furthermore, direct simulation of diffusion is hampered by numerical waves as we shall discuss in Section~\ref{sec:results}, and thus in this work we also infer indirectly about the macroscopic diffusion properties by using wave-filtered velocities as a proxy. Towards this goal we employ a newly computed grid of deep simulations across the effective temperature range 11\,400\,K $\leq$ \teff $\leq$ 18\,000\,K. Numerical details are shown in Table~\ref{tb:deep}. {The simulations detailed with 12\,009\,K $\leq$ \teff $\leq$ 17\,004\,K had nominal input effective temperatures to the nearest 500\,K. The input effective temperatures for the two coolest (warmest) simulations were 11\,400\,K and 11\,600\,K (17\,525\,K and 18\,025\,K). As an indicator of the quality of relaxation of our simulations we find the maximum discrepancy between input effective temperatures and those determined at the surface to be 0.4\%, whilst the majority of the grid (\teff $\geq$ 12\,000\,K) has a maximum difference of 0.1\%.}

These models are similar to the closed box simulations introduced in \citet{tremblay13c,tremblay15} but were extended to deeper layers and are using a larger number of grid points. {The new grid has a vertical and horizontal resolution of 30--50\,m and 50--80\,m, respectively. This is slightly less resolved than in the previous grid, which had resolutions of $\sim$20\,m and 30--50\,m in the vertical and horizontal directions, respectively; a compromise for probing deeper layers.} Furthermore, the grid was extended to significantly cooler and hotter temperatures, with our coolest closed bottom simulation at $T_{\rm eff} \approx 11\,400$\,K overlapping with an open lower boundary simulation in \citet{tremblay13c}. All these simulations are sufficiently deep to contain the entirety of the convectively unstable layers and deeper overshoot layers, providing a unique opportunity to constrain convective velocities in layers deeper than previously done with CO$^5$BOLD, similarly to what has been performed in \citet{kupka18}.

\begin{table}
	\centering
	\caption{Extended grid of CO$^5$BOLD simulations with pure-H composition and \logg = 8.0. All simulations have 200$^3$ grid points, closed bottom boundaries and non-grey opacities \citep{tremblay13a}.}
	\label{tb:deep}
	\begin{tabular}{cccccc}
		\hline
		\teff  & log \tautop & log \taubot & $\Delta H_{p}$ & $x$,$y$ & $z$ \\
		$[\mathrm{K}]$ &  &  &  & [km] & [km] \\
		\hline
		11445    & $-$10.1 &  4.55 & 4.21 & $ 10.1$ & $10.0 $    \\
		11651    & $-$9.24 & 4.38  & 5.18 & $ 16.9$ & $ 9.8  $     \\
		12009    & $-$6.99 & 4.44  & 7.36 & $ 16.8$ & $ 10.6 $     \\
		12514    & $-$6.34 & 4.03  & 7.49 & $ 14.9$ & $ 8.7 $     \\
		13005    & $-$6.20 & 3.63  & 6.94 & $ 15.0$ & $ 7.1 $     \\
		13503    & $-$5.29 & 3.34  & 6.52 & $ 15.0$ & $ 6.1 $     \\ 
		14000    & $-$5.06 & 3.19  & 6.37 & $ 15.0$ & $ 5.8 $     \\
		14498    & $-$5.10 & 3.18  & 6.82 & $ 15.0$ & $ 6.2 $     \\
		15000    & $-$5.01 & 3.15  & 7.01 & $ 15.0$ & $ 6.4 $    \\
		15501   & $-$5.01 & 3.12  & 7.23 & $ 15.0$ & $ 6.7 $     \\
		16002   & $-$5.04 & 3.09  & 7.48 & $ 15.0$ & $ 6.9 $     \\
		16503   & $-$5.06 & 3.05  & 7.67 & $ 15.0$ & $ 7.0 $     \\
		17004   & $-$4.97 & 3.02  & 7.92 & $ 15.0$ & $ 7.1 $     \\
		17524   & $-$5.11 & 2.80  & 7.58 & $ 15.0$ & $ 6.4 $     \\
		18022   & $-$5.10 & 2.56  & 7.34 & $ 15.0$ & $ 5.7 $     
		\\
		\hline
	\end{tabular}\\
	\justify
	{Notes: The number of pressure scale heights from $\tau_{\mathrm{R}}=1$ and the simulation base is given by $\Delta H_{p}$. \tautop and \taubot are geometrical averages of the Rosseland optical depth at the boundary layers {of the simulation box. Further details on the columns are given in the caption to Table \ref{tb:main}.}}
\end{table}

\section{Simulating Macroscopic Diffusion}
\label{sec:Dover}
The motivation for our research is to characterise the macroscopic diffusion of trace metals in polluted white dwarfs for regions beneath the convection zone, where 1D models currently predict no such mixing. In particular, we aim at characterising the depth dependence of the efficiency of macroscopic mixing in these overshoot regions. {We do this by implementing a statistical, ensemble averaged model for the mixing caused by overshoot in a finite region directly beneath the convectively unstable layers, which we outline in the following.}

The concentration of a fluid, $\phi$, can be described by the well known diffusion equation, or Fick's second law, in the absence of a source term and with a spatially dependent diffusion coefficient as
\begin{equation}
\frac{\partial \phi({\bf r},t)}{\partial t} - \nabla (D({\bf r})\nabla \phi({\bf r},t)) = 0~,
\label{eq:Fick2}
\end{equation}
where $D$ is the diffusion coefficient. We neglect any time dependence of the diffusion coefficient as we are working with a system in a statistically steady state - at least concerning transport properties in the vertical direction. Given the spherical symmetry of a white dwarf surface we exclude the possibility of horizontal dependence in the diffusion coefficient for the purposes of this study.  This allows the horizontal information from our 3D simulations to be averaged, providing a robust statistical characterisation of trace element distributions and ultimately giving a one dimensional diffusion problem.  

For a system undertaking a true diffusion process in one dimension the mean displacement, $\langle z \rangle$, of an ensemble of particles, representing the distance the concentration has travelled, is expected to evolve according to 
\begin{equation}
 \langle z^2 \rangle = 2 D t
\label{eq:msd}
\end{equation}
where $t$ is the time over which the system evolves. The objective of this study is to calculate the depth dependent diffusion coefficients associated with trace particles beneath the convectively unstable layers.

We use two methods to quantify the mean displacement of diffusing particles, both of which will be described in the following sections. Method I utilises a module built into \cobold, designed for the study of dust formation in stellar and solar environments, whilst Method II follows more closely the methodology of \citet{freytag96} using tracer particles and serves as an independent test of diffusion.

\subsection{Method I: Tracer Density Arrays in CO$^5$BOLD}
\label{sec:dust}
Utilising the dust module included in CO$^5$BOLD \citep{freytag12} we can add extra passive scalar density arrays to relaxed simulations. These density fields have no mass, no opacity, and should thus only be advected by the local velocity fields. We note, however, that CO$^5$BOLD will adjust the time steps so that the additional density fields can be reconstructed adequately, indirectly influencing the overall numerical scheme. Microscopic diffusion velocity is orders of magnitude smaller than macroscopic velocity fields, hence it is ignored in these experiments. In any case we are only interested in the stellar layers above which microscopic diffusion takes over. Via this method we do not have actual tracer particles, but rather we store horizontally-averaged number density distributions, providing significant advantages for data storage and handling. 

Additional and non-interacting density arrays are initially added to relaxed simulations. To provide the most localised estimation of the diffusion coefficient, a tracer density is inserted as a delta function (or horizontal slab) in the z-direction such that 
\begin{equation}
 \rho_{\mathrm{t}}(z) = \begin{cases}
    10^{5} ~\mathrm{cm^{-3}}, & \text{where $z=z_0$}.\\
    10^{-6} ~\mathrm{cm^{-3}}, & \text{where $z \neq z_0$}.
  \end{cases}
\end{equation}
where $\rho_{\mathrm{t}}(z)$ is the tracer density. In our setup, Eq.~\eqref{eq:Fick2} is linear in $\phi$ so a horizontal slab is ideal for probing vertical mixing. Since these density fields are massless the actual number density is only relevant for the precision of the numerical schemes and rounding errors. The large range between the peak and background tracer densities is chosen to minimise any signature of net downward mixing due to the atmospheric density gradient. Horizontally-averaged tracer densities from simulation C3 are shown in the top panel of Fig.~\ref{fg:mquc_t135} for distributions used at the first and last timesteps of the diffusion coefficient computation. The initial distributions for the same simulation can be seen in the top panels of Fig.~\ref{fg:quc-movie-t135} where a vertical cross section of a relaxed simulation at \T{13\,5} to which the massless tracer density (green) was added is shown.

Our initial experiments showed that standing waves appeared in the base layers of our simulations, driven by convection and trapped between the convectively unstable layer and bottom box boundary. This can be seen in Section~\ref{sec:pathint} {(see Fig.~\ref{fg:pathint_t120})} by the shallowing of the gradient at depths of $z\lessapprox -1.5 \mathrm{\,km}$ (grey shaded regions) in the vertical velocity components of the simulation. We find that across all simulations the modes are either above or below the acoustic cut-off frequency, suggesting both p-modes, oscillations with pressure as the primary restoring force, and g-modes, oscillations with gravity as the primary restoring force, can be excited. In particular for the cooler models ($T_{\mathrm{eff}} \leq $12\,000 K), in the region beneath the unstable layers, we favour the interpretation of g-modes \citep{freytag10}. {Similar wave effects have also been observed in deep white dwarf simulations by other groups \citep{kupka18} where the authors found evidence to suggest the presence of both p- and g-modes.} A comprehensive discussion on the treatment of these waves and extracting a conservative estimate of the diffusion coefficient in this region can be found in Section \ref{sec:filtering}.

Preliminary results showed that the tracer densities were susceptible to enhanced mixing in the wave region. Multi-dimensional simulations of AGB stars have provided evidence that g-modes may be capable of mixing material \citep{freytag10}. In order to ascertain whether the additional mixing is borne of waves or numerical diffusion a comprehensive study of the wave properties would be required, something beyond the scope of this work. Hence for the purposes of direct diffusion experiments we prioritise probing the region directly above the top of the wave region. The mixing timescale in the convectively unstable region, with convective velocities of $\sim$ 1\,km s$^{-1}$, is expected to be extremely short compared to timescales of accretion - effectively providing instantaneous mixing. As such, we prioritise probing the region beneath this, where $z < 0$~km. The convention used throughout this work is that our coordinate system is defined such that the lower Schwarzschild boundary lies at $z=0$~km.

We focus our attention on the seven simulations detailed in Table \ref{tb:main}, all deep enough to fully enclose the convectively unstable layers and at least 4 pressure scale heights beneath the convectively unstable layers, defined by the region where the entropy gradient with respect to depth is negative, i.e., \zst\ $> z >0$~km. The tracer density method is implemented on all simulations in the table, from which diffusion coefficients are directly derived. Simulations A1, B2 \& C3, corresponding to \teff $= 12\,000, 13\,000\,\, \& \,13\,500 \,\mathrm{K}$, respectively, provide results for macroscopic diffusion across a range of temperatures with a fixed grid size ($250^3$). As an independent test of the results, these three simulations are analysed using the method of path integration, the methodology of which will be discussed in detail in Section~\ref{sec:pathint}. 

Simulations C1 \& C2, with \T{13\,5}, are identical to C3, with the exception of grid resolution - and any numerical scheme adjustments, such as time steps, arising from this change in grid size. They are both analysed with the tracer density and path integration methods. The box sizes of $150^3$ and $250^2 \times 150$, respectively, serve as a convergence test of the results with respect to spatial resolution. Simulations B1 \& B2, with \T{13\,0} and grid sizes $150^3$ \& $250^3$, respectively, provide a similar test at a different effective temperature. Finally, simulation C1-2 is identical to C1, except for an extension in the x and y directions by a factor of two. This can provide a further convergence test and also inform us on the nature of the standing waves present in the bottom of the box. 

\subsubsection{Tracer Density Temporal Evolution}
A given tracer density array, inserted as a horizontal slab, is rapidly ($\sim \mu s$) smoothed by the velocity field, where the less deep and more vigorously convective layers promote this smearing to a greater extent. This is demonstrated in Fig.~\ref{fg:quc-movie-t135} for simulation C3 (see Table~\ref{tb:main}) at \T{13\,5}. Here we show snapshots of a 2D vertical slice through 8\,s of simulation time from the beginning of the diffusion experiment. Tracer densities (green) inserted 1.2\,km (left) and 1.0\,km (right) beneath the unstable layers can be seen to be advected by the convective flows (blue). The figure shows a range in tracer density of $1\leq\rho_{\mathrm{trace}}/\mathrm{[cm^{-3}]}\leq10^{5}$ with darker green corresponding to a higher tracer density. Clearly some of the distribution which began closer to the boundary of instability becomes incorporated into the unstable layers ($z \geq 0$\,km) within $t \approx 8\,\mathrm{s}$, though as the tracer density is depicted logarithmically, the positional mean of the distribution remains near the initial position. We show the same evolution in Figs.~\ref{fg:quc-movie}~\&~\ref{fg:quc-movie-t120} for simulations B1 and A1 (13\,000 K and 12\,000 K), respectively.

From the analytic solution of Eq.~\eqref{eq:Fick2}, in an ideal diffusion scenario an initially narrow distribution is expected to evolve in the form of a Gaussian for times $t > 0$. The spatial extent of the density fields in our simulations is quantified by the density-weighted standard deviation of the tracers, \zstarms, which is given by
\begin{equation}
 z^{*}_{\mathrm{rms}} = \left(\frac{n_z}{n_z-1}\sum_{z}\left(\rho_{\mathrm{t}}(z)(z - \langle z^* \rangle)^{2}\right)\middle/ \sum_{z} \rho_{\mathrm{t}}(z)\right)^{1/2}
\end{equation}

where $\rho_{\mathrm{t}}(z)$ is the horizontally-averaged, depth dependent tracer density, $n_z$ is the number of cells in the vertical dimension and $\langle z^* \rangle$ is the density-weighted mean depth of the ensemble given by
\begin{equation}
 \langle z^* \rangle = \left(\sum_{z} z \rho_{\mathrm{t}}(z)\right) \cdot \left(\sum_{z} \rho_{\mathrm{t}}(z)\right)^{-1}
\end{equation}

\begin{figure}
\centering
\subfloat{\includegraphics[width=1.0\columnwidth]{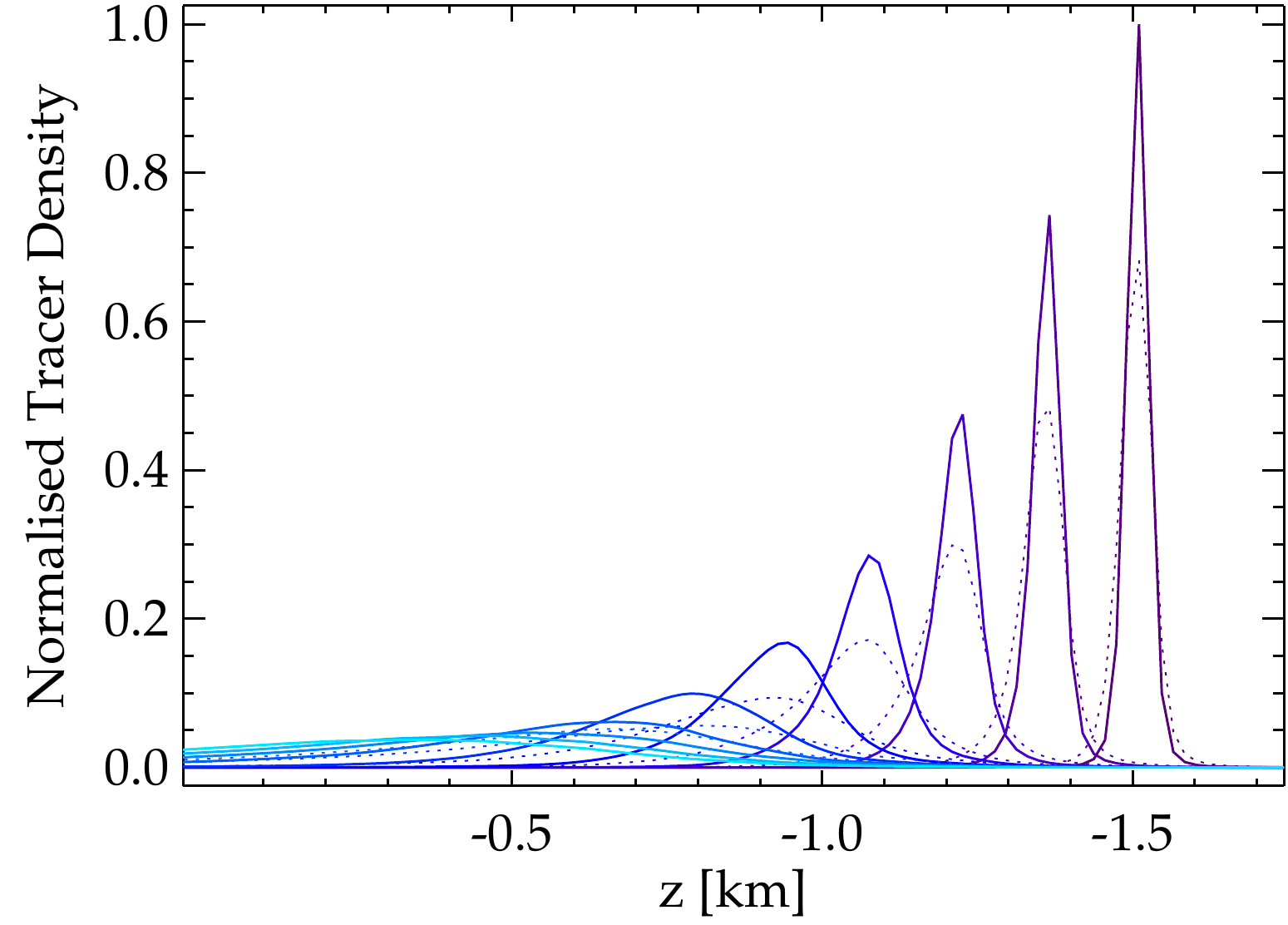}}\\
\subfloat{\includegraphics[width=1.0\columnwidth]{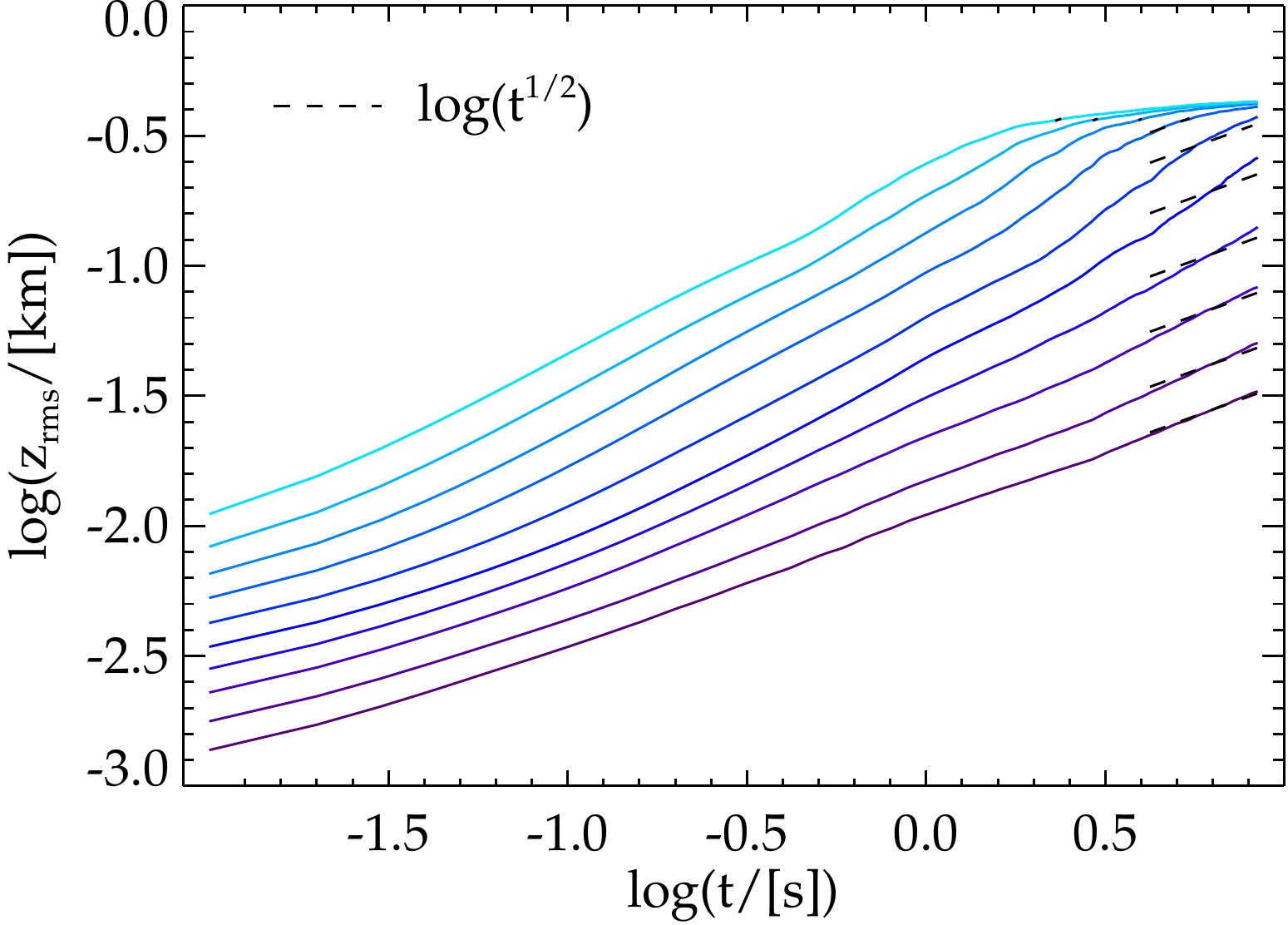}} \\
\subfloat{\includegraphics[width=1.0\columnwidth]{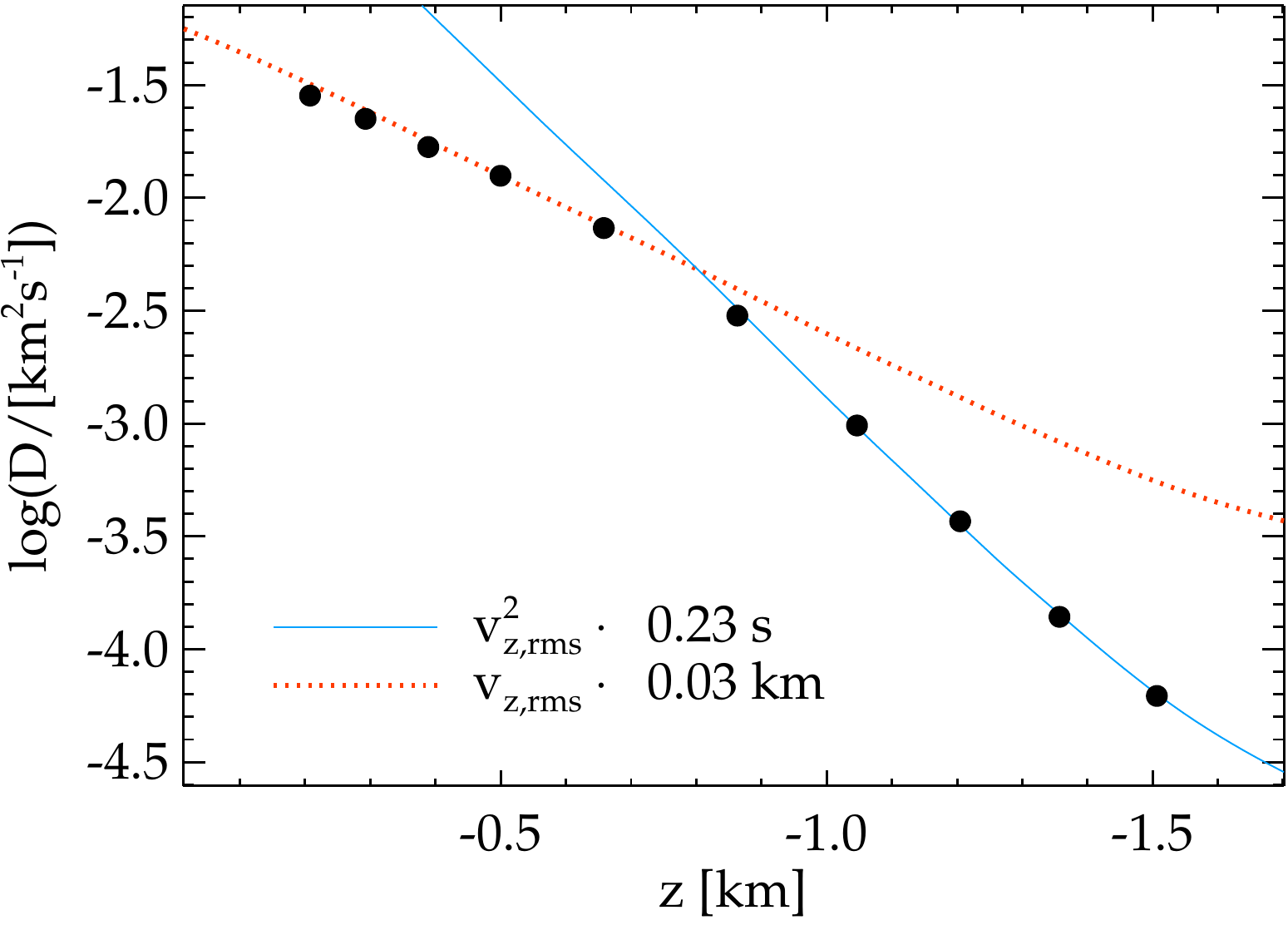}} \\

\caption{Analysis of massless density arrays implemented with the \cobold \, tracer density for a simulation at \teff = 13\,500\,K (Table~\ref{tb:main}; C3) with \logg = 8.0 and box size $250^3$. {\it Top:} Horizontally-averaged tracer number density profile at first (solid) and last (dashed) time step used in \thalf fitting (see middle panel). Colours correspond to tracer densities added at different depths. {\it Middle:} Evolution of the ensemble spread characterised by the tracer density-weighted standard deviation, \zstarms. Gaussian spread (solid) and a best fit of \thalf (dashed). {\it Bottom:} Diffusion coefficients (circles) computed using Eq.~\ref{eq:fwhm-D}. Vertical velocity profiles \vzrms\ and \vsq\ are shown in orange (dashed) and blue (solid), respectively. The depths on the $x$-axis have been adjusted such that the lower Schwarzschild boundary lies at $z=0$~km. We note that the full extent of the $x$-axis is considered convectively stable under the Schwarzschild criterion, yet considerable mixing is observed.} 
\label{fg:mquc_t135}
\end{figure}

\begin{figure*}
\centering
\subfloat{\includegraphics[width=0.48\textwidth]{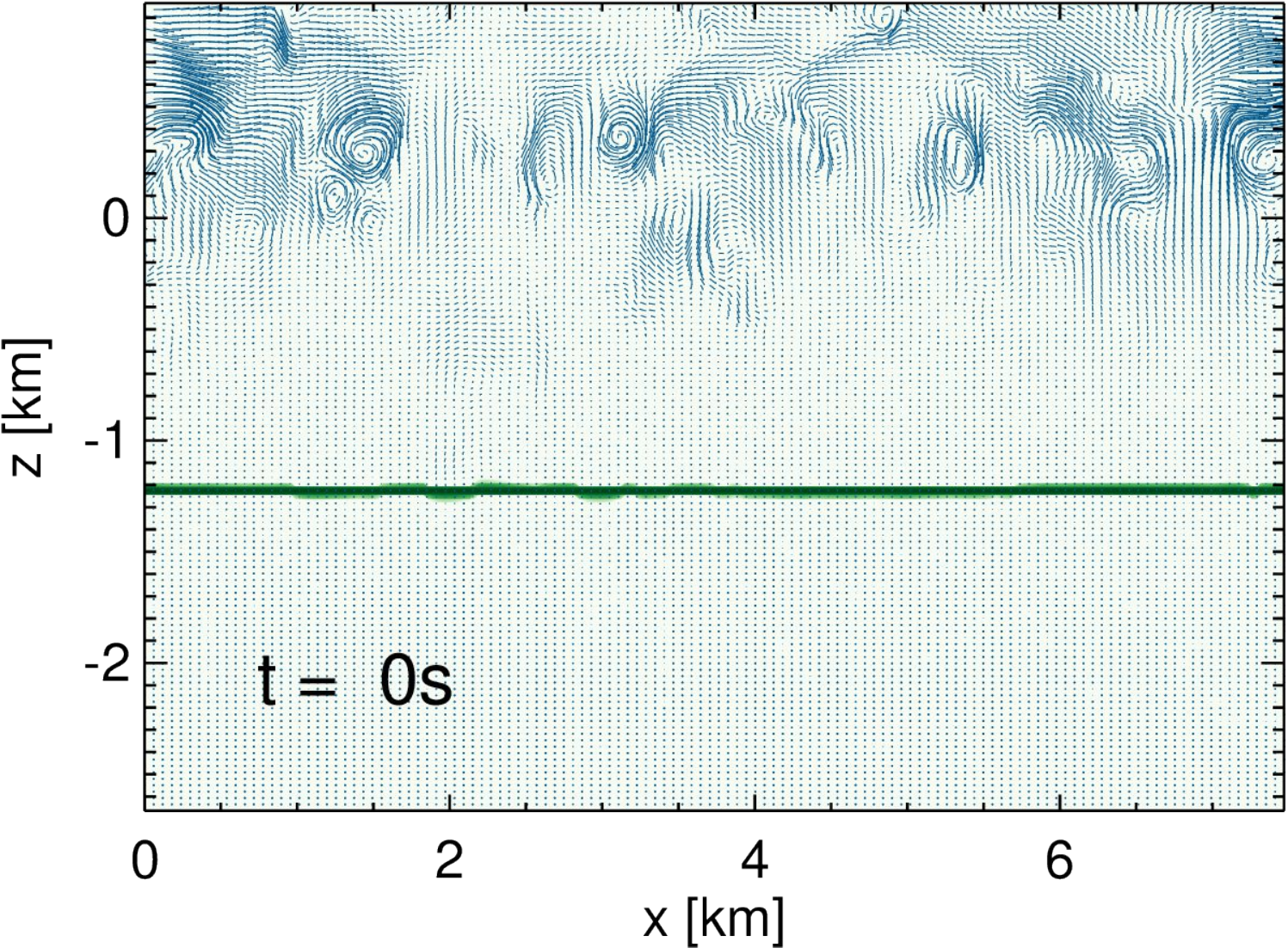}} \hfill
\subfloat{\includegraphics[width=0.48\textwidth]{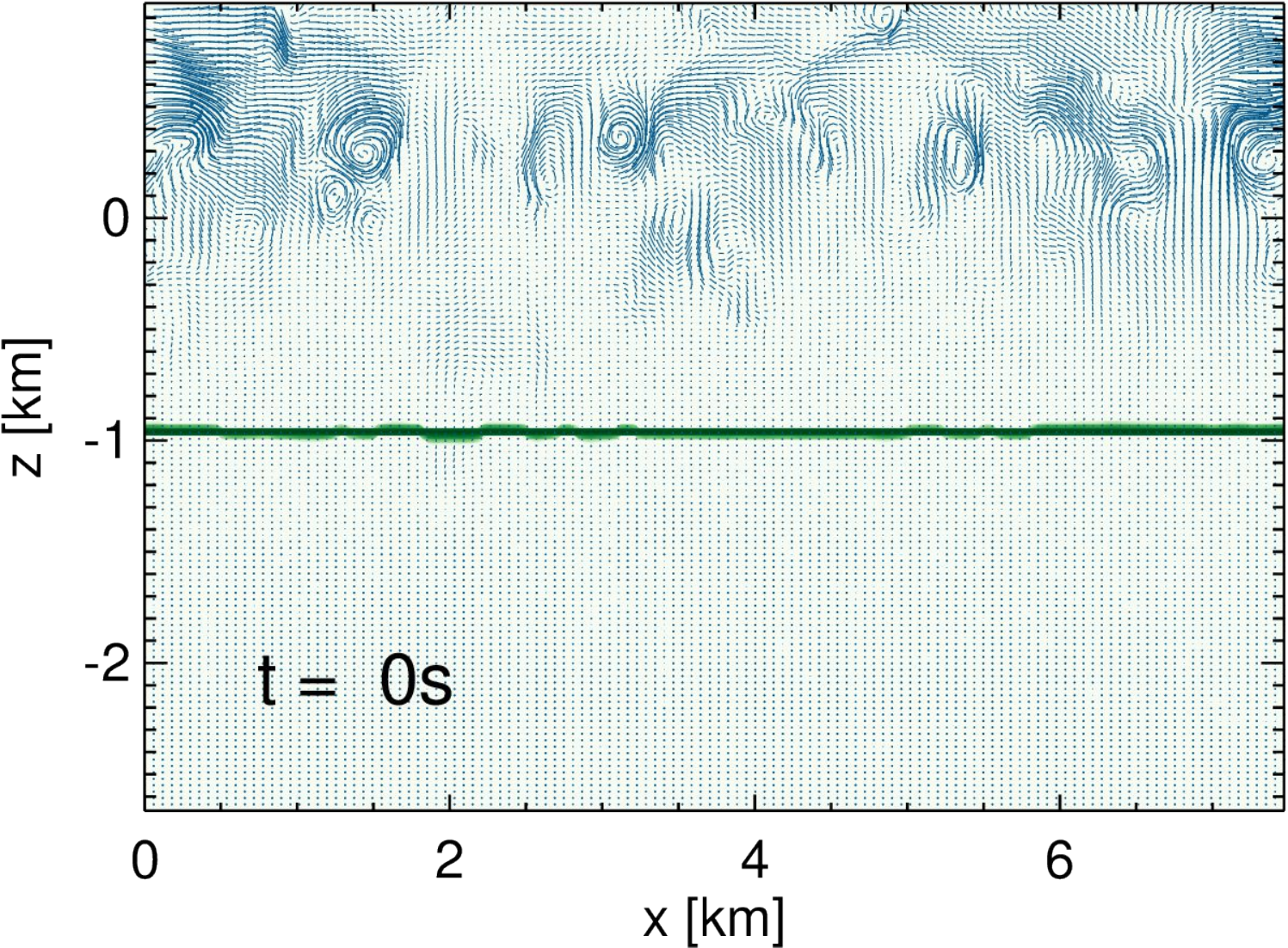}}
\\
\subfloat{\includegraphics[width=0.48\textwidth]{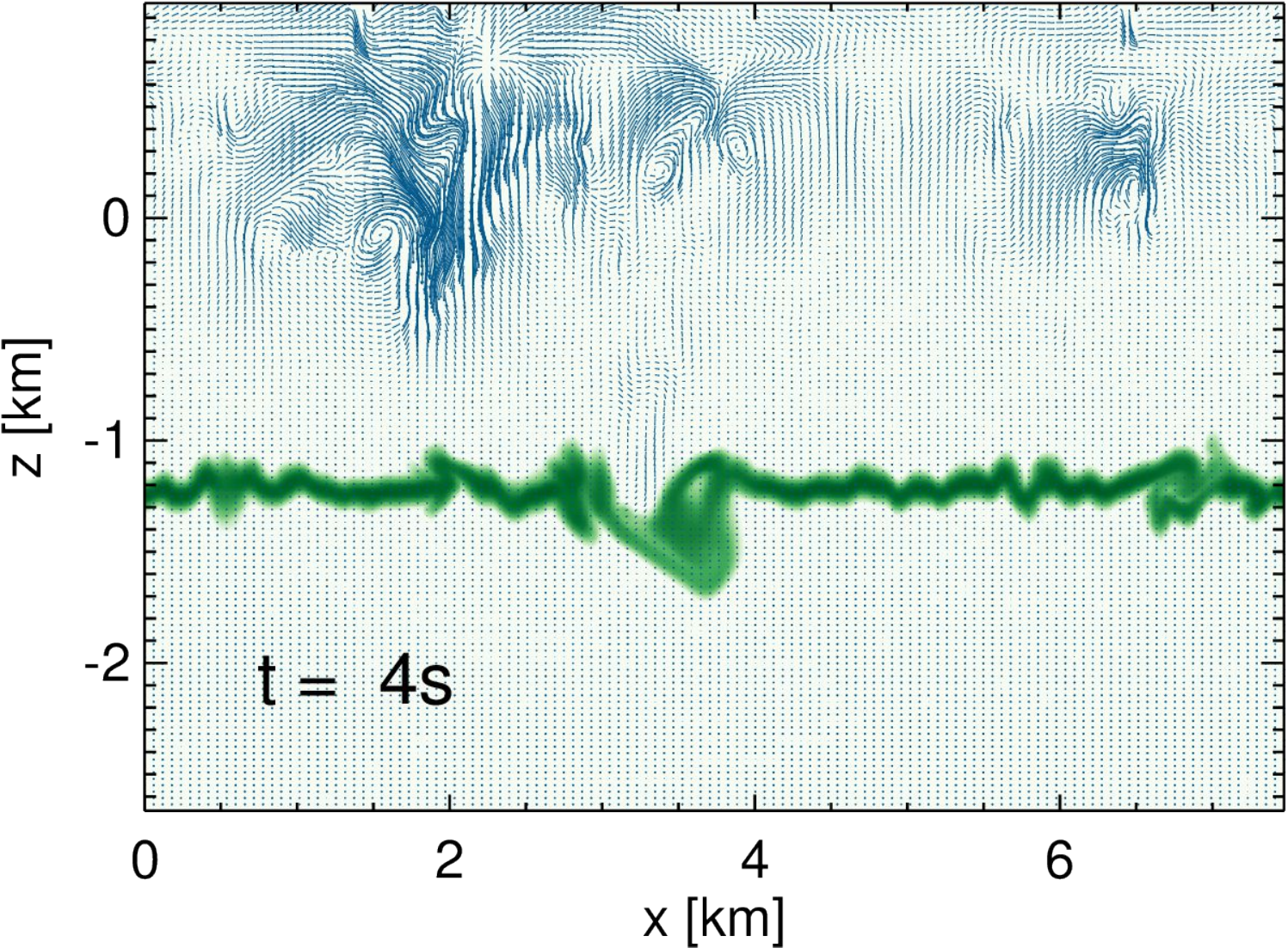}} \hfill
\subfloat{\includegraphics[width=0.48\textwidth]{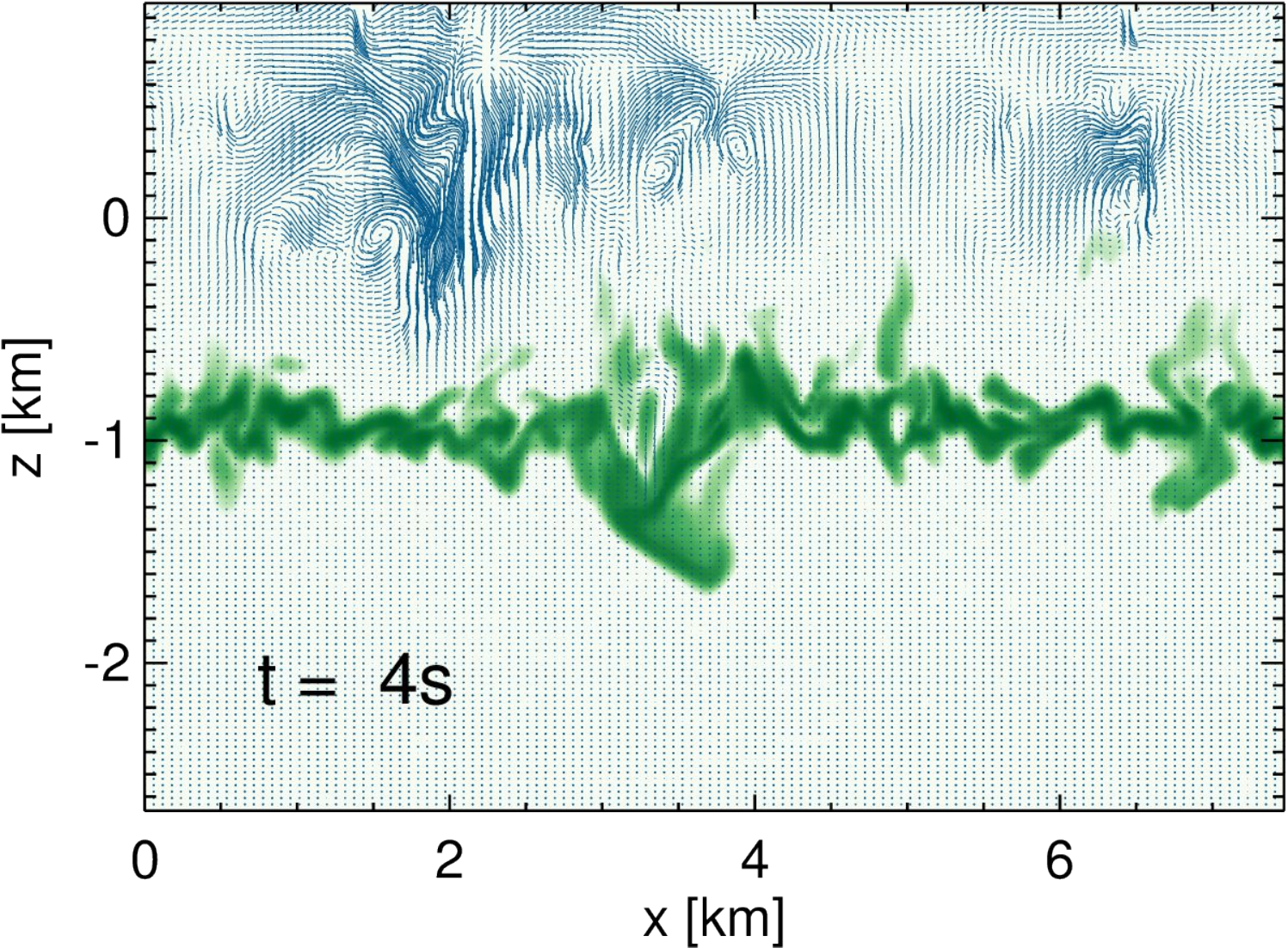}}
\\
\subfloat{\includegraphics[width=0.48\textwidth]{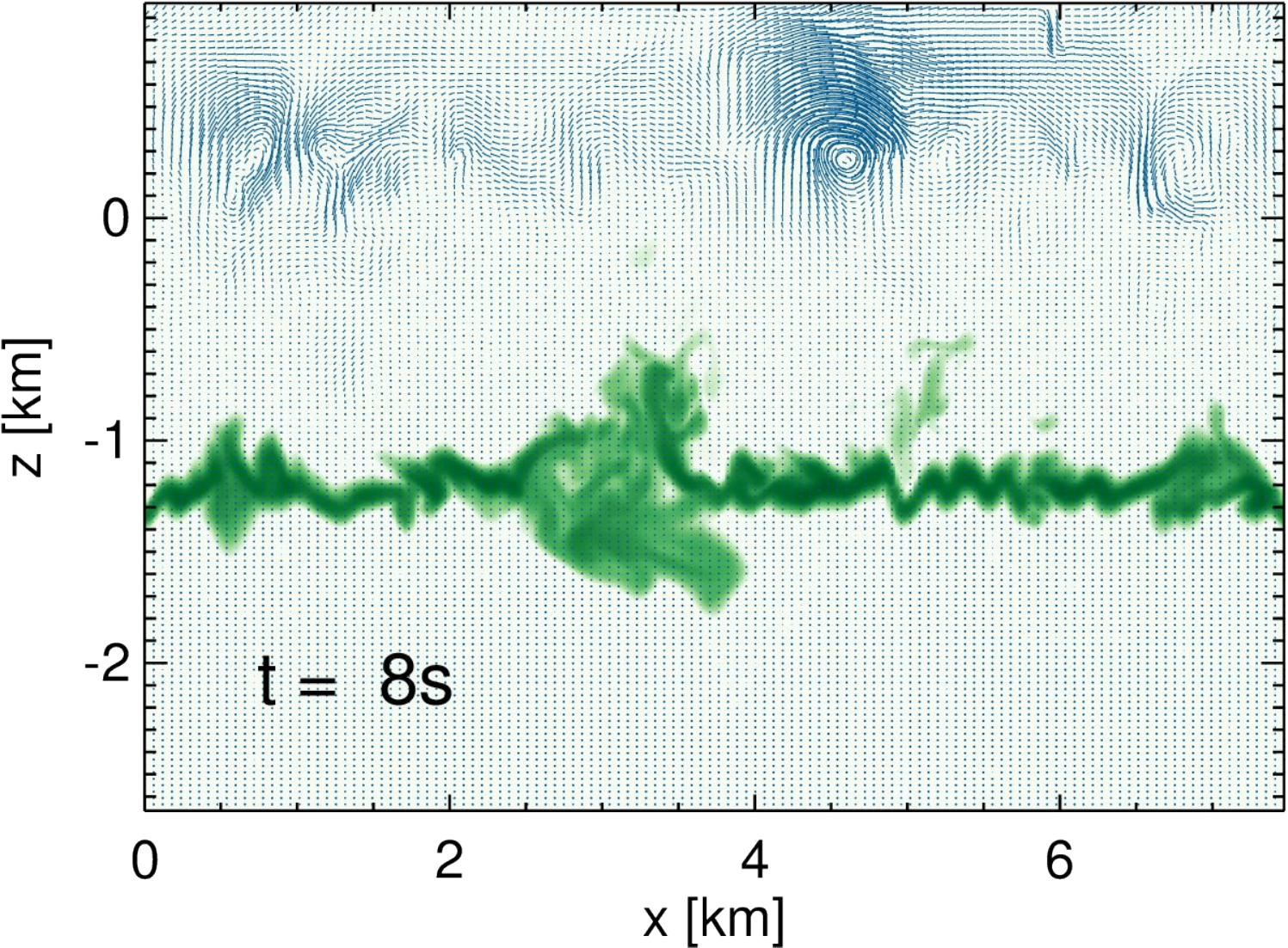}} \hfill
\subfloat{\includegraphics[width=0.48\textwidth]{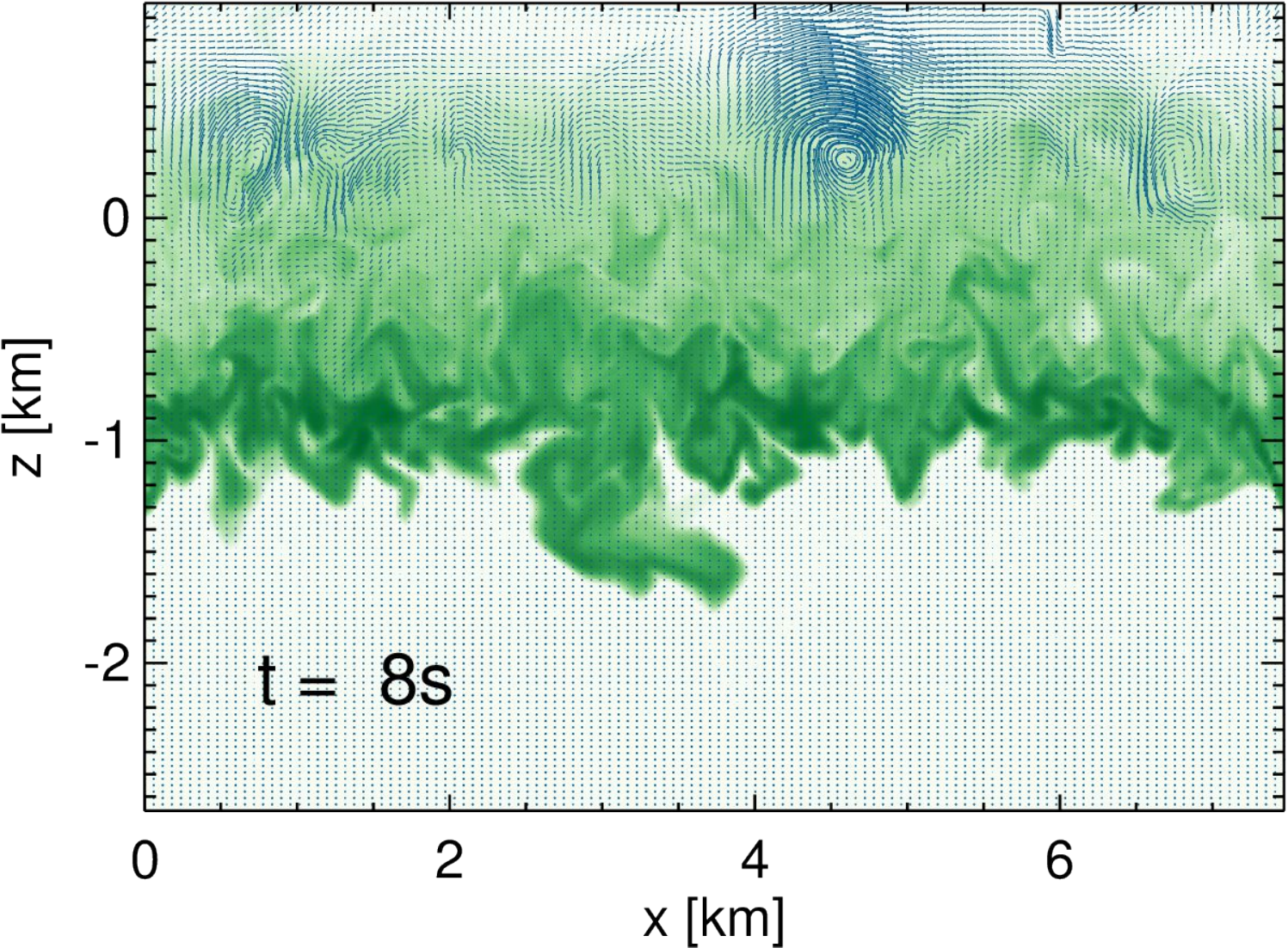}}
\\
\caption{Demonstration of the tracer density implementation with \cobold\ for a simulation at \teff = 13\,500\,K (Table \ref{tb:main}, C3) with \logg = 8.0 and box size $250^3$. Snapshots of a 2D vertical slice through the simulation show the logarithmic tracer density (green) for two (left and right) of the multiple density arrays added to the simulation. Only values of $\rho_{\mathrm{trace}}\geq1$\,cm$^{-3}$ are shown. Convective velocities are shown in blue, where the magnitude is linear with line length. The depths on the $y$-axis have been adjusted such that the lower Schwarzschild boundary lies at $z=0$~km. Animated versions of this figure are currently available at https://warwick.ac.uk/fac/sci/physics/research/astro/people/cunningham/movies/}

\label{fg:quc-movie-t135}
\end{figure*}

\begin{figure*}
\centering
\subfloat{\includegraphics[width=0.48\textwidth]{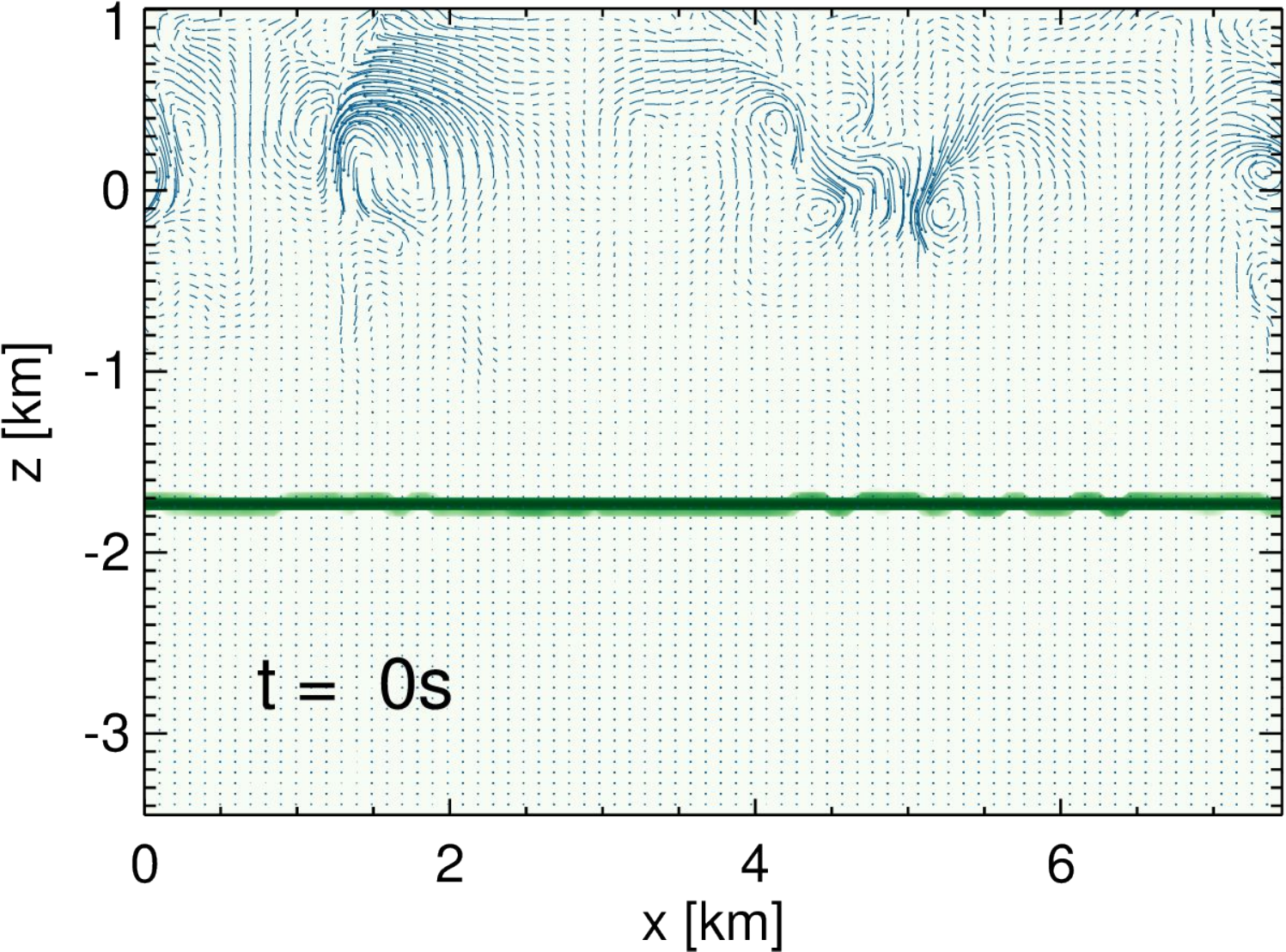}} \hfill
\subfloat{\includegraphics[width=0.48\textwidth]{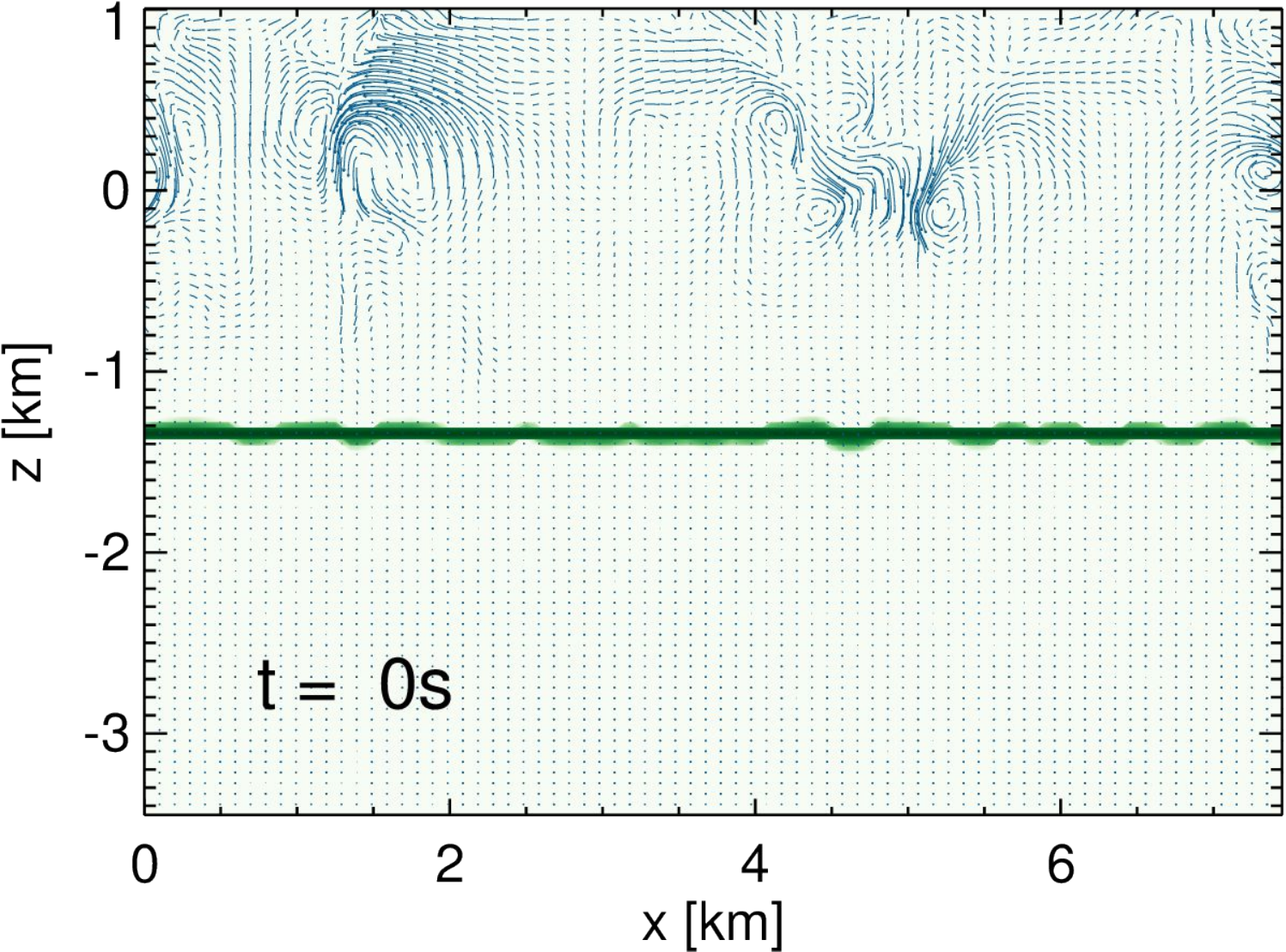}} \\

\subfloat{\includegraphics[width=0.48\textwidth]{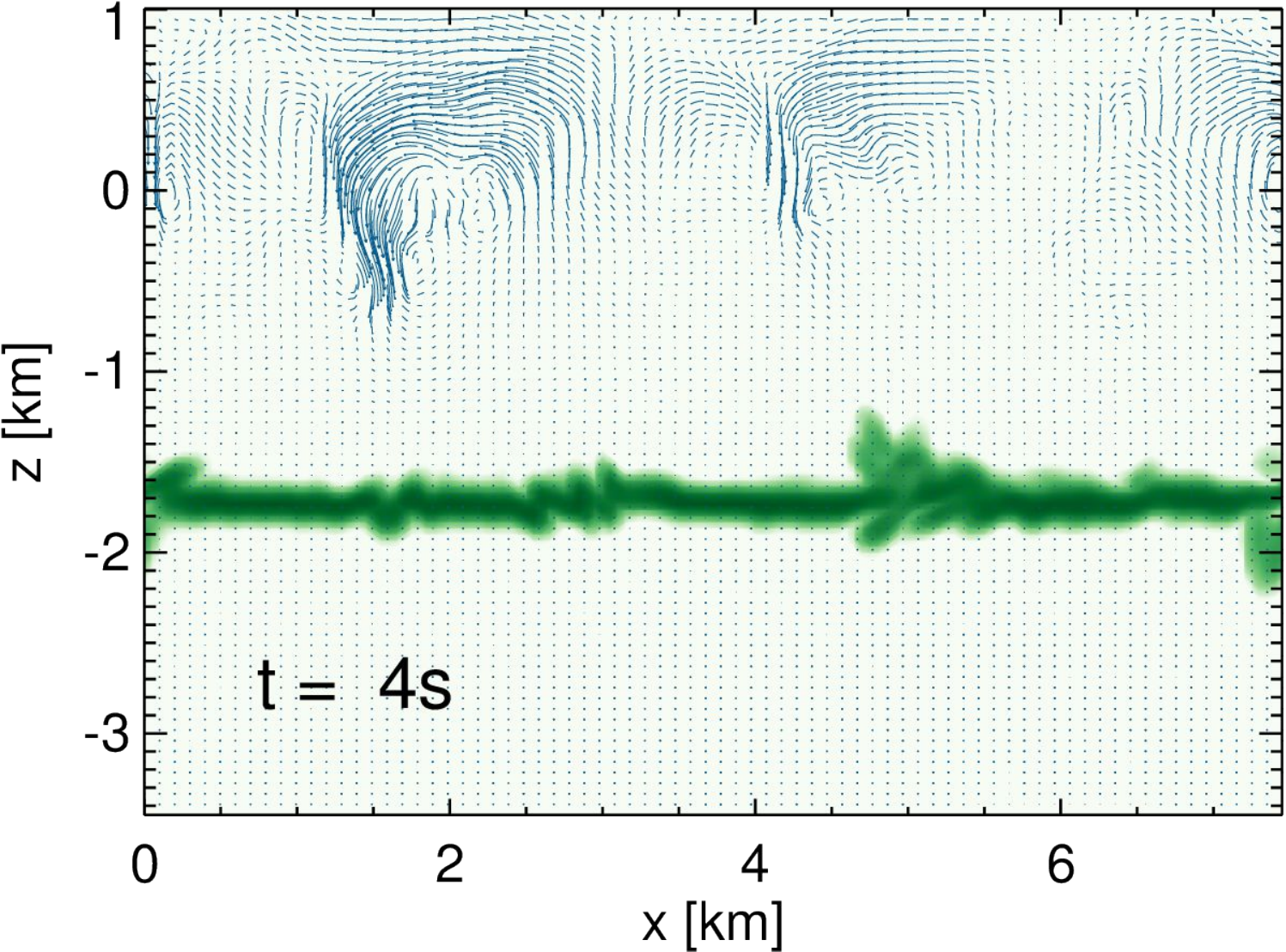}} \hfill
\subfloat{\includegraphics[width=0.48\textwidth]{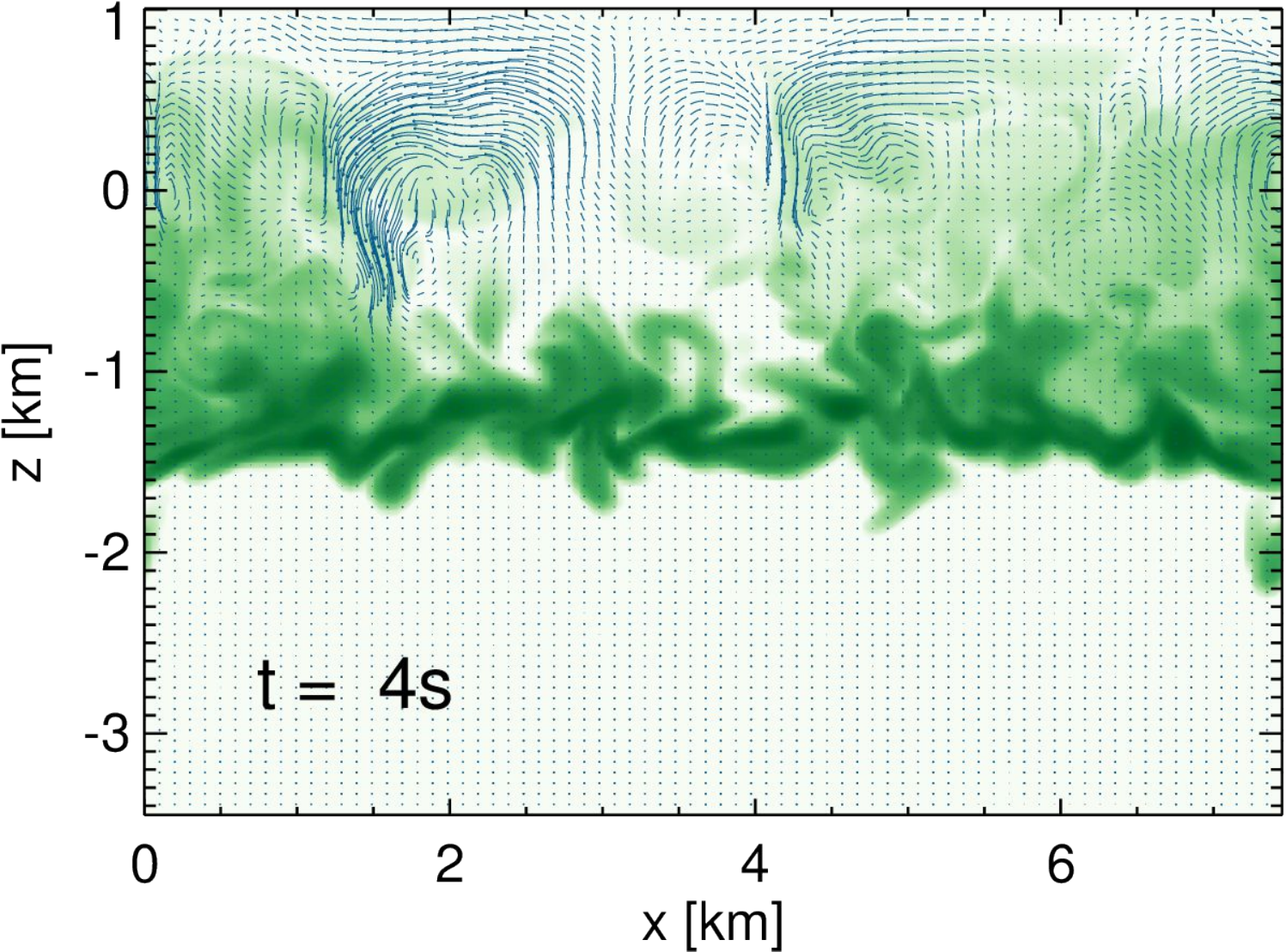}} \\

\subfloat{\includegraphics[width=0.48\textwidth]{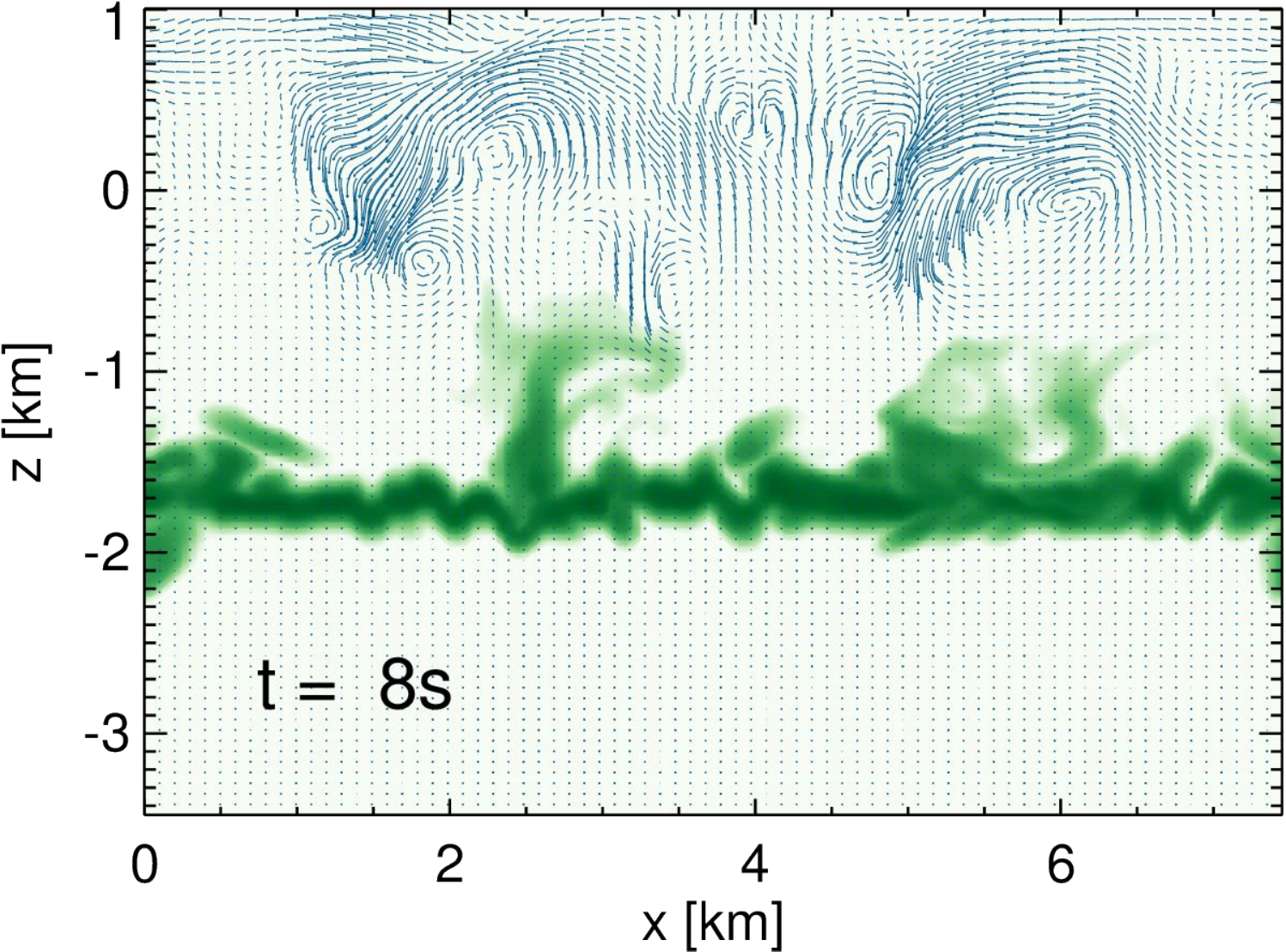}} \hfill
\subfloat{\includegraphics[width=0.48\textwidth]{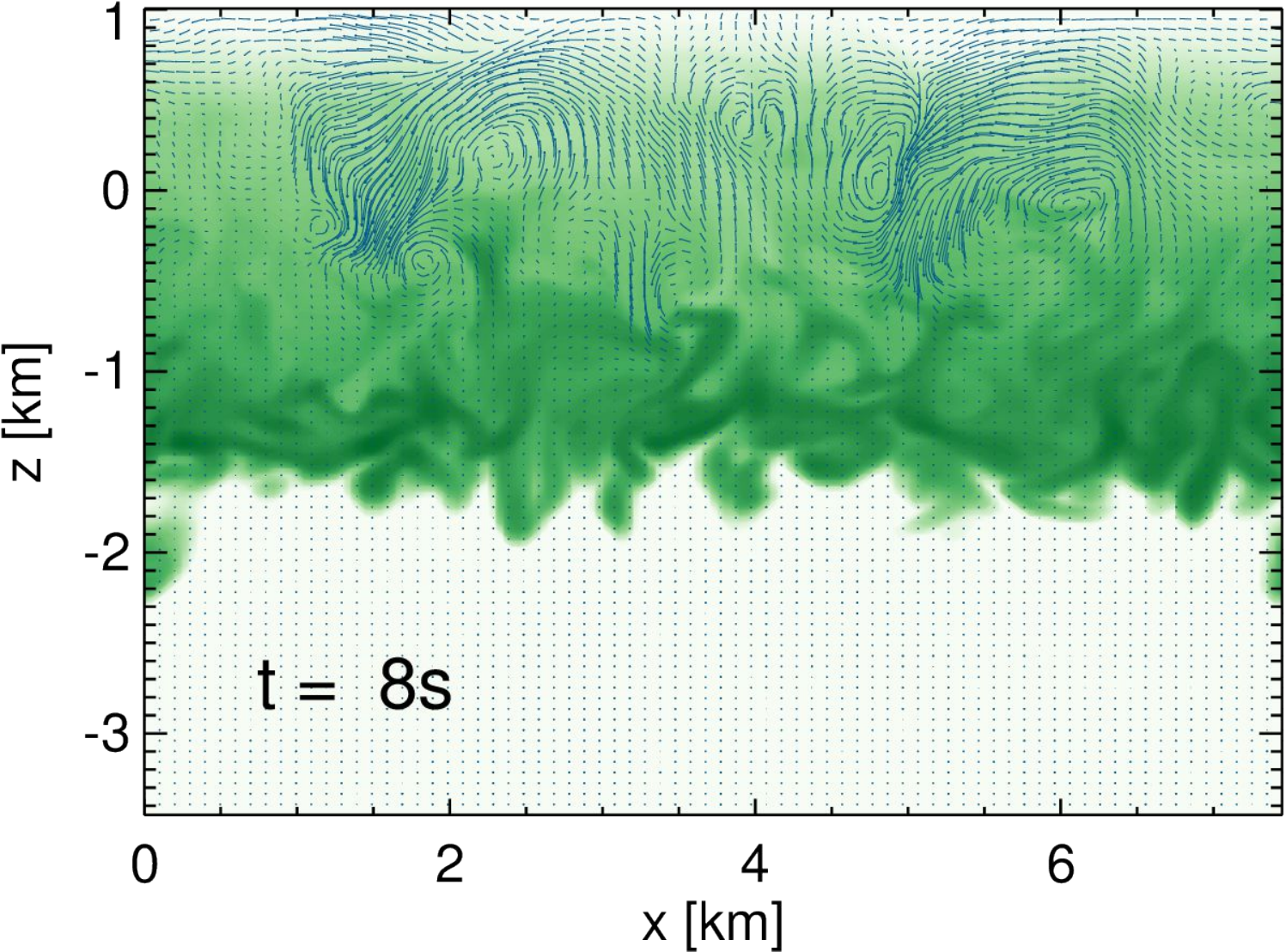}}\\

\caption{Similar to Fig.~\ref{fg:quc-movie-t135} for a simulation at \teff = 13\,000\,K (Table \ref{tb:main}, B1) with \logg = 8.0 and box size $150^3$.}

\label{fg:quc-movie}
\end{figure*}

\begin{figure*}
\centering
\subfloat{\includegraphics[width=0.48\textwidth]{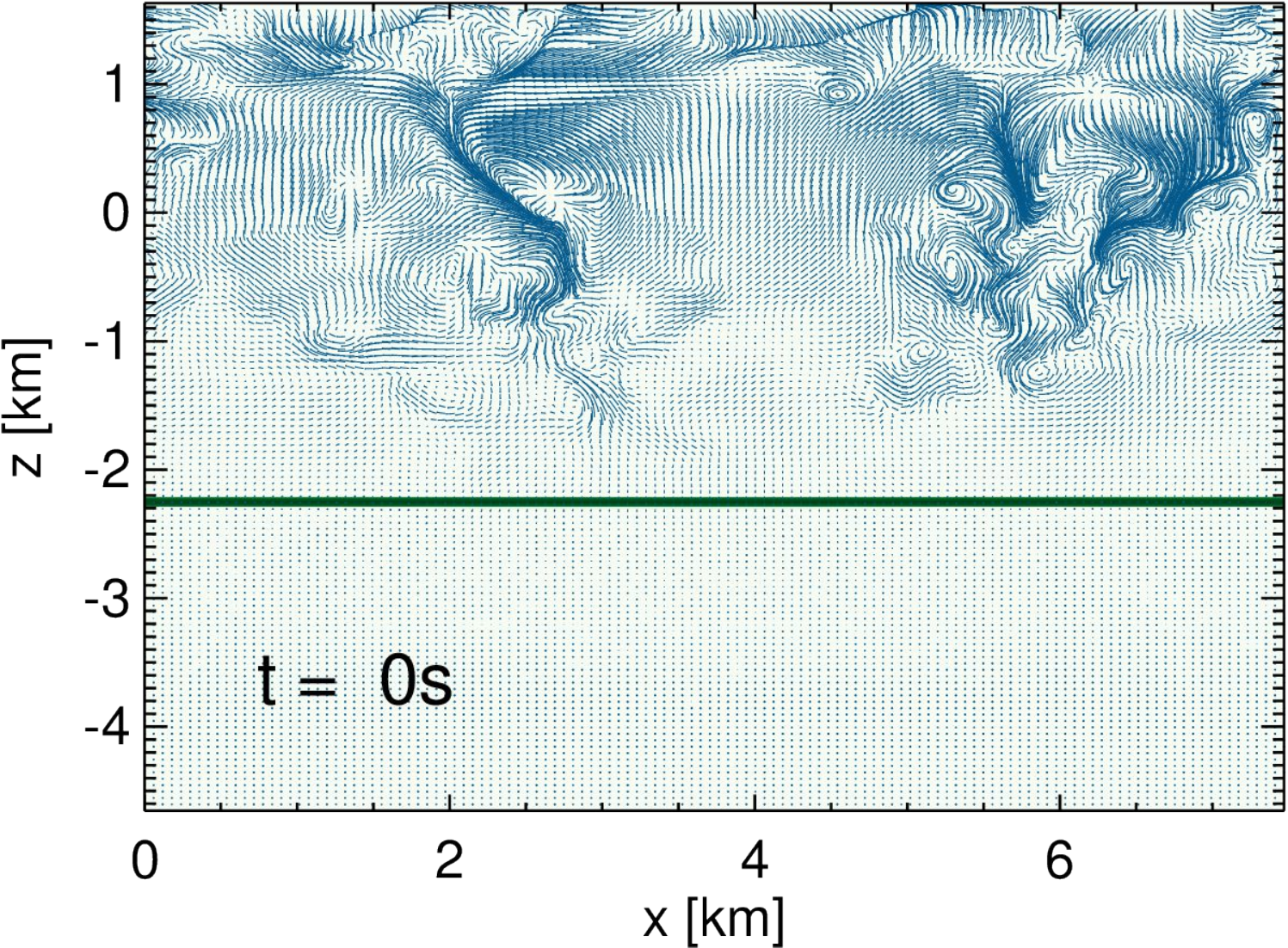}} \hfill
\subfloat{\includegraphics[width=0.48\textwidth]{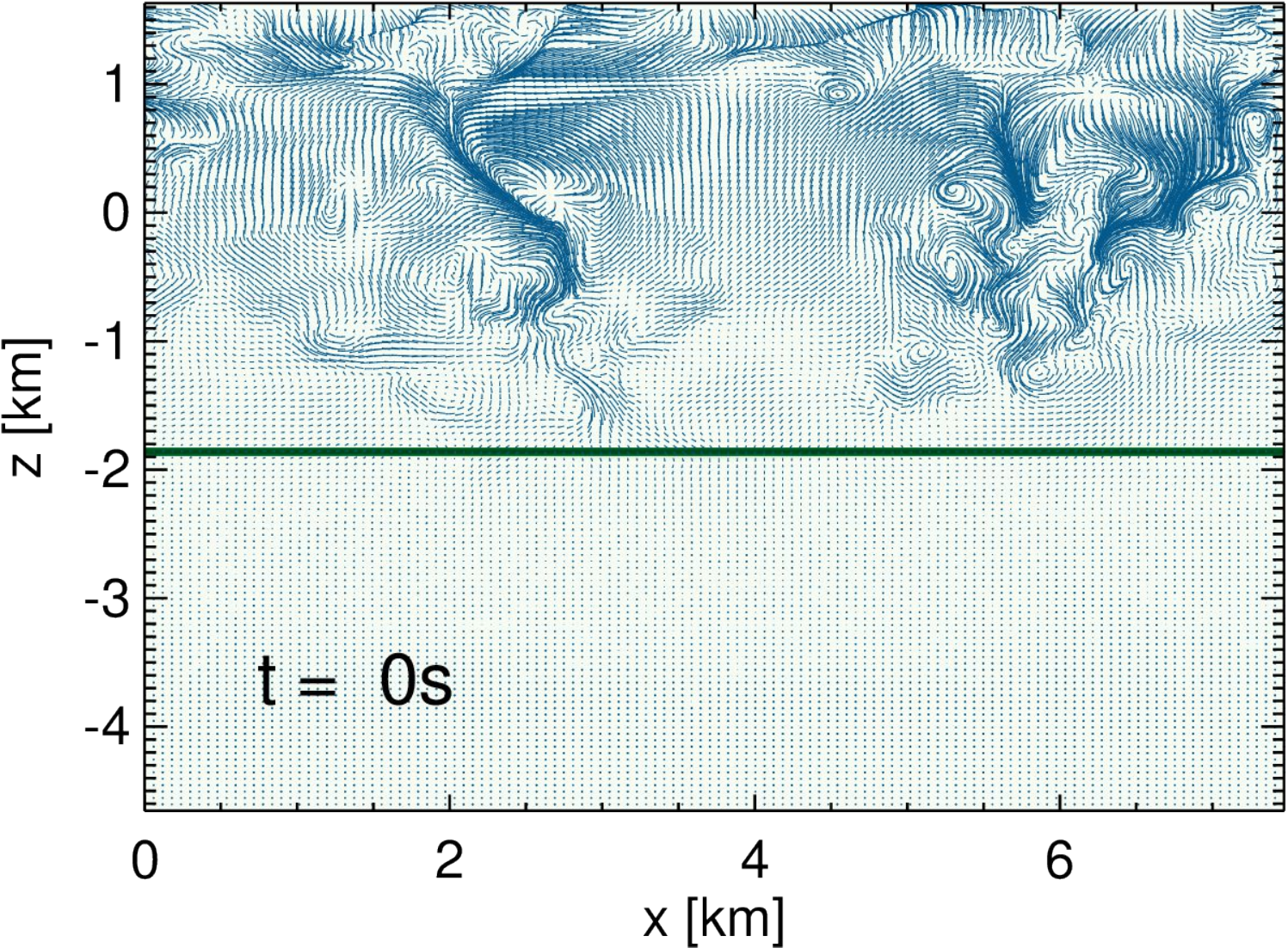}} \\
\subfloat{\includegraphics[width=0.48\textwidth]{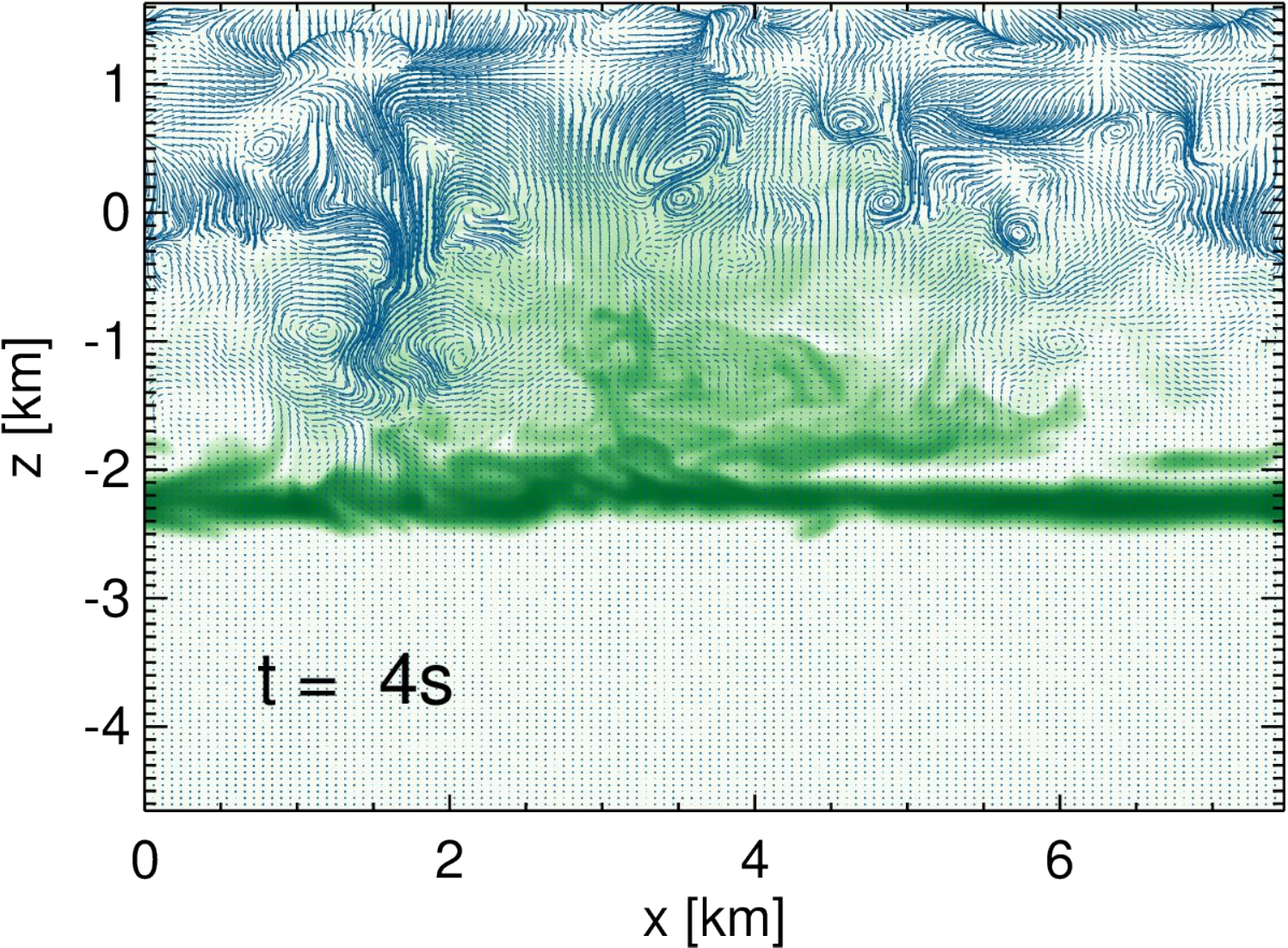}} \hfill
\subfloat{\includegraphics[width=0.48\textwidth]{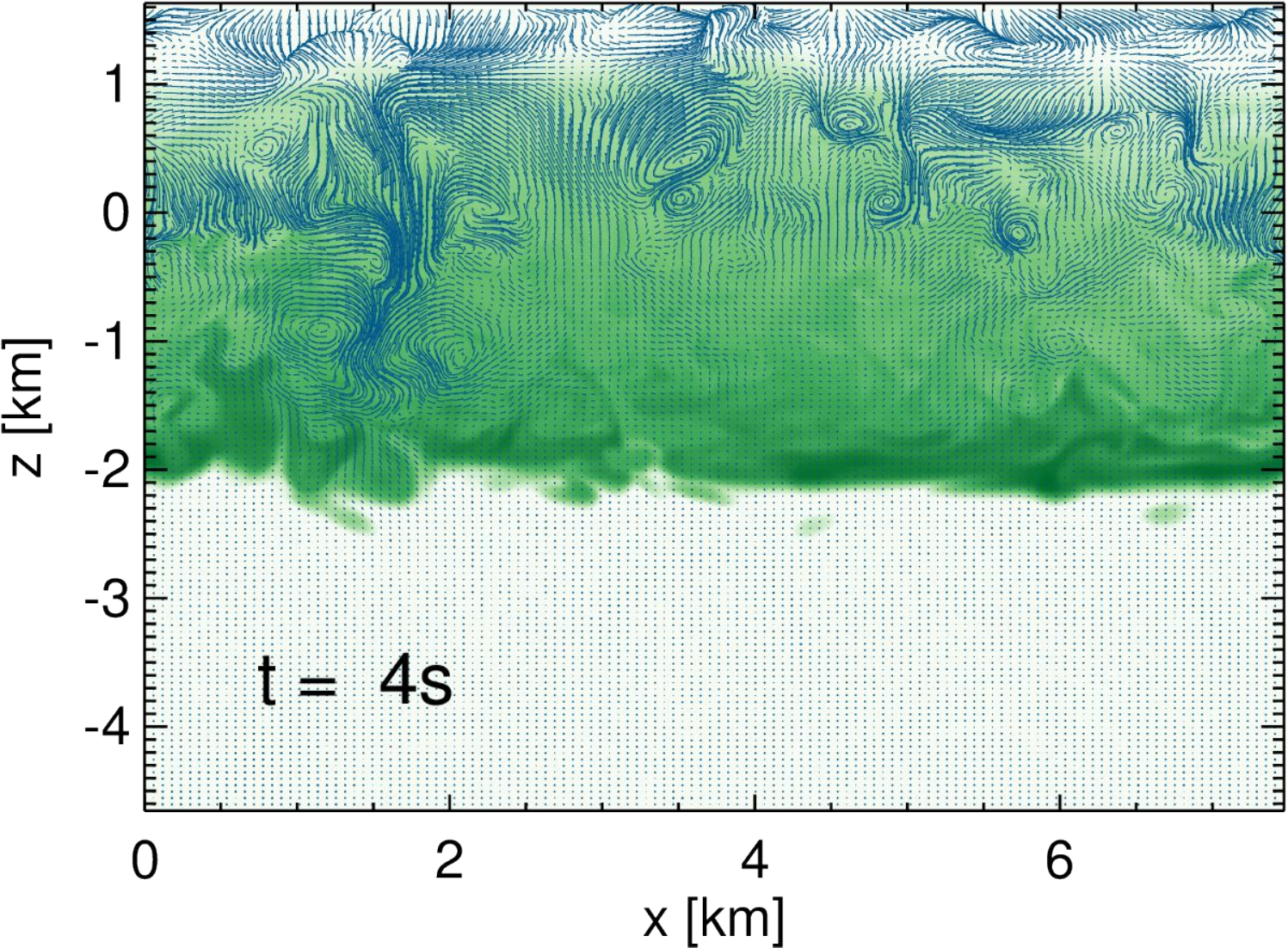}} \\
\subfloat{\includegraphics[width=0.48\textwidth]{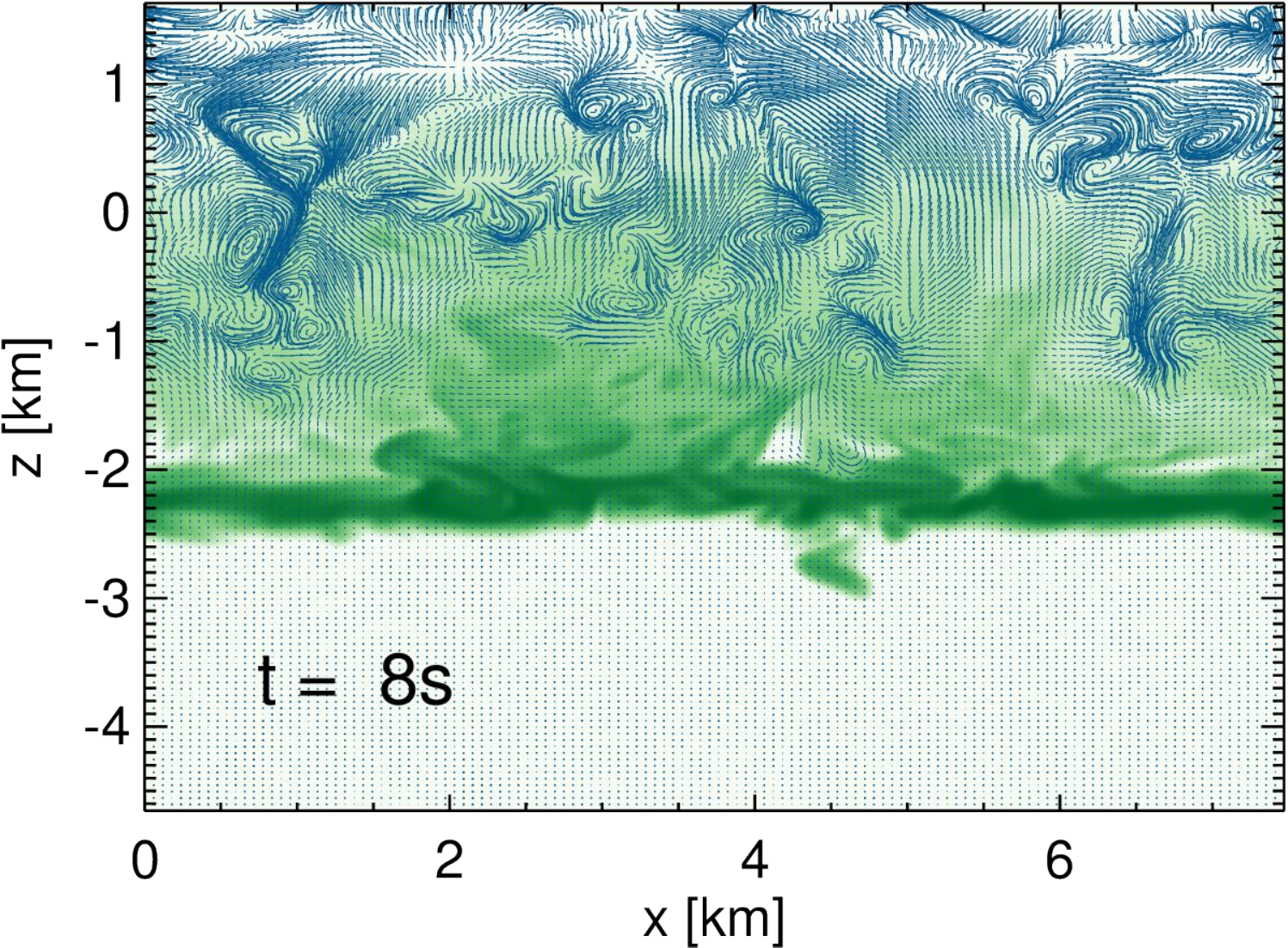}} \hfill
\subfloat{\includegraphics[width=0.48\textwidth]{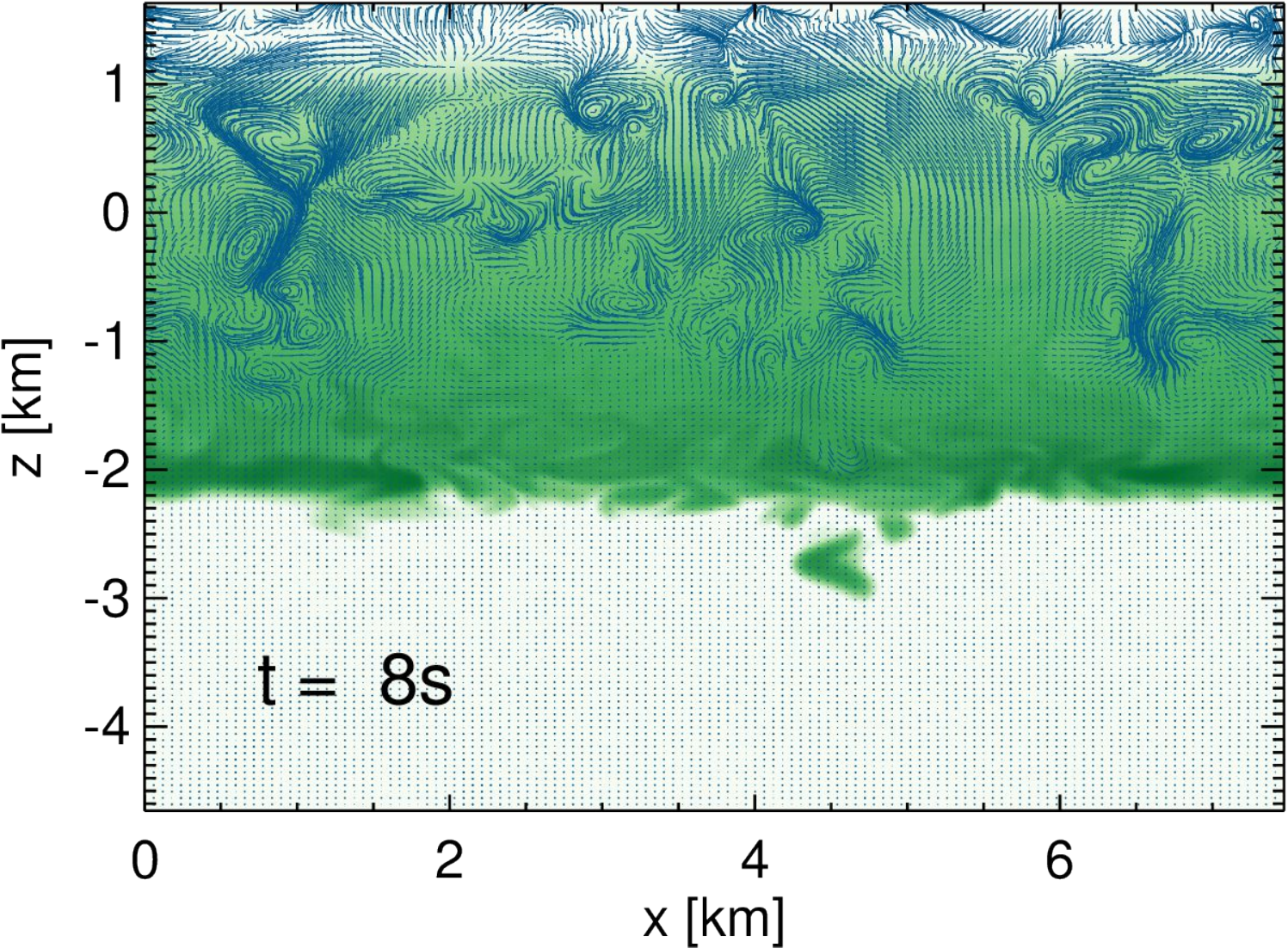}} \\
\caption{Similar to Fig.~\ref{fg:quc-movie-t135} for a simulation at \teff = 12\,000\,K (Table \ref{tb:main}, A1) with \logg = 8.0 and box size $250^3$.}

\label{fg:quc-movie-t120}
\end{figure*}

To ascertain whether the ensemble of particles is evolving through a true diffusion process we plot the density-weighted standard deviation, \zstarms, as a function of time (see Fig.~\ref{fg:mquc_t135}; middle panel). It is expected that for true diffusion, the square of the ensemble spread will evolve proportionally with time and we show the \thalf fits made to the \zstarms\ evolutions with dashed lines. The depth dependent diffusion coefficient, $D(z)$, is thus derived from Eq.~\eqref{eq:msd} to be 
\begin{equation}
 2\log_{10}(z^{*}_{\mathrm{rms}}(z,t)) = \log_{10}(t) + \log_{10}(2D(z))~.
 \label{eq:fwhm-D}
\end{equation}

The bottom panel of Fig.~\ref{fg:mquc_t135} shows the derived diffusion coefficients (circles) for the region extending 1.5~km beneath the lowest convectively unstable layer for simulation C3. Also shown are root mean square vertical velocity profiles - \vzrms\ (dotted, orange) and \vsq\ (solid, blue) - time-averaged over the duration for which \thalf fits are made (middle panel - dashed). It can be seen that the decline of the diffusion coefficient as a function of depth follows $D(z)=$ \vzrms\,$\cdot$ 0.03~km in the near-overshoot region, extending 0.8~km beneath the lower Schwarzschild boundary. For deeper layers, where $z<-0.8$~km, the diffusion coefficient is well described by $D(z)=$ \vsq\,$\cdot$ 0.23~s.

\begin{figure*}
\subfloat{\includegraphics[width=0.48\textwidth]{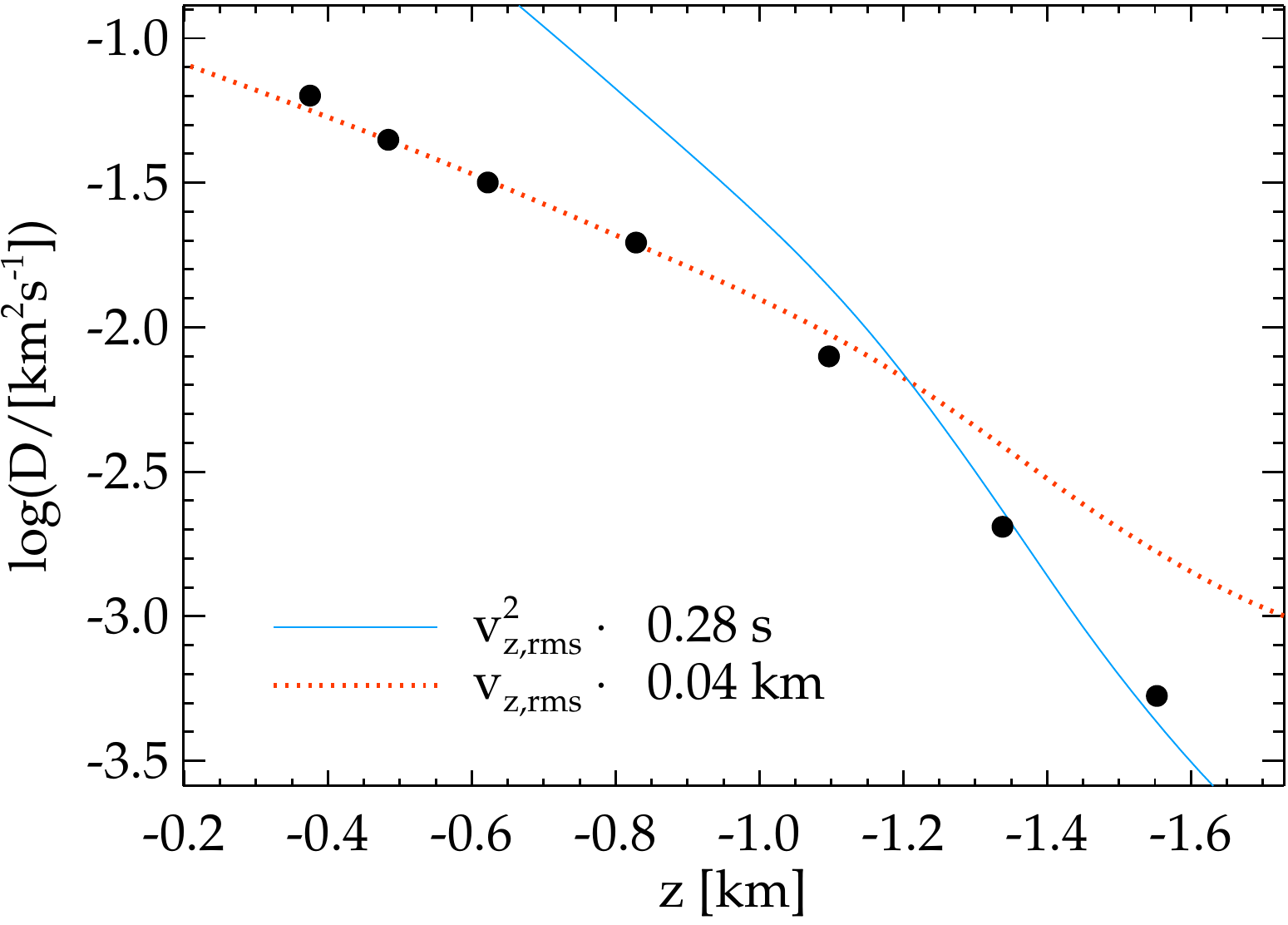}}
\hfill
\subfloat{\includegraphics[width=0.48\textwidth]{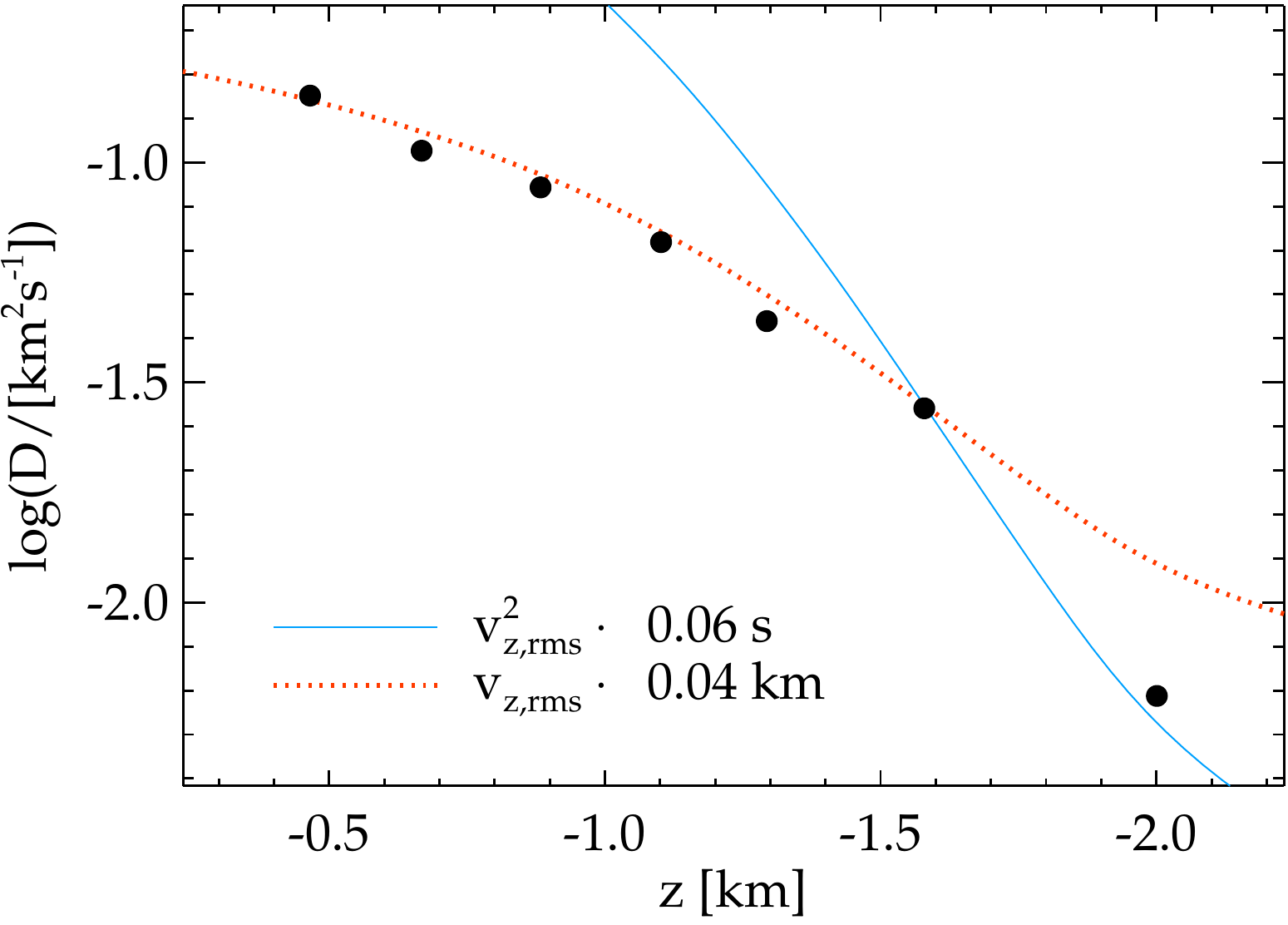}}

 \label{fg:dover_t120}
 \caption{Similar to lower panel of Fig.~\ref{fg:mquc_t135} for simulations at \teff = 13\,000\,K (left, Table~\ref{tb:main}, B2) and \teff = 12\,000\,K (right, Table~\ref{tb:main}, A1), both with \logg = 8.0.}
\label{fg:mquc_t130/t120}
\end{figure*}

Fig.~\ref{fg:mquc_t130/t120} shows diffusion coefficients derived via this method for simulations B2 (left) at \T{13\,0} and A1 (right) at \T{12\,0}. We find that both of these simulations exhibit similar behaviour to the \T{13\,5} simulation (Fig.\,\ref{fg:mquc_t135}), with the depth dependence of the diffusion coefficient being described by $D(z)=$ \vzrms\,$\cdot d_{\mathrm{char}}$ in the near-overshoot region and $D(z)=$ \vsq\,$\cdot t_{\mathrm{char}}$ in the deeper layers. 

We find the transition from \vzrms\ to \vsq\ behaviour occurs at $z=$ $-$0.8, $-$1.2 and $-$1.6~km for simulations C3, B2, and A1 (with $T_{\mathrm{eff}}=$ 13\,500, 13\,000 and 12\,000~K), respectively. The proportionality of transition depth with effective temperature is likely due to the inverse proportionality of convective velocity, or kinetic energy, and effective temperature in the range of these simulations (see Fig.~\ref{fg:conv_vel}). We find that the characteristic distance required to fit the near-overshoot region is relatively unchanging across the three temperatures, with $d_{\mathrm{char}}\approx0.03$~km, whilst the characteristic time required to fit the far-overshoot region varies between $0.06 \leq t_{\mathrm{char}} /[\mathrm{s}] \leq 0.28$. 

{The vertical resolutions of simulations C3, B2 and A1 are 10--20\,m, 10--30\,m and 14--43\,m, respectively, whilst the horizontal resolutions are all 30\,m. As a word of caution we point out that the characteristic distances invoked in the near-overshoot region, $d_{\mathrm{char}}\approx0.03$~km, are of the same order as the grid resolution. At this stage we can not currently rule out the possibility of some impact from numerical diffusion in the near-overshoot region. This is independent of the diffusion coefficients derived with a characteristic timescale in the far-overshoot region, which bares the most significance for the ultimate determination of the mixed mass.}

Previous results have shown that this characteristic time could be estimated as $t_{\mathrm{char}} \sim H_{p}/v_{\mathrm{z,rms}}$ (i.e., Eq.~(9) of \citealt{freytag96}) where the pressure scale height, $H_p$, and RMS vertical velocity are evaluated at the base of the convection zone. From an examination of Table~\ref{tb:main}, which provides these two quantities evaluated at the lower Schwarzschild boundary ($H_{p,\mathrm{S,bot}}$ and $v_{\mathrm{z,S,bot}}$), we find estimates of a characteristic timescales of $t_{\mathrm{char}}=0.12$, 0.14 and 0.23~s for simulations A1, B2 and C3, respectively. These agree with the characteristic times which rendered the best fit to the \vsq\ lines to within a factor 0.5--2.0.

We find that all three simulations have sufficient motion from convective overshoot to mix material for at least 1.5\,km beneath the lowest formally convective unstable layer, corresponding to at least 2.5 pressure scale heights. We observe in all cases two distinct behaviours of the overshoot diffusion coefficient. In the near-overshoot region, which extends 0.8--1.6~km beneath the unstable layers, the diffusive efficacy decays with \vzrms$\cdot d_{\mathrm{char}}$. In deeper layers the mixing efficacy decays more rapidly, following instead a \vsq$\cdot t_{\mathrm{char}}$ profile.

The locations of the derived diffusion coefficients in Figs.~\ref{fg:mquc_t135}~\&~\ref{fg:mquc_t130/t120} can be compared with the y-axis positions in Figs.~\ref{fg:quc-movie-t135}--\ref{fg:quc-movie-t120} where snapshots show the evolution of tracer densities (green). In Fig.~\ref{fg:quc-movie-t135} the tracer densities were placed in the far-overshoot region of simulation C3 (13\,500~K) at $z\approx-1.0$~km and $z\approx-1.2$~km. It is evident that overshoot plumes are capable of penetrating this layer and mixing material in deeper layers ($z < -1.2$~km). These deeper layers correspond to the region which feels the effect of waves trapped in the base of our simulations. The tracer density method is susceptible to artificial diffusion driven by the increased velocities here, typically leading to an overestimation of the diffusion coefficients in this region. The increased velocities manifest as the upward inflexion visible at {$z\approx-1.9$~km} in the right panel of Fig.~\ref{fg:mquc_t130/t120}. This provides a lower limit on the depth at which a diffusion coefficient can be directly derived via the method of tracer density. In Section \ref{sec:filtering} we will discuss an approach which provides a conservative estimate of the decay of mixing efficacy toward the stellar interior and beyond the region directly simulated.

\subsubsection{Effects of Numerical Resolution}
\label{sec:res_test}

The physical structures we wish to characterise include narrow overshoot plumes which require grid points to be spaced sufficiently close. Furthermore, numerical diffusion could adversely impact our results. We perform a convergence test on the dependence of our results on spatial resolution by using the tracer density to analyse simulations (C1, C2 \& C3) at three different resolutions ($150^3$, $250^2 \times 150$ and $250^3$) and \T{13\,5}. We note that the time steps are changed accordingly to respect the Courant condition \citep{freytag12}.

\begin{figure}
\centering

\subfloat{\includegraphics[width=1.0\columnwidth]{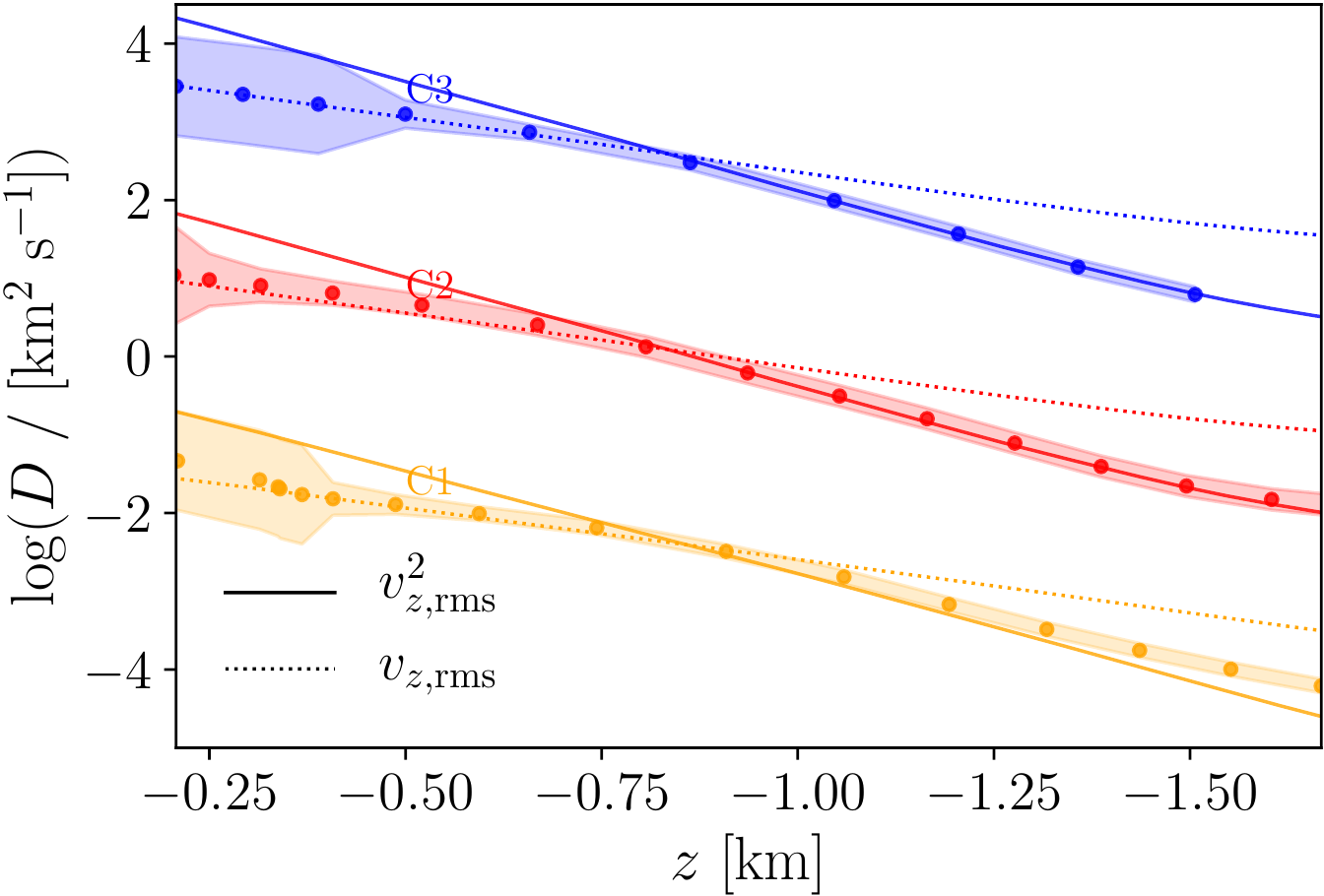}}
\caption{Spatial resolution sensitivity test using tracer density analysis for simulations C1, C2 and C3 and grid sizes $150^3$ (orange), $250^2\times 150$ (red) and $250^3$ (blue), respectively, and \T{13\,5} and \logg = 8.0. The time-averaged vertical velocity profiles \vzrms (dashed) and \vsq (solid) are shown for each simulation with the diffusion coefficient extracted from each simulation (circles). The filled error bars represent the standard deviation of the \thalf fits. Simulations C2 and C3 have been offset by $\log D$ = 2.5 and 5 for clarity.}

\label{fg:t135_resolution}
\end{figure}

Fig.~\ref{fg:t135_resolution} shows diffusion coefficients derived for all three simulations and, qualitatively, we find that in the near-overshoot region the correlation of the diffusion coefficient with the vertical velocity profile, \vzrms, is insensitive to changes in spatial resolution for a fixed box geometry. {We do find a slight increase in the near-overshoot diffusion coefficients for simulation C2, with grid size $250^2\times 150$. It is possible that this is due to the increase in vertical grid spacing compared to the $250^3$ simulation, meaning we cannot rule out that numerical diffusion plays a role in this region and much higher resolution simulations would be needed to comprehensively test this.}

We observe that in the far-overshoot region (i.e., $z<-0.8$~km) the diffusion coefficients tend to the \vsq\ profile as the horizontal resolution increases. No significant change is seen for an increase in the vertical resolution. We conclude that a resolution of $250^2\times 150$ is adequate to resolve the physical processes involved with macroscopic diffusion driven by deep convective overshoot for an atmosphere with  \T{13\,5} and \logg= 8.

\begin{figure}
\centering
\subfloat{\includegraphics[width=1.0\columnwidth]{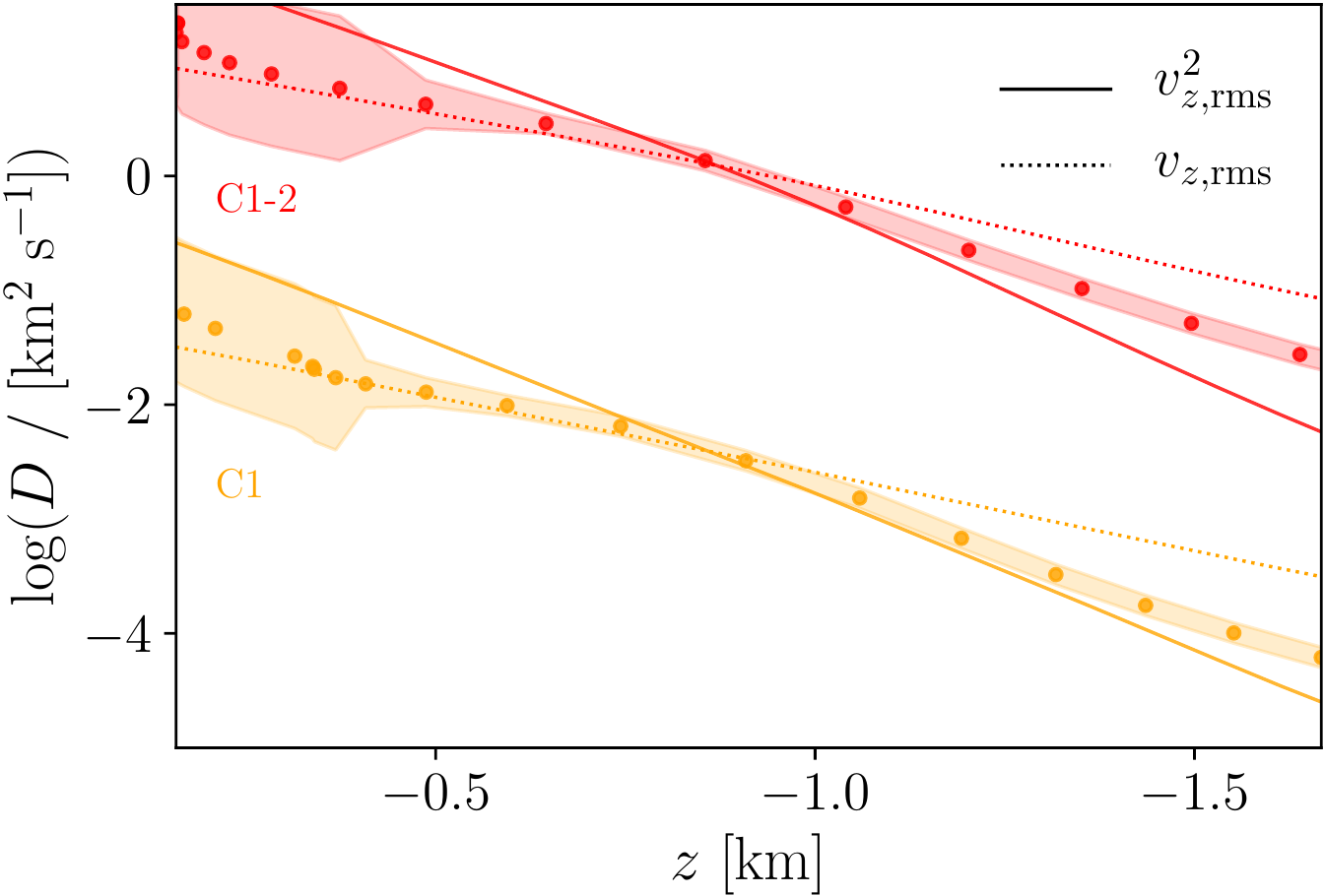}}

\caption{The derived diffusion coefficients for simulations C1 and C1-2 from Table \ref{tb:main} with \T{13\,5}, \logg = 8.0 and grid size $150^3$. The simulations differ only in horizontal size where, compared to simulation C1 (orange), the geometric extents in the $x$ and $y$ direction have been increased by a factor of two for simulation C1-2 (red). The time-averaged vertical velocity profiles \vzrms (dashed) and \vsq (solid) are shown for each simulation with the diffusion coefficient extracted from each simulation (circles). The filled error bars represent the standard deviation of the \thalf fits. Simulation C1-2 has been offset by $\log D$ = 2.5 for clarity.}
\label{fg:mquc-width-test}
\end{figure}

We now discuss a complimentary test to the above where the grid size is fixed at $150^3$ and the box geometry is allowed to vary. The two simulations C1 and C1-2 from Table~\ref{tb:main} have horizontal extents in the $x$ and $y$ direction of 7.5 and 15.0~km, respectively. This tests the spatial resolution in the horizontal plane and as the vertical extent is kept constant across both simulations this also serves as a test of varying aspect ratio. It can be seen from Fig.~\ref{fg:mquc-width-test} that for layers deeper than 1~km beneath the unstable layers the diffusion coefficient exhibits similar behaviour, though decays less rapidly. Within 1~km of the unstable layers the diffusion coefficient behaviour is unchanged for an augmentation in the horizontal directions by a factor of two. {As a word of caution we point out that these convergence tests are not exhaustive and, to be so, simulations with a significantly higher resolution would be required. Were it for this tracer density method alone, these results would be tentative, but a comparison with the path integration results - presented in the following section - provides evidence that our estimates of the size of the mixed region are physically robust.}

\subsection{Method II: Path Integration}
\label{sec:pathint}
 
We discuss in the following the derivation of diffusion coefficients using a method where the path of individual tracer particles is followed directly through the simulation. 
In principle, path integration should give the same results as the tracer density method presented in the previous section if we
were to horizontally average tracers, with any difference caused by numerical schemes. Ostensibly, the way in which the diffusion coefficient is derived from either method, path integration or tracer density, is also much the same. We still look for a true diffusion process based on the spread evolving with the square root of time.

Without an implemented tracer particle module, the path integration is performed using the velocity field of pre-computed CO$^5$BOLD snapshots. This method is closer to the one presented in \citet{freytag96} and it allows a more direct comparison to their results.

To ensure that the path of the seeds is fully interacting with the physical processes within the simulation we are careful to sample velocity information sufficiently frequently. To account for temporal changes in the velocity field the sampling rate should be high enough that changes in the velocity field are small. Spatial changes in the velocity field, which dictate the path taken by the massless tracers, are handled by the integration time step being sufficiently small. 

\begin{figure}
	\centering
	\includegraphics[width=\columnwidth]{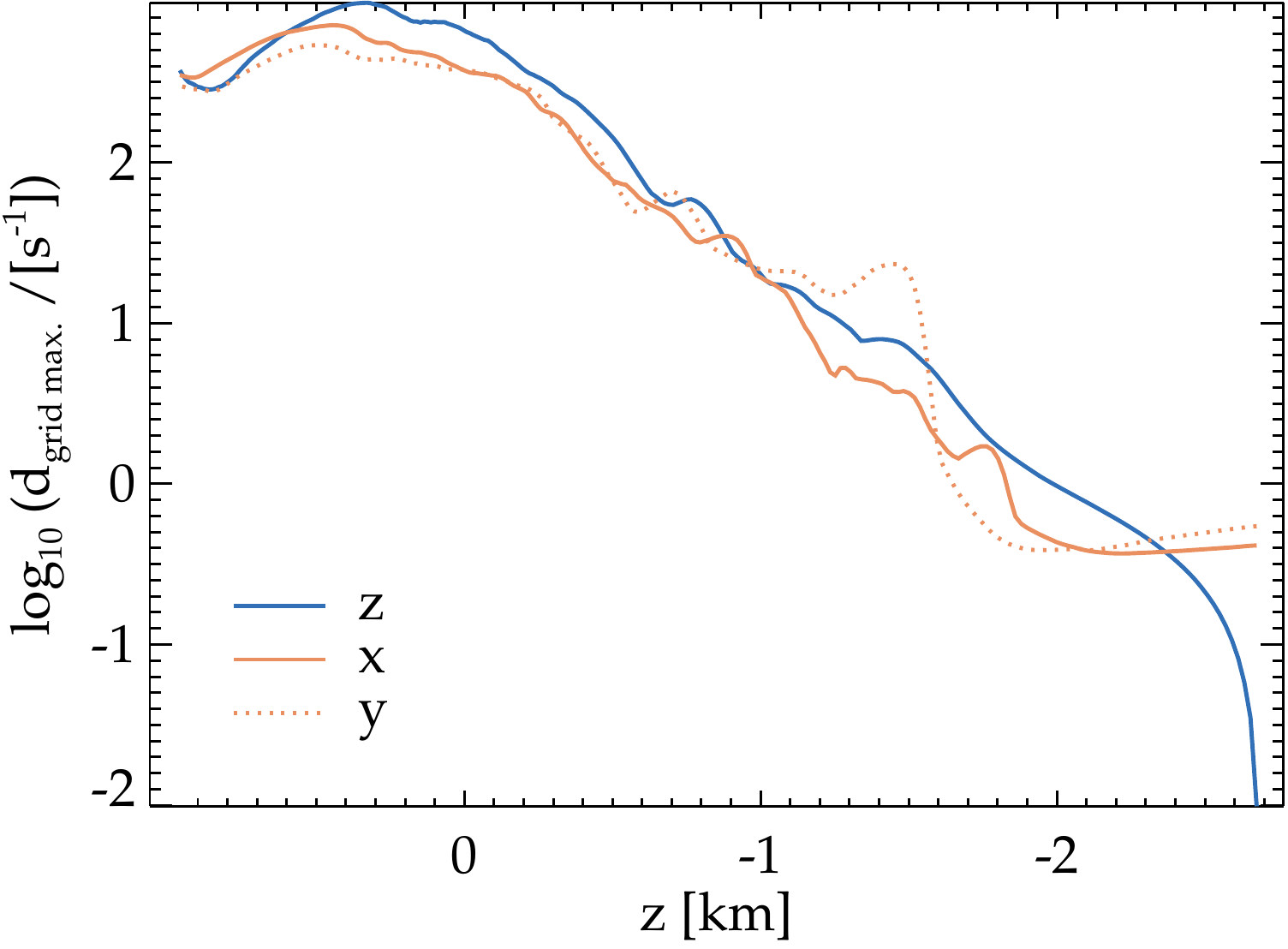}
	\caption{Depth dependence of maximum grid point displacement per second for simulation C3 from Table \ref{tb:main} with \T{13\,5} and \logg = 8.0. Vertical (blue) and horizontal (orange) displacements are shown independently. Displacements have been maximised over the final 25 ms of the diffusion experiment.}
	\label{fg:calc_tres}
\end{figure}

Fig.~\ref{fg:calc_tres} shows the depth dependent maximum velocity in units of grid points per second for simulation C3, where the y-axis can be interpreted as the minimum sampling frequency for any given depth. Ideally data would be sampled such that massless particles are unlikely to travel much further than a few grid points in one sampling period. The figure shows that to capture fully the dynamical processes at every layer, velocity information would be required at intervals of $\Delta t \sim 10^{-3}$\,s, though this is only necessary if we want to probe the most vigorous convective layers. It can be seen in Fig.~\ref{fg:calc_tres} that an order of magnitude longer sampling period, $\Delta t \sim 10^{-2}$\,s, is sufficient and this also represents the chosen integration period.

\begin{figure}
	\centering
	\includegraphics[width=\columnwidth]{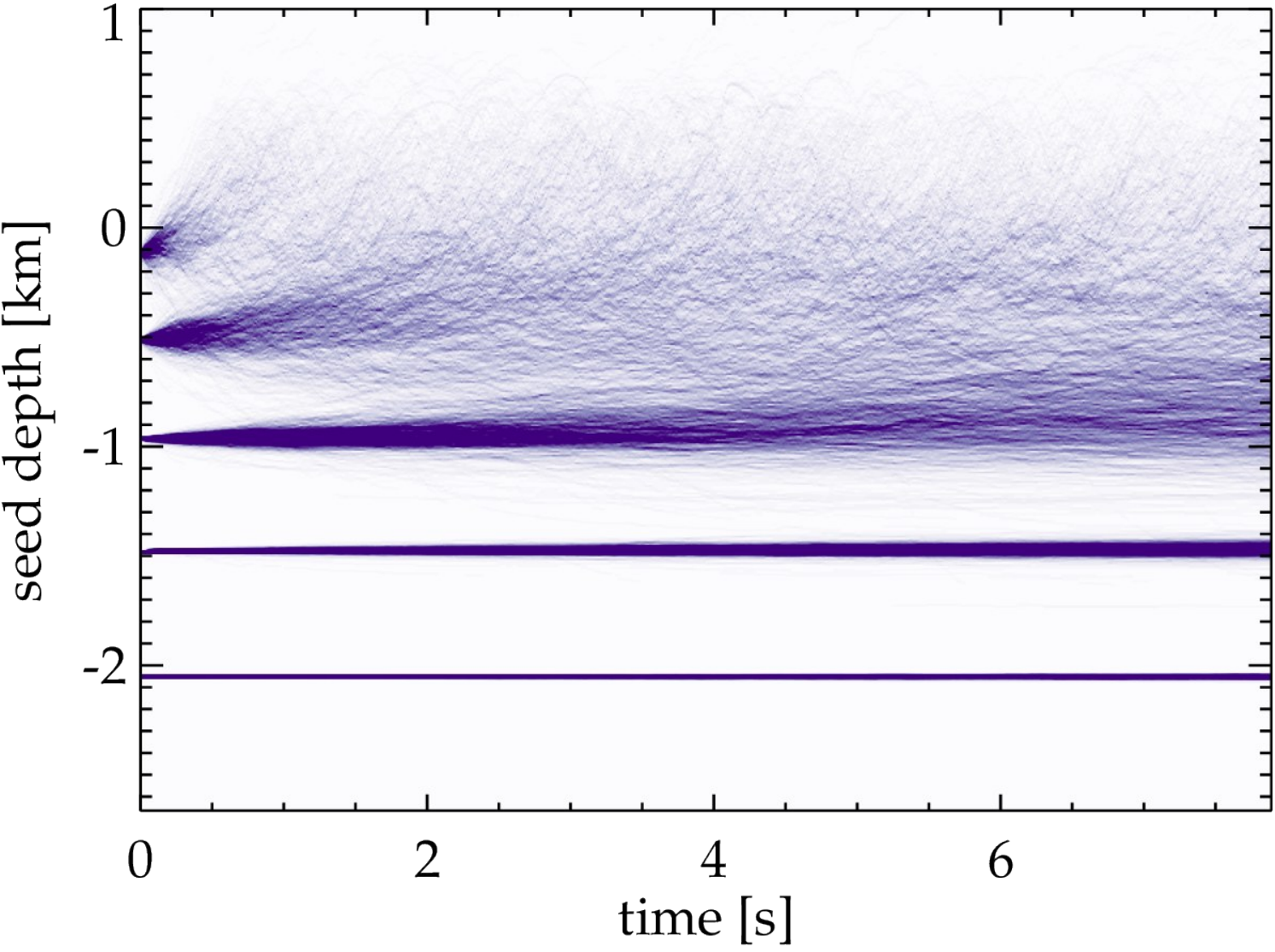}
	\caption{Tracer particle evolution for simulation C3 with \T{13\,5} and \logg = 8.0. Vertical displacement of $25 \times 25$ seeds (placed at every tenth horizontal grid point for clarity), each placed at five different atmospheric depths in the overshoot region. The colour intensity increases linearly with seed density.}
	\label{fg:pathint_t135_trace}
\end{figure}

Fig. \ref{fg:pathint_t135_trace} illustrates the vertical displacement of seeds for the simulation C3 at \T{13\,5} with grid size $250^3$ over 8 seconds. At five depths beneath the unstable layers, $25\times25$ seeds were traced, at every tenth grid point. This lower number of particles helps to clearly demonstrate the depth dependence of the spread of particles, with those in deeper layers spreading less. For the path integration analysis proper, which we will discuss in the next section, we use at least two orders of magnitude more seeds at every depth.

For a given depth we place seeds at every grid point in a horizontal slice, which for the highest resolution simulations presented here (A1, B2 and C3) corresponds to 62500 seeds per depth. The three dimensional position, $r_{x,y,z}(t)$, of a tracer beginning at grid point $r = [x_0,y_0,z_0]$ is given by

\begin{equation}
 r_{x,y,z}(t) = r_{x_{0},y_{0},z_{0}}(t_0) + \int v_{x,y,z}(t)~dt~,
 \label{eq:pathint}
\end{equation}
where $v_{i}$ represents the $i^{\mathrm{th}}$ component of the velocity. This integration is performed using a forward Euler method, where $dt = \Delta t_{\mathrm{diff~exp}}$ is the time interval between consecutive snapshots of extracted velocity information (see Table~\ref{tb:main}). Although the quantity of interest is the depth dependence of diffusive efficacy it is necessary to track the three-dimensional position of each tracer so that all the physical processes present in our simulations have the opportunity to influence the trajectory of the particles. 

For the high resolution simulations, Eq.~\eqref{eq:pathint} is solved 62500 times at each depth and time step such that we follow $\approx 10^7$ tracers per simulation. As our primary focus is the region beneath the unstable layers we present here tracers placed at every second vertical grid point for the region defined by $z < 0$\,km, corresponding to $\approx 10^{6.5}$ tracer particles.

As discussed in Section \ref{sec:dust}, to derive a diffusion coefficient we need to quantify and fit the ensemble spread. With the method of path integration we track the actual position of seeds and thus are able to access the mean square displacement, $z_{\mathrm{rms}}$, given by
\begin{equation}
 z_{j,\mathrm{rms}}(t) = \left(\langle(z_j(t) - \langle z_j(t) \rangle_{x,y})^2\rangle_{x,y}\right) ^{1/2}~,
 \label{eq:zrms}
\end{equation}
{\noindent}where $z_j(t=0)$ refers to the initial vertical position of a 'tray' of seeds placed at depth $j$ and $\langle z_j \rangle_{x,y}$ to the mean vertical position of the seeds.

\begin{figure}
\subfloat{\includegraphics[width=\columnwidth]{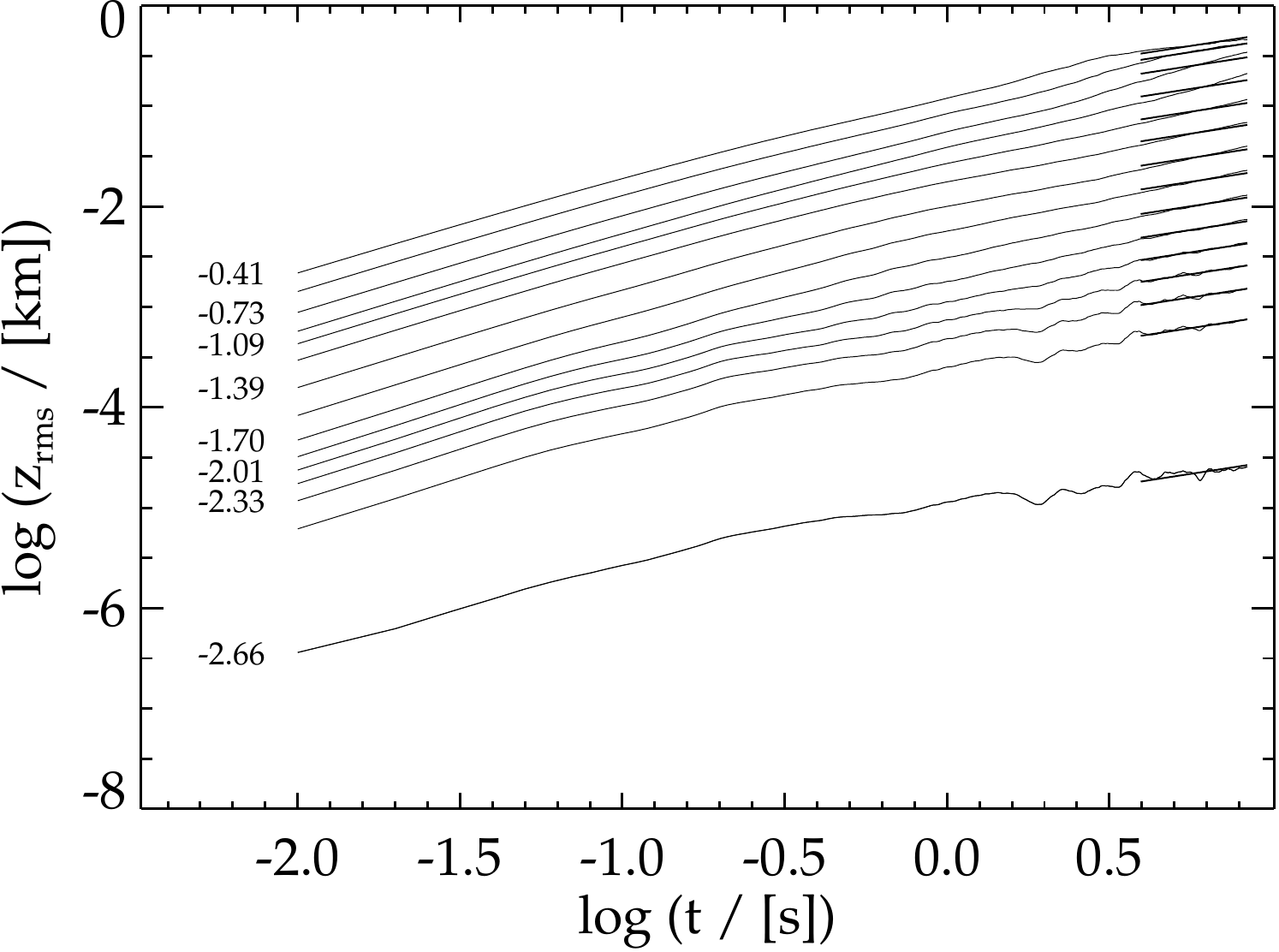}}\\
\subfloat{\includegraphics[width=\columnwidth]{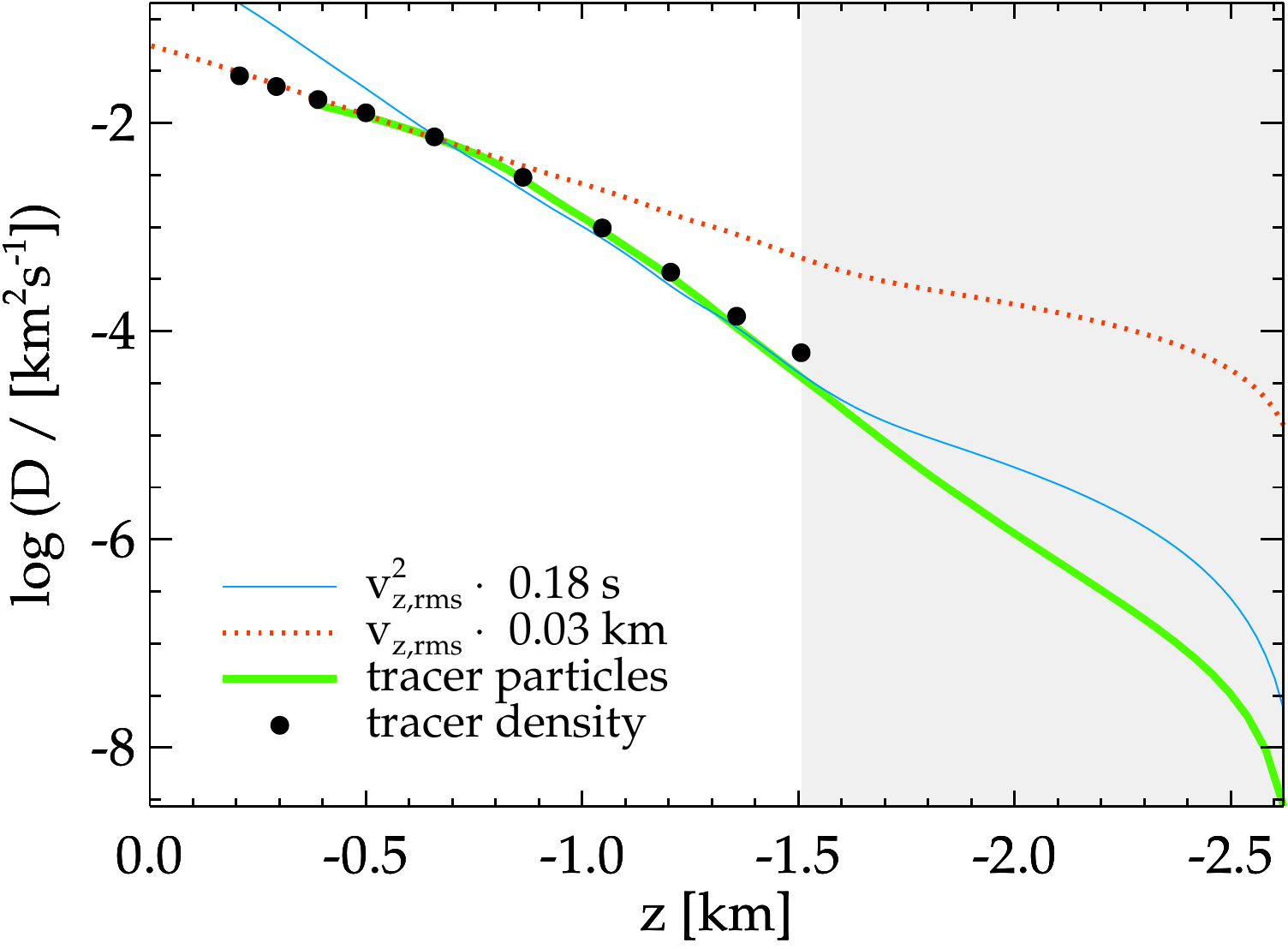}}\\
\label{fg:pathint_t135b}
\caption{Path integration analysis of simulation C3 with \T{13\,5} and grid size 250$^3$, with plots akin to the two lower panels of Fig.~\ref{fg:mquc_t135}. {\it Top:} Standard deviation and \thalf fit (thick) for 15 of the 62 unique depths at which seeds were placed. The mean initial depth of each set of $249 \times 249$ seeds is indicated, for every other distribution plotted. {\it Bottom:} Diffusion coefficients computed from the method of path integration (solid green) are shown with vertical velocity profiles (\vzrms, dotted-orange, and \vsq, solid-blue). Also shown are the results from the tracer density experiment for the same simulation (black circles) from the bottom panel of Fig.~\ref{fg:mquc_t135}. The layers impacted by waves (see Section~\ref{sec:filtering}) are indicated by the grey shaded region. No tracer density results are shown in this region due to the artificial enhancement of the diffusion coefficients (see Section~\ref{sec:dust}).} 
\label{fg:pathint_t135}
\end{figure}

Eq.~\eqref{eq:zrms} is the standard deviation of the ensemble, making the retrieval of a meaningful spread computationally simple and statistically robust. We can then derive a diffusion coefficient from Eq.~\eqref{eq:msd} as
\begin{equation}
 2\log_{10}(z_{\mathrm{rms}}(z,t)) = \log_{10}(t) + \log_{10}(2D(z))~.
 \label{eq:zrms-D}
\end{equation}
The evolution of this quantity is shown in the top panel of Fig.~\ref{fg:pathint_t135} for simulation C3. The final half of the total evolution time is fitted with lines of $z_{\mathrm{rms}}\sim t^{1/2}$ (thick). All of the spread evolutions shown correspond to tracers which began beneath the unstable layers, with the distance beneath the lower Schwarzschild boundary indicated on the figure. 

\subsubsection{Characterising Diffusion}
\label{sec:results-pathint}

Our experiments with path integration allow us to derive diffusion coefficients that can be compared to the method of density arrays (Section \ref{sec:dust}). We first discuss the results for simulation C3 with \T{13\,5} and grid size $250^{3}$ in Fig.~\ref{fg:pathint_t135}. For the region within the convectively unstable layers ($0.5 > z / [\mathrm{km}] \geq 0.0$) it is expected that seeds are mixed rapidly \citep{freytag96} with the spread increasing linearly with time. Just below the convectively unstable region ($z < 0.0$\, km) seed ensembles are likely to couple with the convectively unstable layers over a short timescale, $t<1$\,s. This manifests in the top panel of Fig. \ref{fg:pathint_t135} as a plateau in ensemble spread, \zrms. {Once the upper edge of a seed ensemble reaches the convectively unstable region, particles are carried in bulk by convective velocities and the statistical, ensemble averaged model of diffusion is no longer justified. Deriving a diffusion coefficient for the upper layers would require fitting the region in time before the seeds couple with the convectively unstable layer. There is significantly less data in this region and for this reason no diffusion coefficients are computed here for $z > -0.2$~km.}

The smoothness of the $z_{\mathrm{rms}}$ evolution demonstrates that the resolution of this simulation provides a more than adequate sample of tracers from which robust statistics are drawn. The fits by \thalf functions are excellent, implying that the average distribution of these tracers, whilst moved by convective overshoot, can unequivocally be described as a diffusion process. The diffusion coefficients computed from this evolution via Eq.~\eqref{eq:zrms} are shown in the lower panel of Fig.~\ref{fg:pathint_t135} in green. This plot shows a strong correlation of the diffusion coefficient with \vzrms\, in the region $-0.3 > z/\mathrm{[km]} > -0.8$, which is in agreement with the results from Method I (black circles) for the same simulation (Fig.~\ref{fg:mquc_t135}; lower panel). The diffusion coefficients computed by the method of path integration are also in agreement in the far-overshoot region ($z<-0.8$~km) where a strong correlation with \vsq\ is evident. This is again in agreement with the prediction made by \citet{freytag96} where the authors used the method of path integration for a simulation at \T{13\,4}. 

\begin{figure}
\subfloat{\includegraphics[width=\columnwidth]{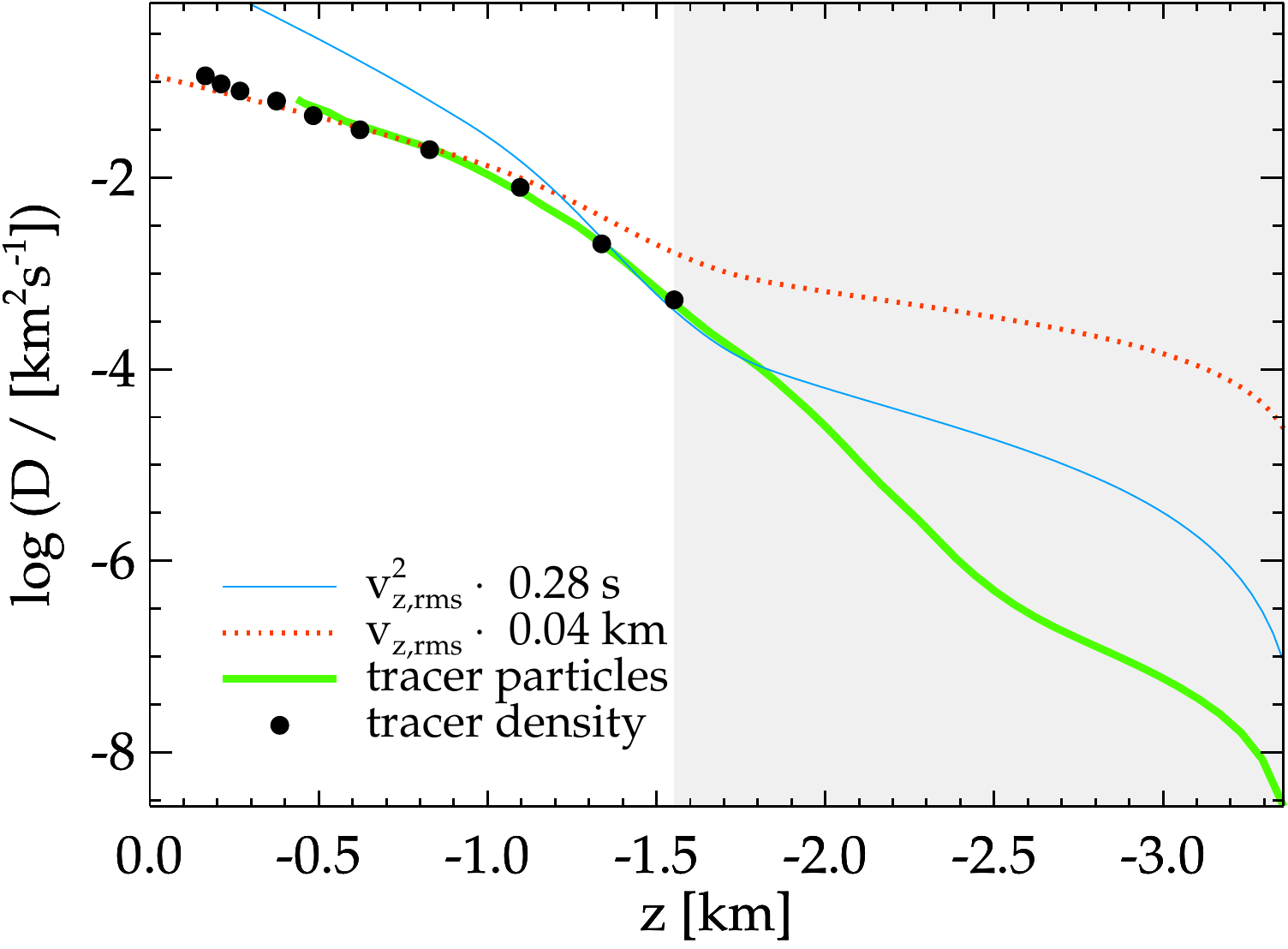}}\\
\caption{Results of path integration analysis of simulation B2 with \T{13\,0} and grid size 250$^3$. Akin to lower panel of Fig. \ref{fg:pathint_t135}.}
\label{fg:pathint_t130}
\end{figure}

\begin{figure}
\subfloat{\includegraphics[width=\columnwidth]{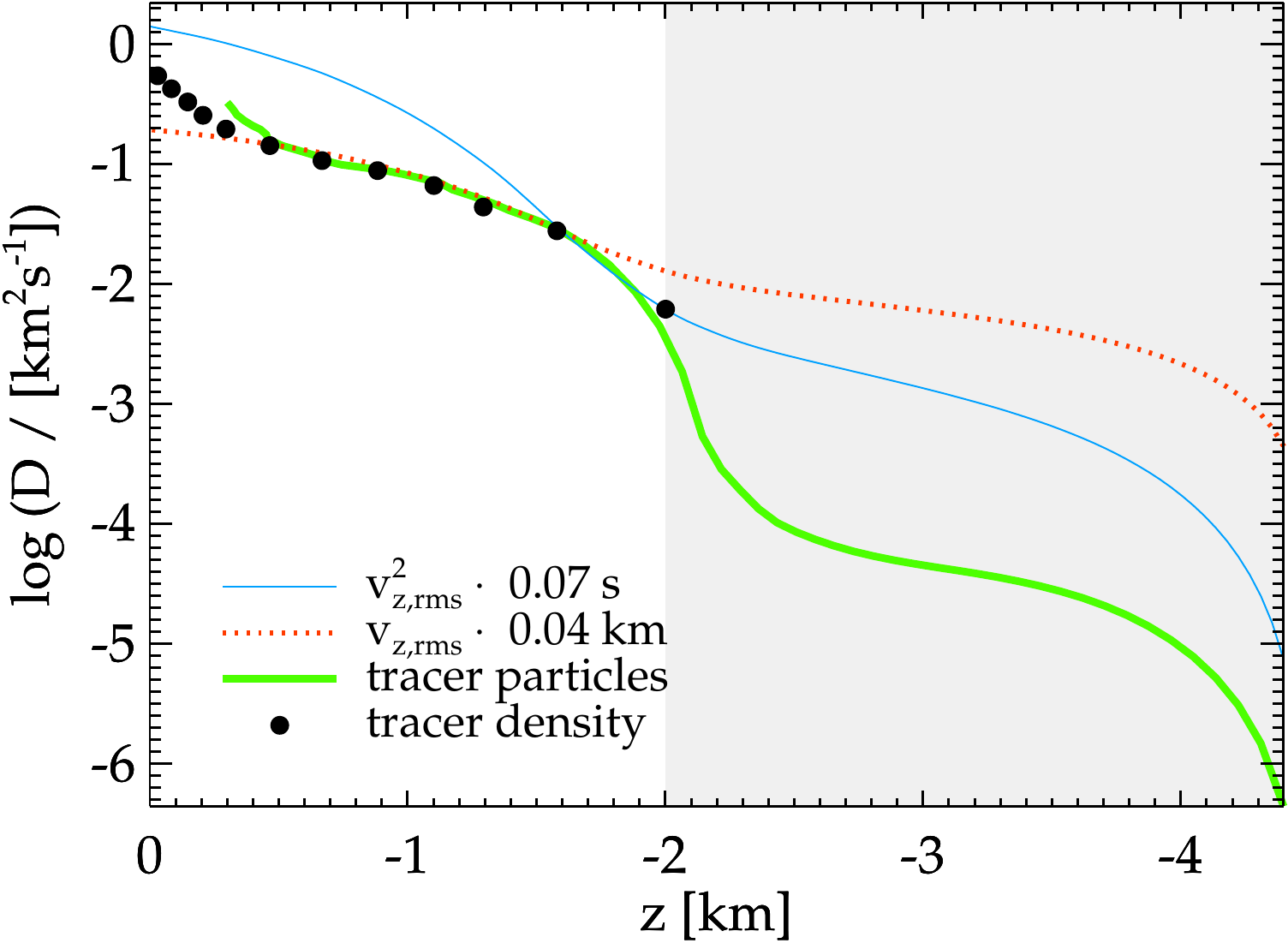}}\\
\caption{Results of path integration analysis of simulation B2 with \T{12\,0} and grid size 250$^3$. Akin to lower panel of Fig. \ref{fg:pathint_t135}.}
\label{fg:pathint_t120}
\end{figure}

Figs.~\ref{fg:pathint_t130}~\&~\ref{fg:pathint_t120} show the results for simulations B2 and A1, respectively, which have the same resolution as C3 and \T{13\,0} and 12\,000~K. It is again clear that the diffusion coefficients derived using the method of path integration are in good agreement with the tracer density method at all depths for which results are available. In the wave-dominated region (grey shaded), where tracer density results proved inaccessible, the method of path integration is able to separate diffusive action of the overshoot motions  from the essentially reversible motions due to waves. This is evident from the flattening out of \vzrms\ profile whilst the diffusion coefficient continues to follow an exponential decay towards the core. This effect becomes more obvious as the effective temperature decreases, with simulation A1 at 12\,000~K showing a sharp drop at the top of the wave region. In simulations A1 and B2 the diffusion coefficient flattens out at $z=-2.4$ and $-$2.5~km, respectively. This likely represents the limit of the path integration method to disentangle the two kinds of motion and further work would be required to ascertain the physical meaning of this behaviour. 

\subsubsection{Effects of Numerical Resolution}
\begin{figure}
\centering
\subfloat{\includegraphics[width=\columnwidth]{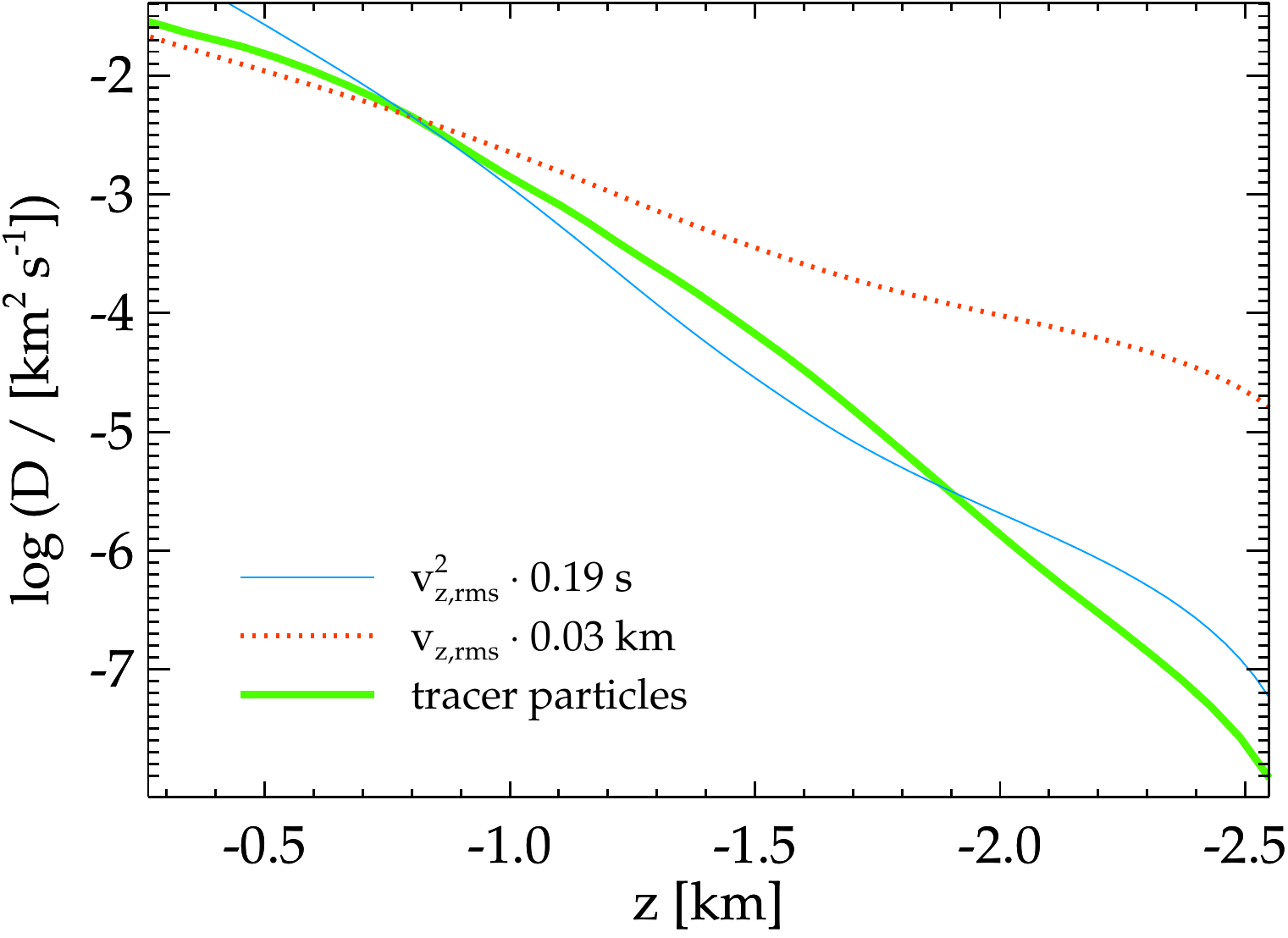}} \\

\subfloat{\includegraphics[width=\columnwidth]{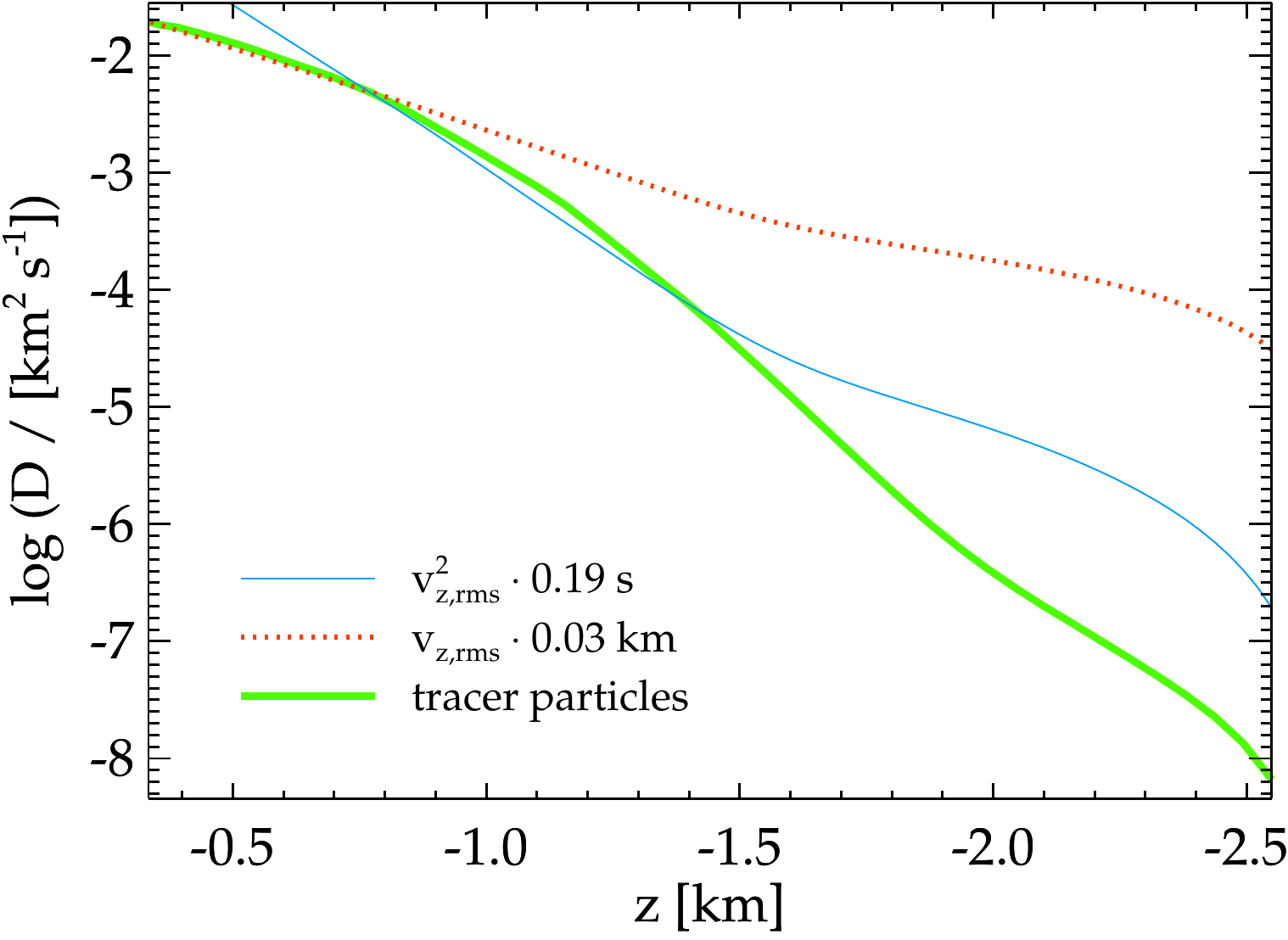}} 

\caption{Similar to lower panel of Fig.~\ref{fg:pathint_t135} for simulations C1 (top) and C2 (bottom) from Table \ref{tb:main} with \teff = 13\,500 K, \logg = 8.0 and resolutions $150^3$ and $250^2\times150$, respectively.}
\label{fg:wide}
\end{figure}

We present here a test of the diffusion coefficient's dependence on spatial resolution derived via the method of path integration. Fig.~\ref{fg:wide} shows the results for the two simulations C1 (top panel) and C2 (bottom panel) which were prepared with grid size $150^3$ and $250^2\times150$, respectively, to compare to the $250^3$ simulation shown in Fig. \ref{fg:pathint_t135} with the same effective temperature. Qualitatively, both simulations retrieved similar behaviour in the diffusion coefficient compared to simulation C3 - with a diffusion coefficient scaling with \vzrms\ for 0.8~km beneath the unstable layers and with \vsq\ for deeper layers with $z < -0.8$~km. 

We observe a similar behaviour to the resolution test performed in Section~\ref{sec:res_test} for the tracer density method with diffusion coefficient tending to \vsq\ in the far-overshoot region as the horizontal resolution increases. Qualitatively we also see no change as the vertical resolution increases and thus conclude the results are derived from sufficiently resolved simulations. 

The three simulations are also in good agreement on the absolute value of the diffusion coefficient which is most readily seen by examining the characteristic distances and times required to fit the \vzrms\ and \vsq\ profiles to the diffusion coefficients.

What is evident from the analysis of the five simulations discussed throughout this section is that the vertical velocity profile is likely to be of great significance to the characterisation of diffusive efficacy beneath the convectively unstable layers, especially if one wishes to not perform a direct diffusion experiment at every temperature of interest. The scaling of the diffusion coefficient with \vzrms\ immediately beneath the unstable layers and with \vsq\ in deeper layers found in both of tracer experiments is in good agreement with the dependence observed by \citet{freytag96}. Both approaches agree in terms of the observable prediction, which is that trace elements are fully mixed in the overshoot region $\approx$ 2.5 -- 3.5 pressure scale heights immediately beneath the convection zone, with mixing timescales still orders of magnitudes {shorter than} those typical of microscopic diffusion. Furthermore, we emphasise that both methods are impacted by standing waves and do not currently allow to directly estimate the full extent of the mixed regions and the total mass of trace metals in a white dwarf, which is therefore the goal of Section \ref{sec:results}.

\section{Results}
\label{sec:results}
In Section \ref{sec:Dover} we detailed the two methodologies used in this study to probe the mixing efficiency of macroscopic diffusion due to convective overshoot, namely the inbuilt \cobold\, tracer density and path integration of seeds. While standing waves near the bottom of the simulations prevented us to simulate macroscopic diffusion over the full convective overshoot region, our results highlighted the possibility of using \vzrms\ and \vsq\ as proxies for macroscopic diffusion. We therefore use the extended grid of CO$^5$BOLD simulations presented in Section~\ref{sec:deep-grid} to study \vzrms\ in detail for a wide range of depths and $T_{\rm eff}$ values. Then, in Section~\ref{sec:filtering} we develop a framework to lessen the contribution of waves. We adopt this approach as fully quantifying the contribution of waves to mixing is beyond the scope of this work. Though there exists some evidence to suggest waves may contribute to the vertical mixing of material \citep{freytag10}, our philosophy is that by removing their contribution we can provide a lower limit on the mixed mass. 

\begin{figure}
\centering
\subfloat{\includegraphics[width=0.8\columnwidth]{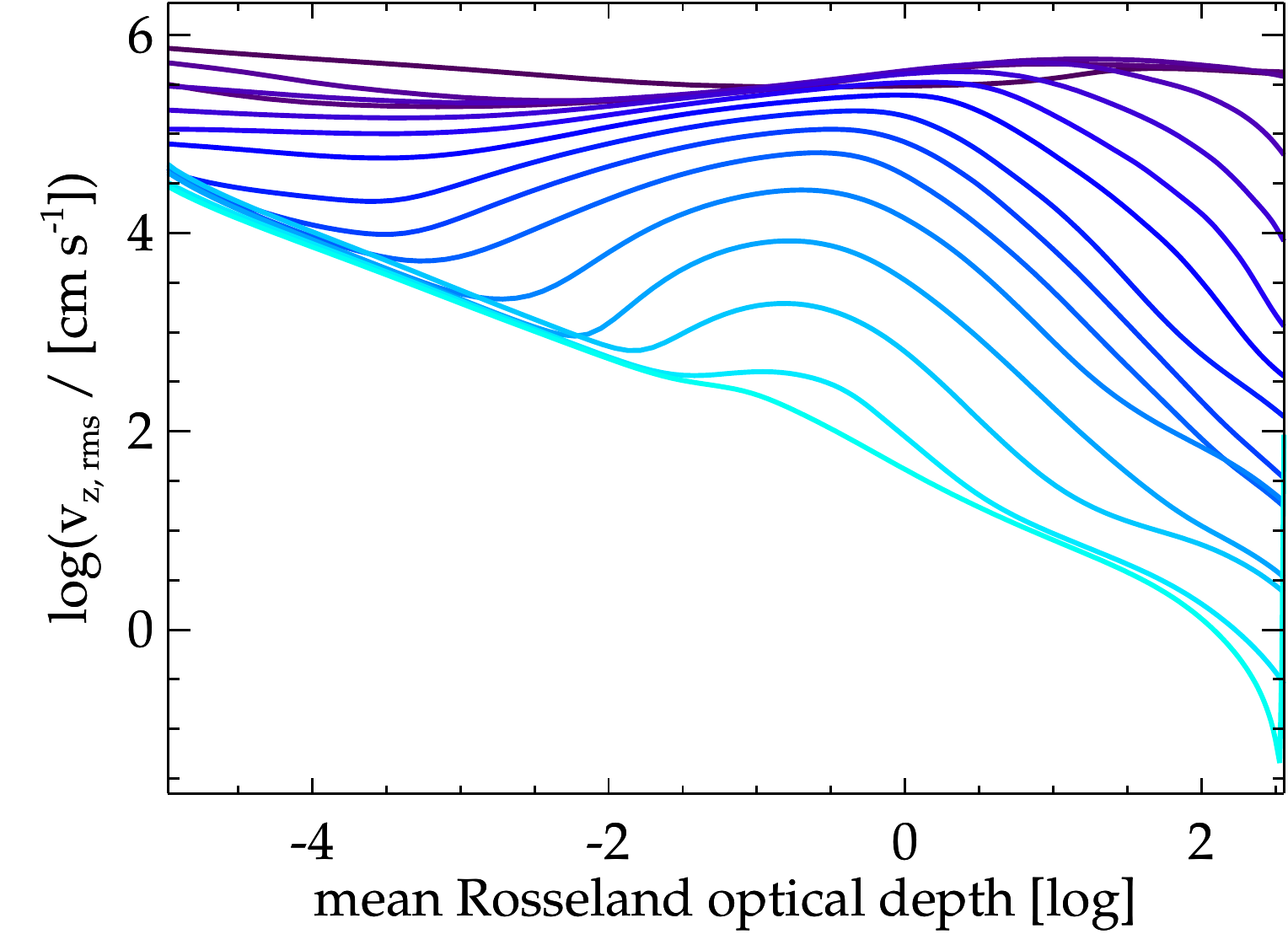}}\\
\subfloat{\includegraphics[width=0.8\columnwidth]{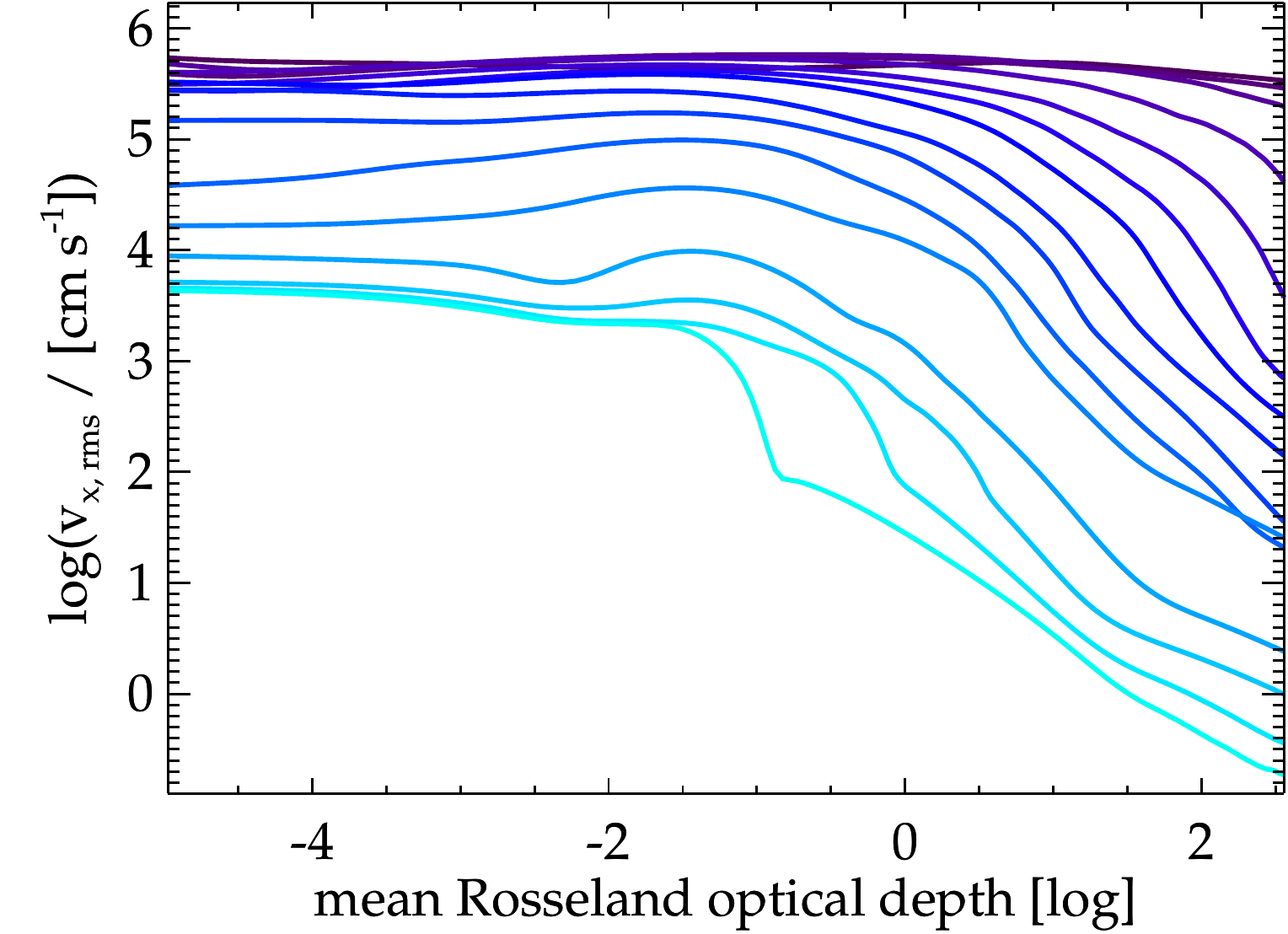}}\\
\subfloat{\includegraphics[width=0.8\columnwidth]{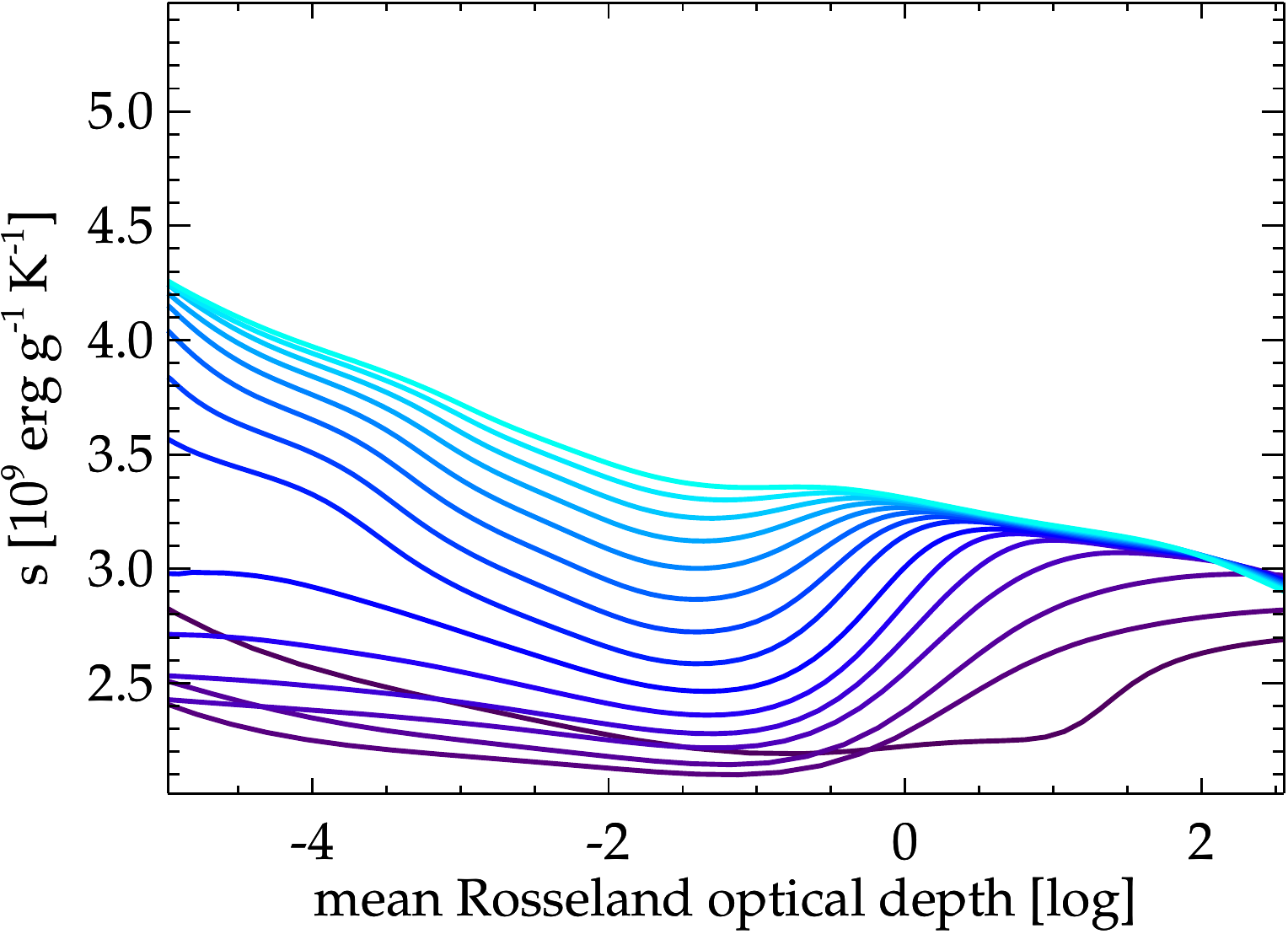}}\\
\subfloat{\includegraphics[width=0.8\columnwidth]{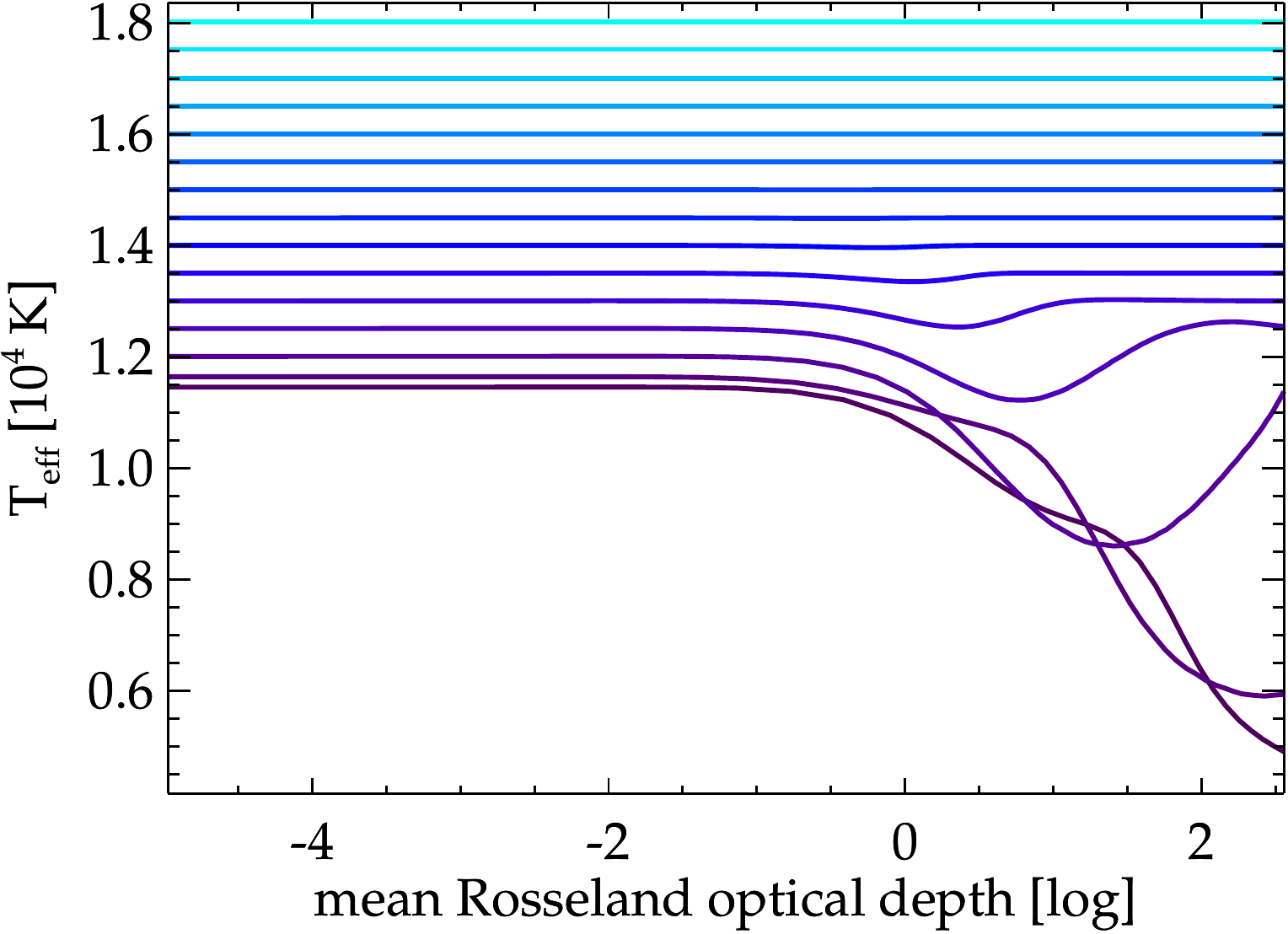}}\\
\caption{Dynamical and thermodynamic quantities extracted from the grid of deep 3D simulations (Table~\ref{tb:deep}) as a function of average Rosseland optical depth. {\it Top to bottom:} Vertical component of convective velocities, horizontal component of convective velocities, entropy and effective temperature. Effective temperature is calculated as $T_{\mathrm{eff}}=\sqrt[4]{F_{\mathrm{rad}}/\sigma}$ where $F_{\mathrm{rad}}$ is the radiative flux and $\sigma$ is the Stefan-Boltzmann constant.}
\label{fg:deep}
\end{figure}

\begin{figure}
\centering
\subfloat{\includegraphics[width=0.8\columnwidth]{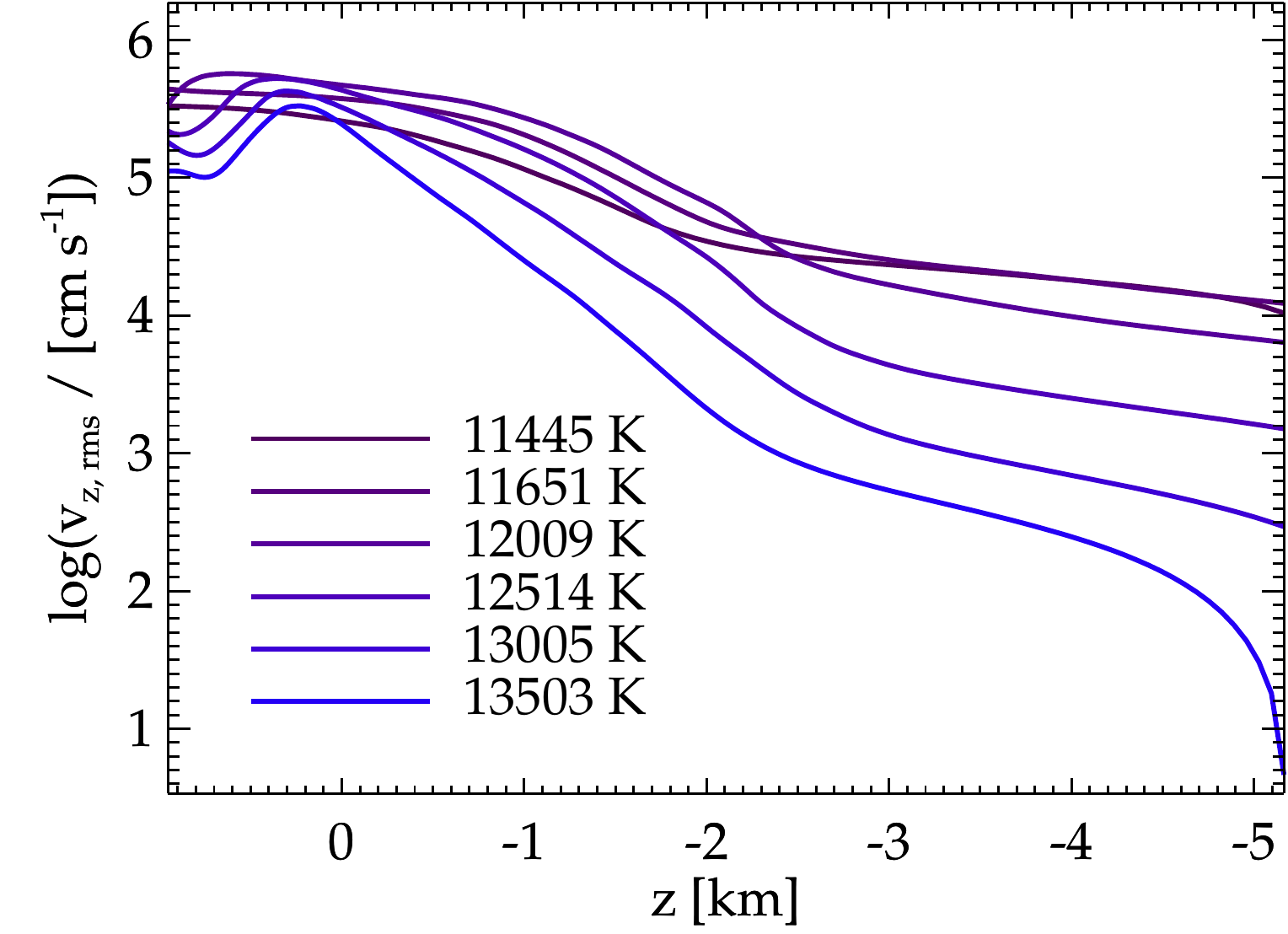}}\\
\subfloat{\includegraphics[width=0.8\columnwidth]{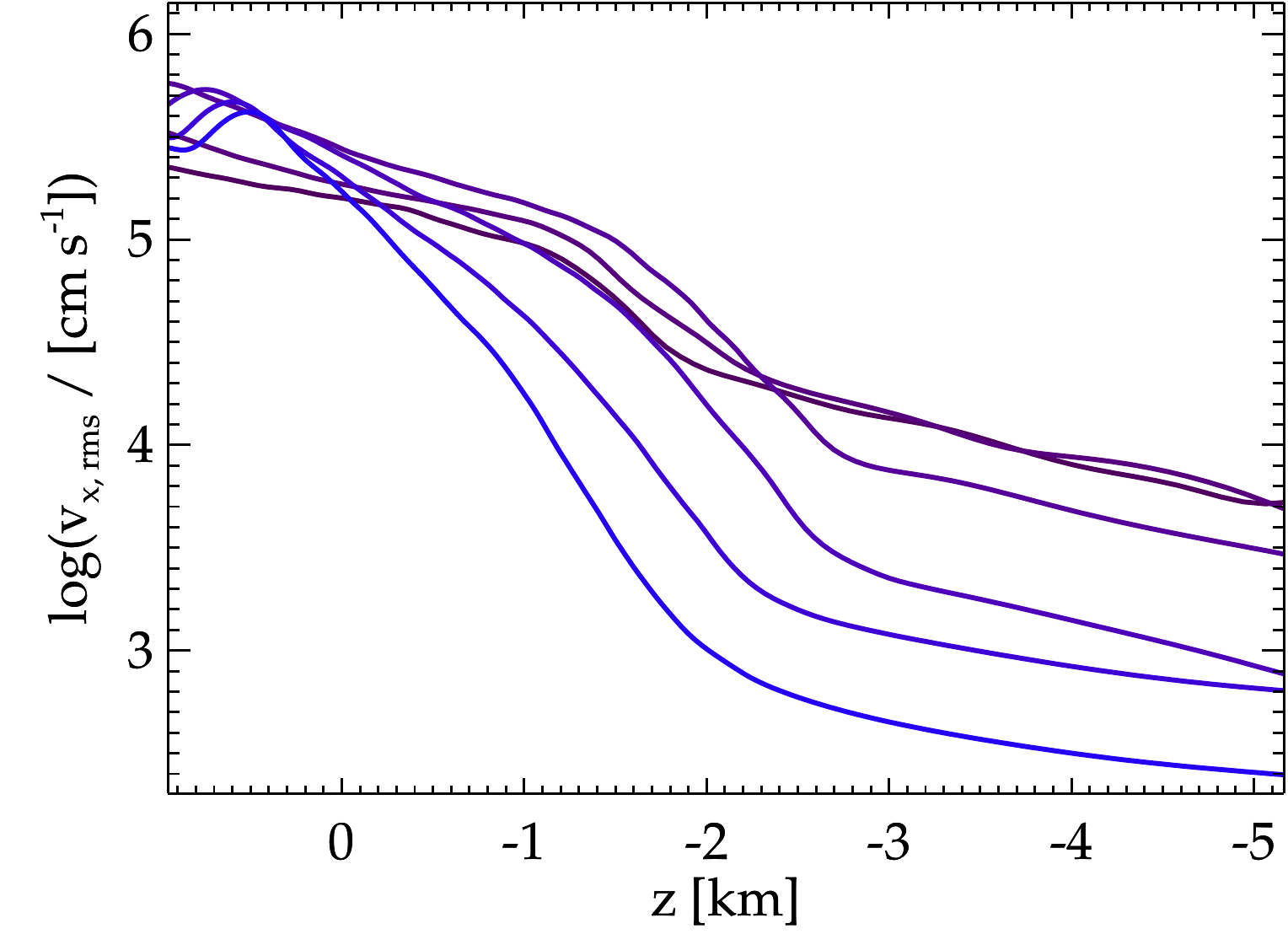}}\\
\subfloat{\includegraphics[width=0.8\columnwidth]{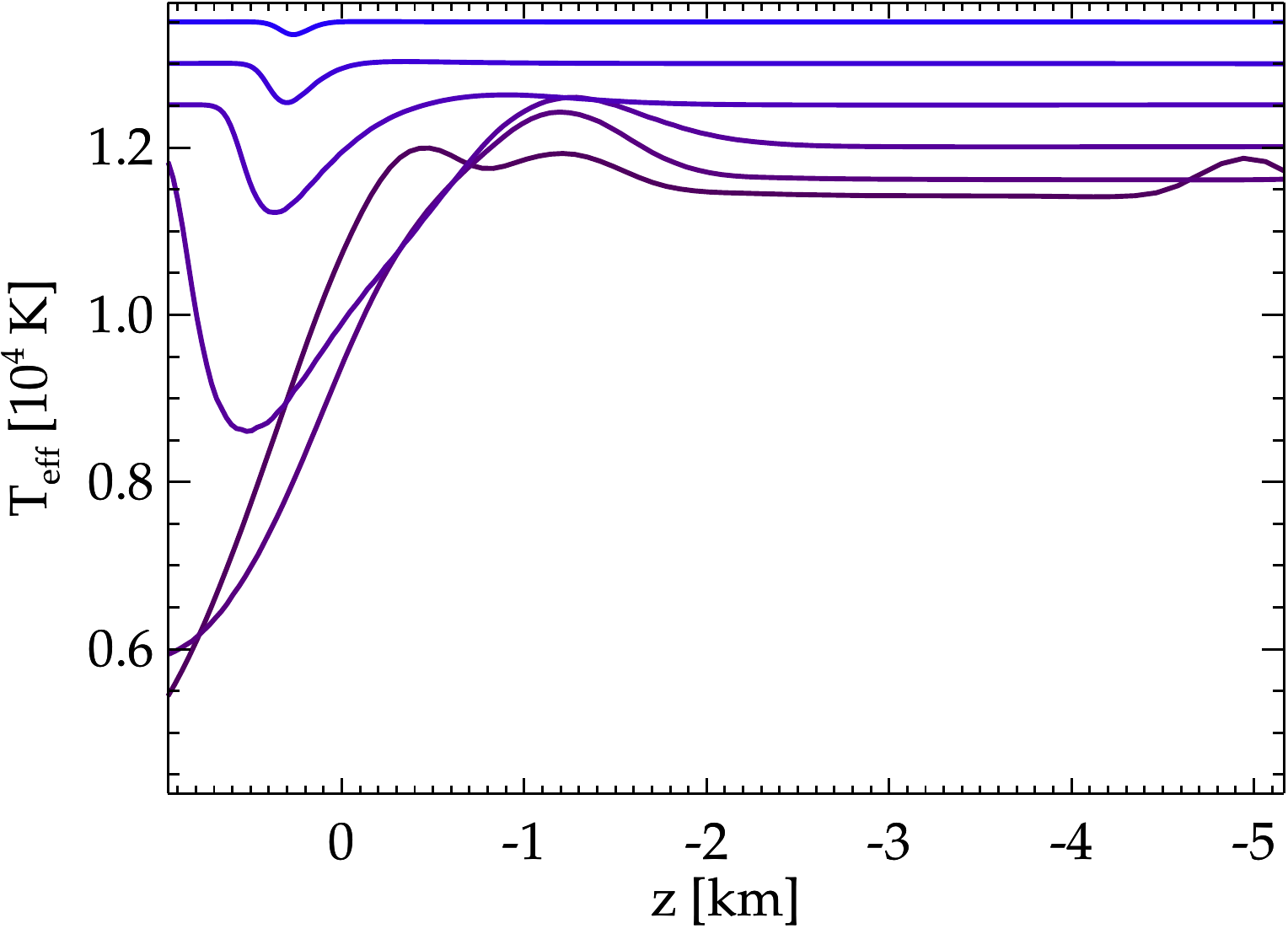}}\\
\caption{Similar to Fig.~\ref{fg:deep} but for models exhibiting sizable convection ($11\,400 \leq$ \teff / [K] $\leq 13\,500$) and as a function of geometric depth.}
\label{fg:cool-only}
\end{figure}

\subsection{Dynamical and Thermal Properties}

We rely on the 15 deep and closed simulations presented in Table~\ref{tb:deep} and described in Section~\ref{sec:deep-grid}, covering the range 11\,400\,K $\leq$ \teff $\leq$ 18\,000\,K. Selected mean quantities are shown in Figs.~\ref{fg:deep}-\ref{fg:cool-only} for all simulations computed in the extended grid.

\begin{figure}
	\centering
	\includegraphics[width=\columnwidth]{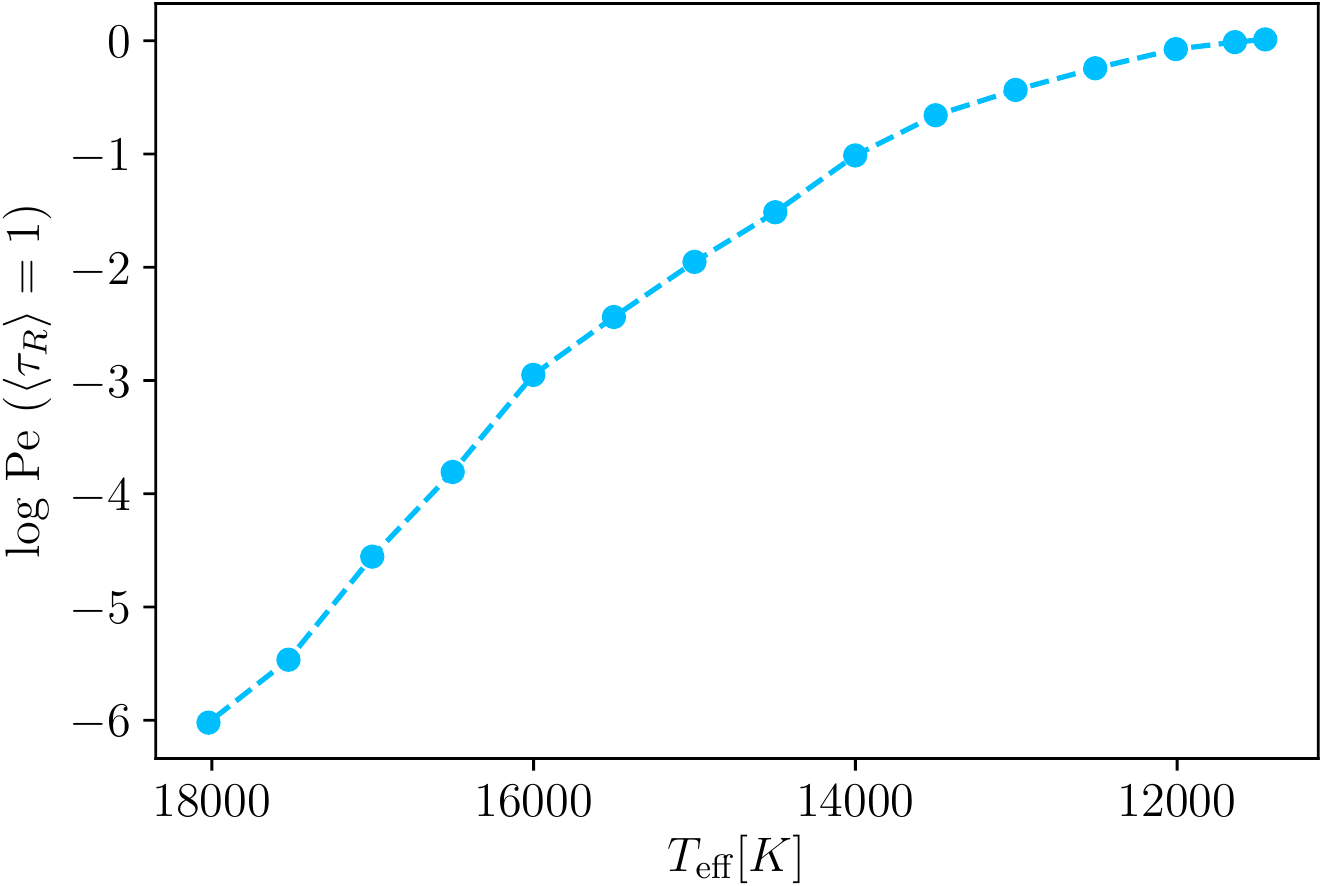}
	\caption{P\'eclet number evaluated at $\langle \tau_{\mathrm{R}} \rangle=1$ as a function of effective temperature and averaged over the final 50\,ms of the deep 3D grid presented in Table~\ref{tb:deep}.}
	\label{fg:TL15fig13remake}
	\label{fg:Peclet} 
\end{figure}

Radiative energy transport becomes fully dominant for DA white dwarf atmospheres above \teff $\approx 14\,000$\,K as illustrated by the bottom panel of Fig.~\ref{fg:deep}. This warmer end of the grid is convectively unstable according to entropy profiles (Fig.~\ref{fg:deep}, third panel from the top) but not convectively efficient i.e., negligible energy is transported in convective flows.  In contrast, the three lowest effective temperature simulations ($T_{\mathrm{eff}} =$~11\,400, 11\,600 and 12\,000~K) can be seen in Fig.~\ref{fg:cool-only} to have deep wells in their radiative flux profiles which implies a large proportion of the energy carried in convective flows. Fig.~\ref{fg:Peclet} shows the time-averaged P\'eclet number, evaluated at \tauR = 1, as a function of $T_{\rm eff}$. The P\'eclet number is given by \citet{tremblay13b} as  
\begin{equation}
\mathrm{Pe} = \frac{t_{\mathrm{rad}}}{t_{\mathrm{adv}}} = \frac{\langle \rho \rangle\langle c_p \rangle v_{\mathrm{rms}}\tau_{\mathrm{e}}}{16\sigma\langle T \rangle^3}\left(1+\frac{2}{\tau^2_{\mathrm{e}}}\right)~,
\label{eq:Peclet}
\end{equation}
where $t_{\mathrm{rad}}$ is the radiative timescale and $t_{\mathrm{adv}}$ the advective, or convective turnover, timescale, $c_p$ is the specific heat per gram, $T$ the temperature, $\sigma$ the Stefan-Boltzmann constant, and $\tau_e = \langle \kappa_{\mathrm{R}} \rangle \langle \rho \rangle H_p$ the characteristic optical thickness of a disturbance with size $H_p$, for $\kappa_{\mathrm{R}}$ the Rosseland mean opacity per gram. It represents the ratio of energy being carried by bulk movement of fluid and that being carried by radiation. For Pe $\ll$ 1, convection will not be responsible for carrying much of the energy. It can be seen from Fig.~\ref{fg:Peclet} that Pe < 1 for \teff $\geq$ 12\,000\,K and thus it describes the transition to inefficient convection at larger temperatures. We emphasise that it does not imply a change in the mixing properties of the atmosphere at that temperature. In fact, \citet[][see figure 13]{tremblay15} have shown that \vzrms, and thus mixing capabilities, remain large for 3D simulations up to \teff$ = 14\,000$\,K, the hottest model in their grid at $\log g = 8.0$. This behaviour is in agreement with the mixing-length theory, which predicts that thermally inefficient but dynamically active convective zones are found in DA white dwarfs up to 18\,000\,K, with the exact \teff value strongly dependent on the assumed mixing-length parameter (ML2/$\alpha$ = 0.7 in \citealt{tremblay15}). Furthermore, the 1D model can only suggest whether or not there is mixing and cannot provide realistic mixed masses due to its neglect of overshoot.

\begin{figure}
	\centering
	\includegraphics[width=\columnwidth]{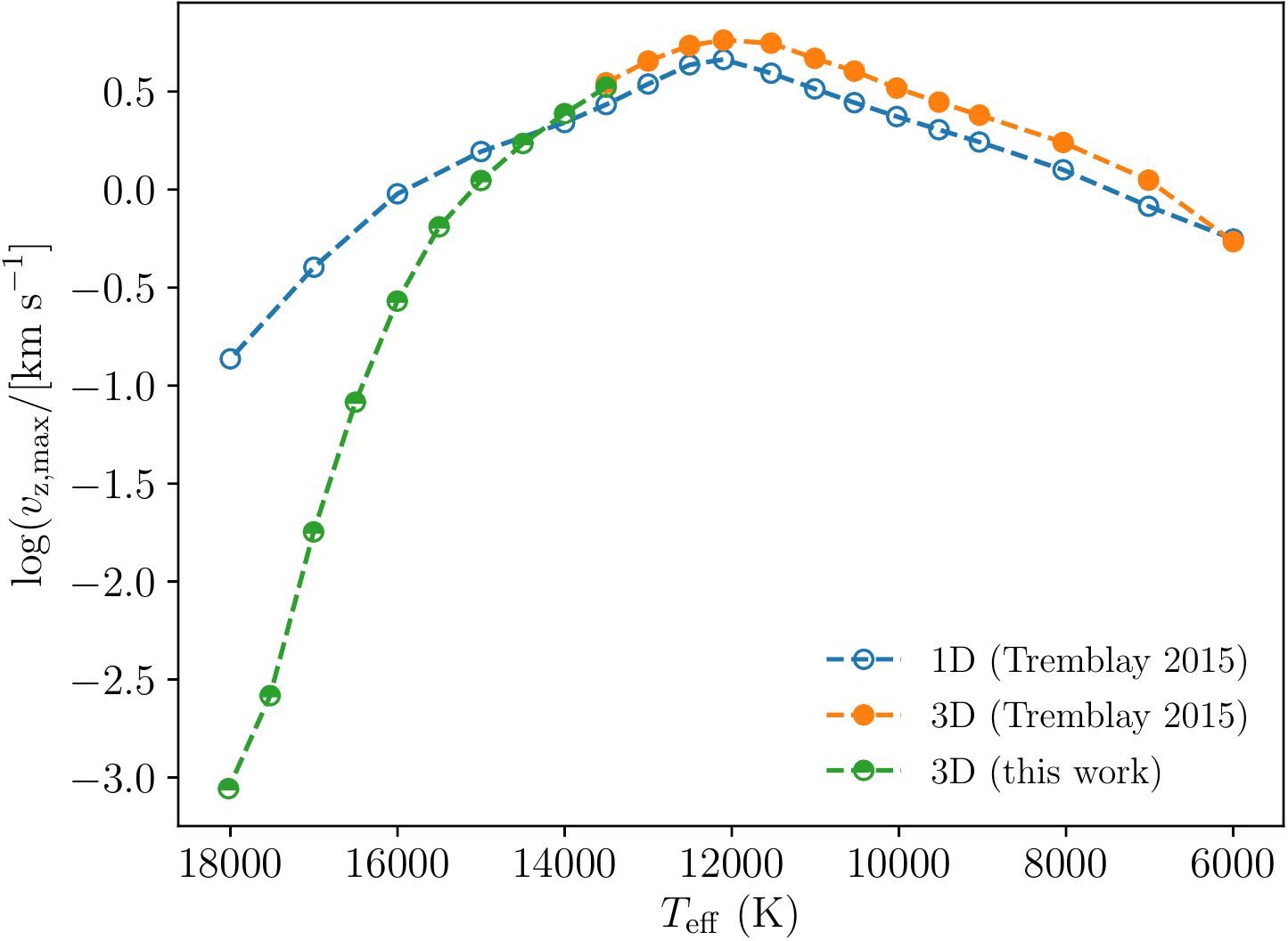}
	\caption{Maximum convective velocities, time-averaged over the final 2.5\,s for simulations in Table \ref{tb:deep} (green, half-filled) with 13\,503~$\leq$~\teff~/~[K]~$\leq$~18\,002. Also shown are previous 3D results (orange, filled) and 1D ML2/$\alpha$ = 0.7 mixing-length theory predictions (blue, open) from \citet{tremblay15}.}
	\label{fg:TL15fig13remake}
	\label{fg:conv_vel} 
\end{figure}

The entropy profiles of our extended grid of 3D simulations in Fig.~\ref{fg:deep} demonstrate that pure-hydrogen atmospheres become unstable to convection for \teff$\lessapprox 18\,000$\, K. We attempted a \teff$ = 18\,250$\,K simulation, not shown on the panel, but it did not develop a convection zone. While this is similar to the 1D prediction of the onset of convection assuming a ML2/$\alpha$ = 0.7 parameterisation \citep{tremblay15}, the 3D simulations provide a more robust threshold. Fig. \ref{fg:conv_vel} shows the maximum convective velocity in the convectively unstable region, time-averaged over the final 2.5\,s, for all simulations in our deep grid (see Table \ref{tb:deep}) as well as \citet{tremblay15}. It can be seen that the grid presented in this work fully connects with the previous, shallower one. It is also observed that 1D MLT under-predicts convective velocities for lower temperatures (\teff$<14\,000$\,K) and over-predicts for higher temperatures (\teff$>14\,000$\,K), although this could be corrected by using a variable mixing-length parameter as a function of $T_{\rm eff}$. The first convective model at \teff = 18\,000\,K is already showing large 1D velocities, illustrating the sharp and unphysical transition between convective and radiative atmospheres. In 3D, the warmer convective models have much lower velocities, in line with a smoother onset of convection in evolving DA white dwarfs. This effect may be enhanced by numerical viscosity which, in the regime where radiative energy transport is significantly more efficient than convection, may reduce the 3D convective velocities.

That the atmosphere is unstable is of critical importance for the capacity for DA white dwarfs in this temperature range to mix material, specifically accreted metals. Fig.~\ref{fg:deep} demonstrates that all simulations have large velocities, although with a contribution both from convection and waves. It therefore requires further investigation to fully characterise the mixing capabilities. 

\subsection{Mixed Mass}
\label{sec:filtering}

The depth dependence of the diffusion coefficient just above the wave region is well described by the square of the RMS vertical velocity, \vsq , (see Section~\ref{sec:Dover}) with the characteristic timescale $t_{\mathrm{char}} \approx 0.18$\,s taken from the \T{13\,5}\ simulation (see Fig.~\ref{fg:pathint_t135}). Ideally our simulations would extend to sufficiently deep layers that convection driven velocities decay enough that microscopic diffusion takes over, bridging the discontinuity between macroscopic and microscopic diffusion highlighted by \citet{tremblay17}. This would be computationally intractable given the typical microscopic diffusion velocities of $v_{\mathrm{diff}} \sim 10^{-7}$\,km\,$\mathrm{s}^{-1}$ and a significant hinderance to this approach is the presence of waves in the base of the simulation.

The upward inflexion in the velocity profile at the bottom of the overshoot region highlighted in earlier figures (see Figs.~\ref{fg:pathint_t135}~\&~\ref{fg:pathint_t120}, and upper panel of Fig.~\ref{fg:cool-only}) is due to these standing waves. Multi-dimensional simulations of brown dwarf atmospheres have provided evidence that waves may be capable of mixing material \citep{freytag10}. However, using \vsq\ in the wave region as a proxy for macroscopic diffusion could lead to an overestimation as standing waves may be amplified by boundary conditions. Alternatively, extrapolating the diffusion coefficient from the top of the wave region, e.g. with an exponential decay as a function of the pressure scale height \citep{tremblay15}, would lead to an extremely shallow decay of macroscopic mixing as a function of depth that may not be realistic \citep{kupka18}. This suggests that more investigation is needed to obtain sensible mixed masses.

\subsubsection{Filtering}

We employ a filtering technique to remove the waves from the velocity profile which has been implemented in convective studies of M-type, main-sequence and pre-main-sequence objects \citep{ludwig06}. 

\begin{figure}
	\centering
	\includegraphics[width=\columnwidth]{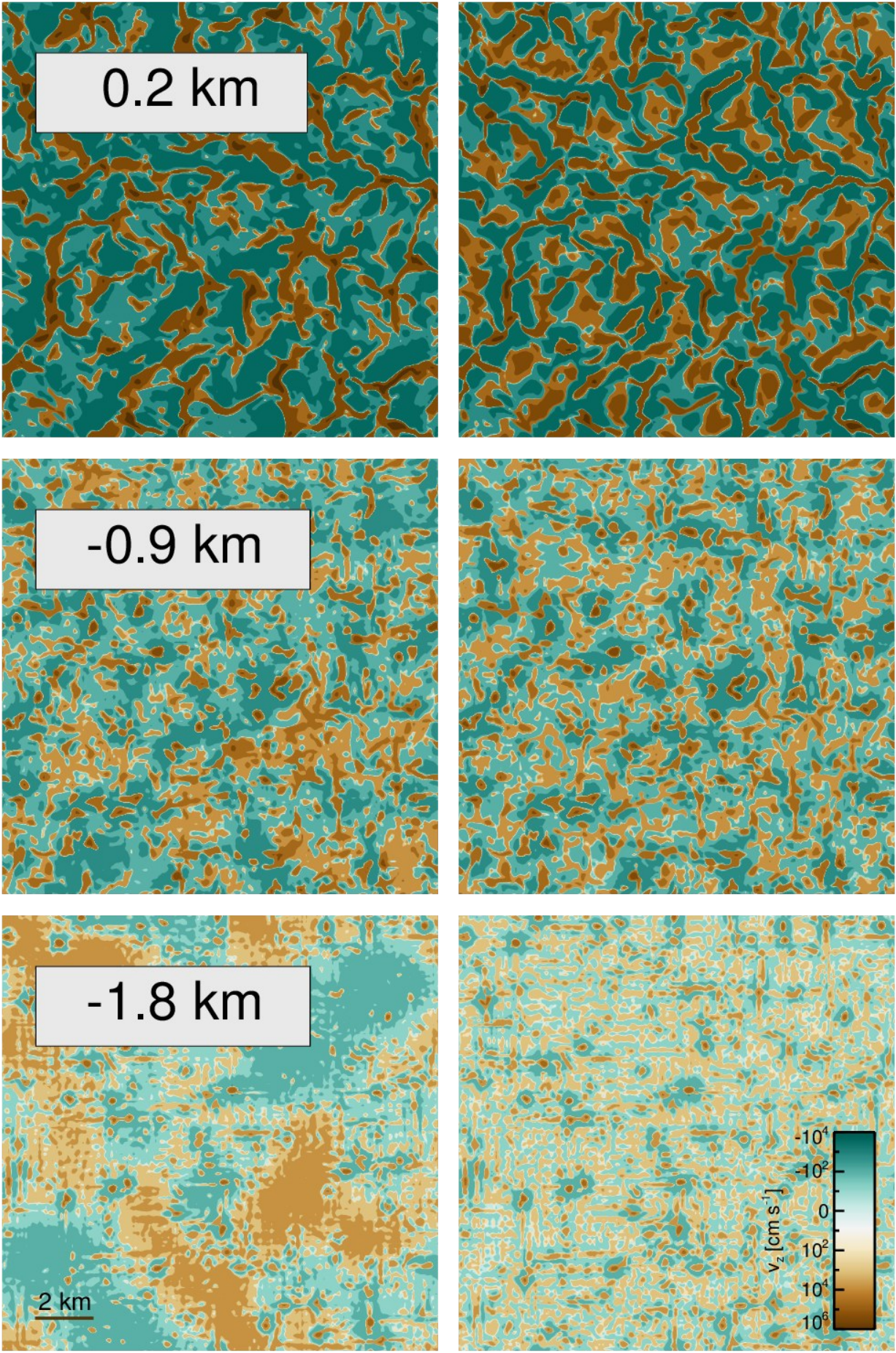}	
	\caption{Logarithmically coloured vertical velocity snapshots with (right) and without (left) spatial frequency filtering for regions demonstrating: convective instabilities (top), overshoot plumes (middle) and transition from plumes to waves (bottom), for simulation B1 with \T{13\,0}, \logg = 8.0 from Table~\ref{tb:main}. The vertical distances from the lower Schwarzschild boundary are indicated in the panels.}
	\label{fg:vpanels}
\end{figure}

Fig. \ref{fg:vpanels} shows the logarithmic vertical velocities at a layer within each of the three regions with distinct physical characteristics; the convectively unstable region (top), overshoot region (middle) and layer where plumes and waves overlap (bottom), for simulation B1 from Table \ref{tb:main} with \T{13\,0}, \logg = 8.0 and grid size 150$^3$. In the convectively unstable region the convective cells, with characteristic size $d_{\mathrm{gran}}\approx 1$\,km, can be seen with upward (teal) and downward (bronze) velocities. Beneath this layer in the overshoot region one can identify around twenty distinct overshoot plumes, which in the horizontal slice manifest as small circles with characteristic size $\sim 100$\,m. These circles are surrounded by large scale patches of size $\sim 5$\,km and relatively low velocity. These patches correspond to the standing waves. In the region below the deepest overshoot plume all that remains is the contribution from periodic oscillations.

\begin{figure}
	\centering
	\subfloat{\includegraphics[width=0.75\columnwidth]{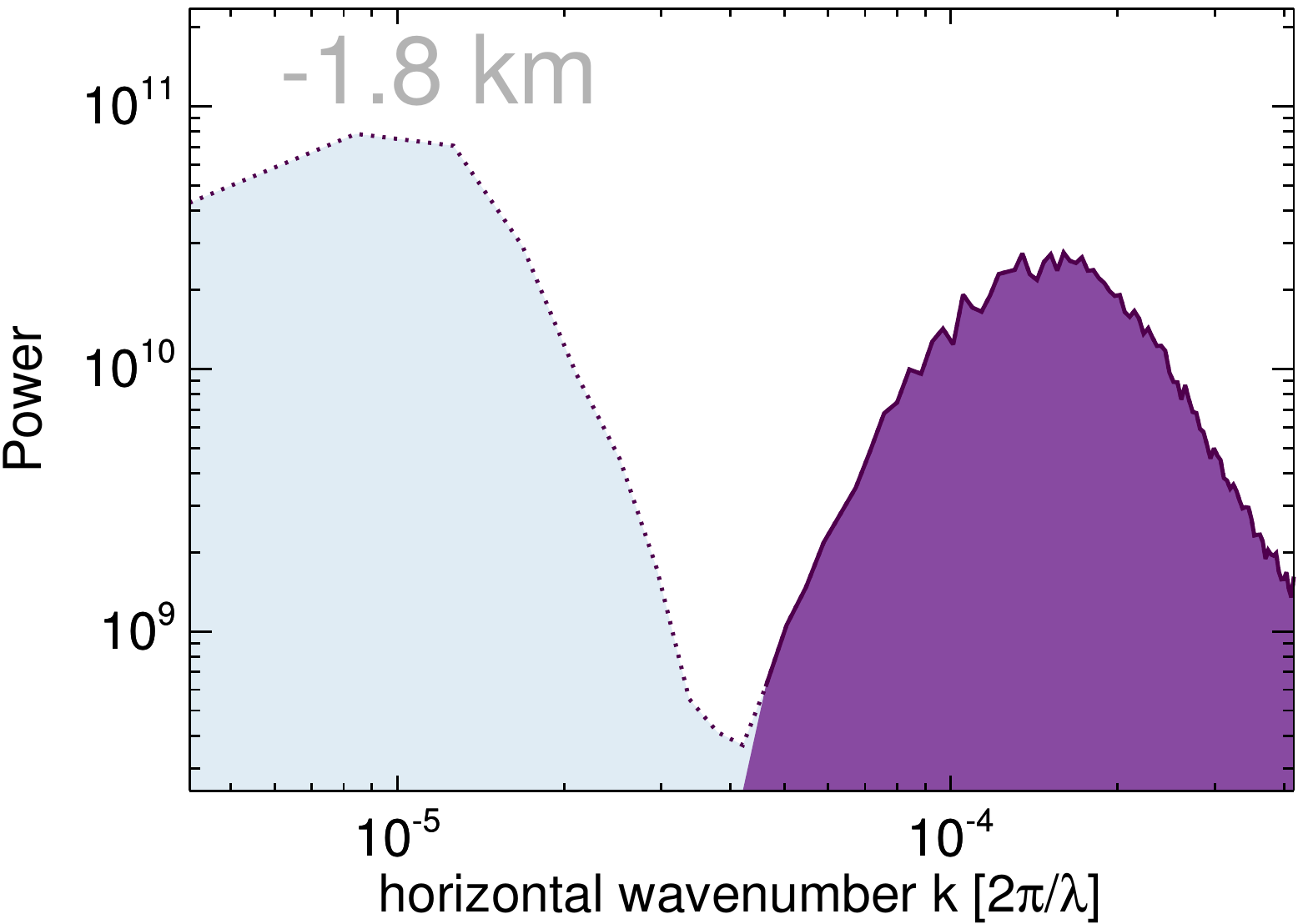}}\\
	\caption{Power spectrum for spatial frequencies present in lower panel of Fig.~\ref{fg:vpanels}. The unfiltered power spectrum (lower left panel of Fig.~\ref{fg:vpanels}) can be interpreted as a combination of the blue and purple regions. The filtered power spectrum (lower right panel of Fig.~\ref{fg:vpanels}) corresponds to the purple region only. The vertical distance from the lower Schwarzschild boundary is indicated in the panel and corresponds to the layer showing a transition from plumes to waves.}
	\label{fg:vpanels_spectrum}
\end{figure}

The contrast in spatial extent of the overshoot plumes and standing waves is of great benefit as the two occupy different regions of the power spectrum in spatial frequency. This is depicted in Fig.~\ref{fg:vpanels_spectrum} where the power as a function of horizontal wavenumber is shown for vertical velocities. The large scale wave features manifest as the blue peak at low wavenumber ($k<4\cdot10^{-5}$), whilst the overshoot plumes correspond to the purple peak at higher wavenumbers. The masking of the blue region (low wavenumbers) allows the removal of the major contribution from waves in frequency space, while leaving the overshoot plumes ostensibly untouched.

\begin{figure*}
	\centering
	\subfloat{\includegraphics[width=\columnwidth]{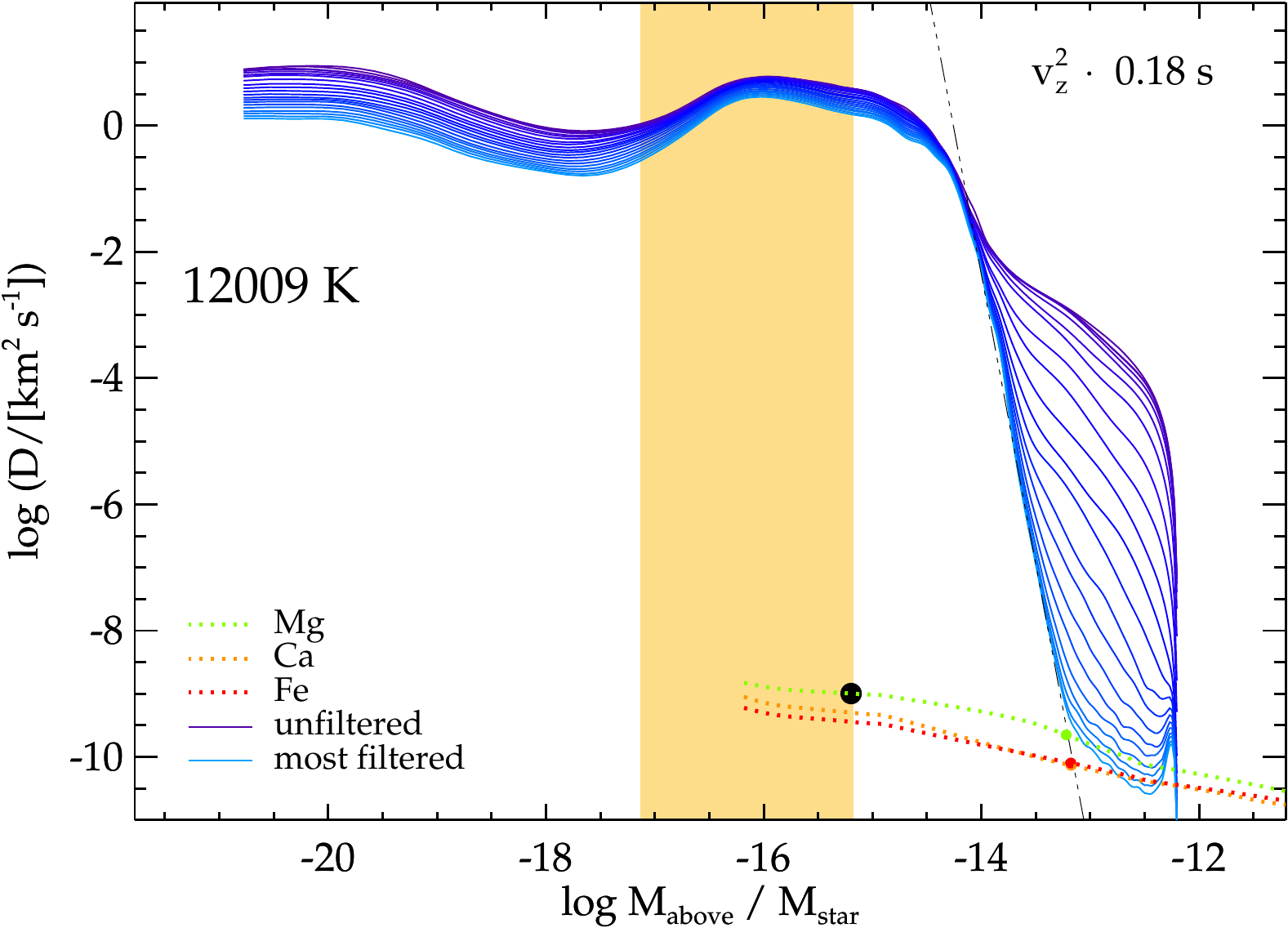}} \hfill
	\subfloat{\includegraphics[width=\columnwidth]{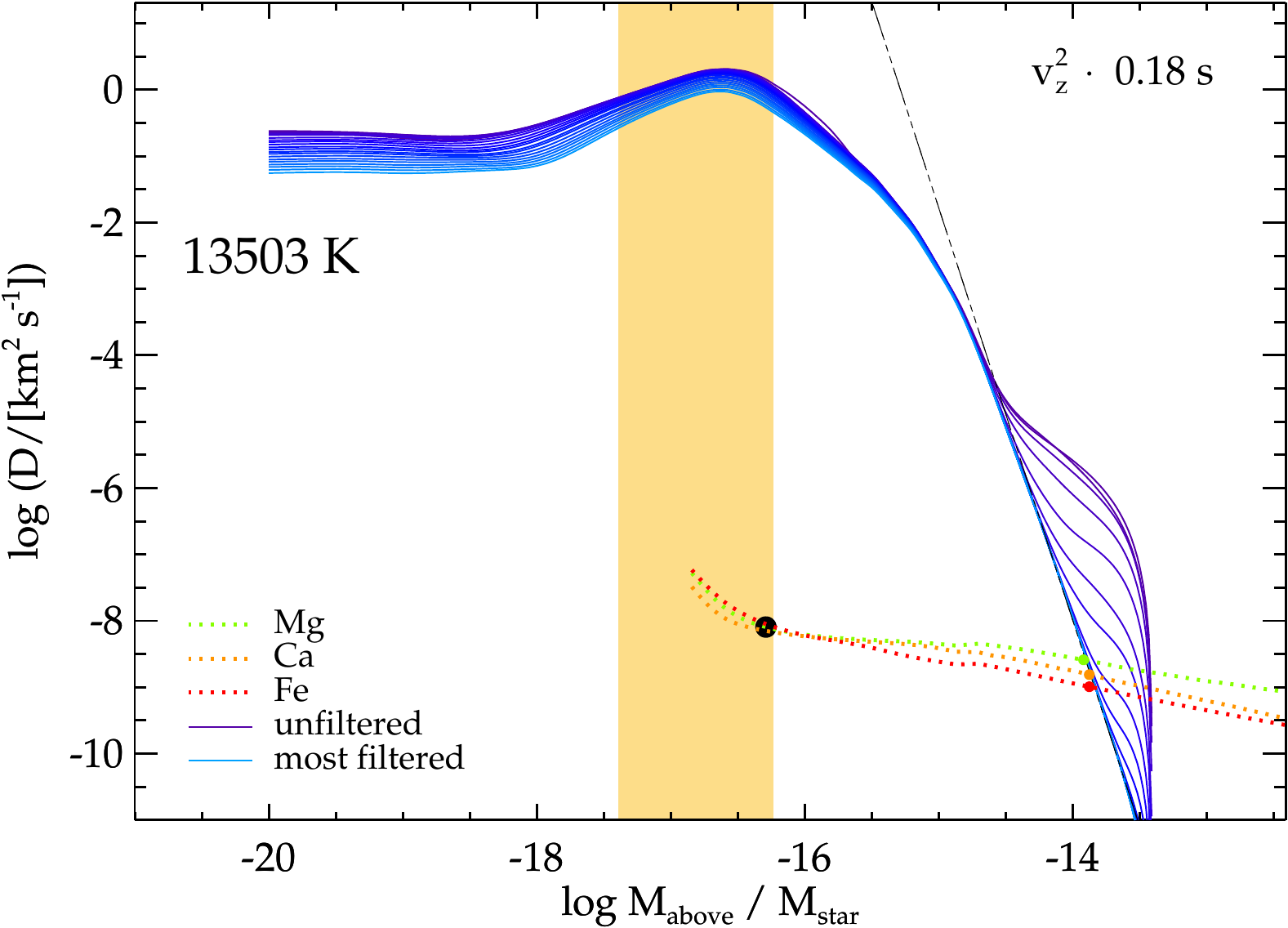}}			
	\caption{Diffusion coefficients as a function of enclosed stellar mass from deep simulations of Table~\ref{tb:deep} with \teff~= 12\,000\,K (left) and 13\,500~K (right). The macroscopic diffusion coefficients computed from the directly simulated \vsq\ profile are shown as a solid purple line. Those computed with an increasingly filtered \vsq\ profile (i.e., increasingly small structures removed) are shown in bluer colours, with adjacent lines increasing the filtering in Fourier space by $\Delta k = 1$ in box size units. The characteristic time scale of 0.18\,s found in Section~\ref{sec:Dover} is employed. Our proposed extrapolation of macroscopic diffusion coefficients is shown as a black dot-dashed line. The convectively unstable region defined by the Schwarzschild criterion is indicated by the orange shaded region. Also shown are the microscopic diffusion coefficients for Mg, Ca and Fe from updated tables of \citet[][dotted lines]{koester09}, taken as characteristic elements. The base of the mixed region is where the macroscopic and microscopic diffusion coefficients are equal. Diffusion coefficients at the base of the mixed regions defined in 1D (black circle) and with our proposed parameterisation of convective overshoot for the three characteristic elements (coloured circles) are also presented.}
	\label{fg:extrapolation_t135}
\end{figure*}

Fig.~\ref{fg:extrapolation_t135} shows the diffusion coefficient computed from the \vsq\ profiles, time-averaged over the final 2\,s, of the simulations from Table~\ref{tb:deep} with \T{12\,0} and 13\,500\,K. The characteristic time required for the computation is taken from the results of simulation C3 (see Fig.~\ref{fg:mquc_t135}). We note that the scaling of the diffusion coefficient with \vzrms\ in the near-overshoot region is of little relevance for deriving the total mixed mass. The effect of removing increasingly small scale structures from the power spectrum is shown by the lines becoming bluer. There is a clear asymptotic behaviour which emerges below enclosed mass $\log M_{\mathrm{above}} / M_{\mathrm{star}} = -14$ in both cases. The simulations also show significant influence from the lower boundary condition which enforces zero mass flux. Clearly, filtering in $k$-space against low spatial frequencies removes a large contribution from waves yet results in little disturbance of the velocities within the convective zone or the upper overshoot region. Any increase in filtering beyond what is shown in Fig.~\ref{fg:extrapolation_t135} leads to a significant loss of velocities within the convection zone which is an unphysical scenario we seek to avoid.

In the study by \cite{ludwig06} the filtering technique was also implemented in the temporal domain, ultimately making a cut in $k$- and $\omega$-space, where $k$ is the horizontal wavenumber. In our white dwarf atmospheres the standing waves have typical periods of 0.1\,s \citep[see also][]{kupka18}, similar to the typical decay time of granulation within the convection zone \citep{tremblay13c}. Therefore, it is difficult to disentangle convection from waves in $\omega$-space. Given the effectiveness of the $k$-space filtering technique discussed above, we simply neglect time filtering in the following.

We find that the asymptotic filtered \vsq\ is linear in $\log D$ - $\log M_{\rm above}/M_{\rm star}$ space, allowing us to perform a linear fit immediately beneath the start of the wave region. We can then extrapolate this behaviour to deeper layers that are out of reach with 3D simulations. Fig.~\ref{fg:extrapolation_t135} demonstrates that the decay of convective velocities in the wave region is much more rapid with depth than in the overshoot region above. Indeed the velocity scale height in the wave region (where overshoot structures still exist) is much less than the pressure scale height, reflecting similar behaviour which is observed in simulations of convection in brown dwarf atmospheres \citep{freytag10}, but we cannot confirm whether this change in convective velocity profile is impacted by waves, e.g. with wave damping convective overshoot. For this reason and also our neglect of waves as a possible contributor to mixing, our estimate of the mixed mass is a lower limit.

We estimate the minimum amount of mixed mass in accreting white dwarfs by looking for the intersection of the asymptotic filtered macroscopic diffusion coefficient with microscopic diffusion coefficients. In Fig.~\ref{fg:extrapolation_t135} we estimate these mixed masses using the magnesium, calcium and iron diffusion coefficients (dotted) from \citet{koester09}; using updated tables.\footnote{\scriptsize{http://www1.astrophysik.uni-kiel.de/$\sim$koester/astrophysics/astrophysics.html}} Here it is important to emphasise that 3D corrections to mixed masses can be read directly from Fig.~\ref{fg:extrapolation_t135}. The crude 1D estimate is simply the intersection of the base of the convection zone with the diffusion coefficient (black circle), and the more robust 3D picture is based on the intersection of the asymptotic filtered macroscopic diffusion coefficient with the microscopic diffusion coefficients (coloured circles). The fact that convective velocities decay rapidly with depth in the wave region implies that while we extrapolate over a sizeable range in velocity space, it corresponds to a fairly modest additional amount in mixed mass. In other words, our direct simulations in Section~\ref{sec:Dover} have robustly shown that a sizeable amount of mass is unambiguously mixed in the overshoot region, and the additional amount added by our filtering procedure is fairly moderate.

\begin{figure}
	\centering
	\includegraphics[width=\columnwidth]{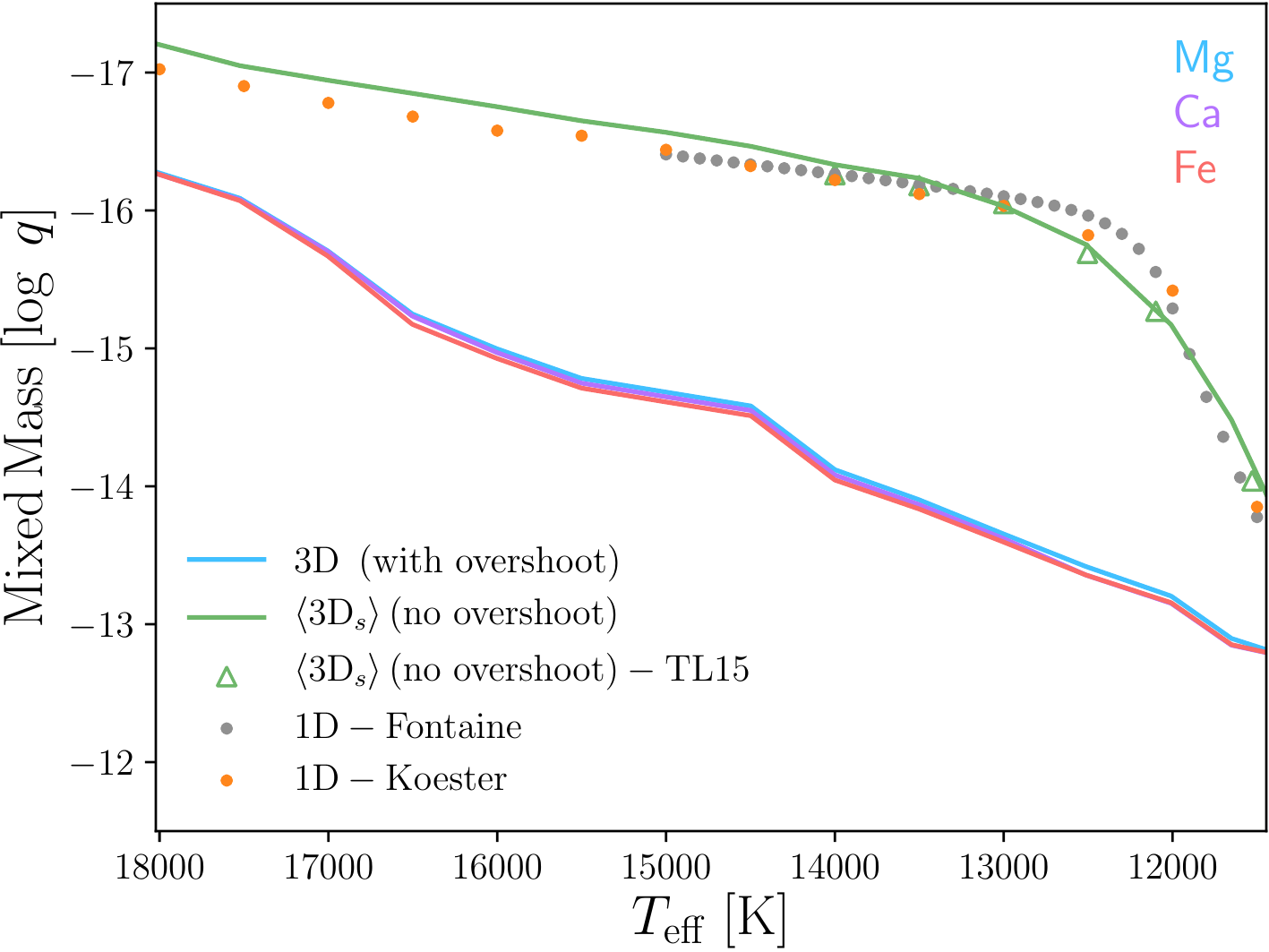}
	\caption{Mixed mass in units of enclosed mass, $\log q = \log M_{\rm above}/M_{\rm star}$, as a function of effective temperature. Mixed masses determined by the intersection of macroscopic and microscopic diffusion coefficients (see Fig.~\ref{fg:extrapolation_t135}) for Mg (blue), Ca (magenta) and Fe (red) are shown in solid for all simulations detailed in Table~\ref{tb:deep}. Also shown is the 3D definition of the unstable region from the grid presented in this study (green, solid) and selected models from the earlier lower resolution grid (green, triangles) from \citet{tremblay15}. This corresponds to the definition of mixed mass without overshoot. The two coolest models plotted from the latter grid are open bottom simulations, and show good agreement with the closed bottom simulations calculated in this work. In comparison, the orange and grey lines show the 1D mixed mass of the unstable layers according to 1D structures of Koester and Fontaine, respectively, using ML2/$\alpha=0.8$.}
	\label{fg:full_1d_vs_3d}
\end{figure}

The convection zone mass after one includes the full extent of the convectively unstable and overshoot regions is shown in Fig.~\ref{fg:full_1d_vs_3d} with solid lines for Mg (blue), Ca (purple) and Fe (red) and described in Table~\ref{tb:3D-mass}. The figure also shows the mass at the base of the convection zone in 3D described from the Schwarzschild criterion for our deep grid of simulations (green line). Those are in fairly good agreement with the results of \citet{tremblay15} (green triangles) and 1D structures (grey and orange points), and this corresponds to the mixed mass in the 1D picture, i.e. neglecting the contribution from overshoot. The lower limit on the 3D correction in mass is at least 1 dex but can reach $\approx$ 2 dex for the intermediate $T_{\rm eff}$ values in our grid. We will go on to consider the implications of this result for the accretion-diffusion picture.

Recently \citet{kupka18} also presented a prediction for the range of mixed mass for a simulation at \T{11\,8}~to be $-12.7 \gtrapprox \log q \gtrapprox -13.9$ based on the plume penetration depth and horizontal to vertical velocity ratios, respectively. An interpolation of Fig.~\ref{fg:full_1d_vs_3d} reveals this to be in good agreement with the predicted mixed mass from our 3D grid which estimates the mixed mass at \T{11\,8}~ to be $\log q = -13.0$. {We also find agreement with the predicted size of the convectively unstable zone that the authors found to be $\log q = -15.2$, which compares well with the green line in Fig.~\ref{fg:full_1d_vs_3d}.}

\section{Discussion}
\label{discussion}
Chemical abundances in polluted white dwarf atmospheres as well as characterisation of circumstellar dust and gas disks for a fraction of those are the main observational properties with which we can constrain evolved planetary systems. On the one hand, the chemical properties and time variability of circumstellar material is independent of the white dwarf convection model. These systems can be used to understand the variability of predicted accretion rates onto the white dwarfs \citep{wyatt14}. On the other hand, it is the accretion-diffusion scenario, currently under the 1D MLT assumption of no convective overshoot, that is used to predict parent body chemical compositions, the total accreted mass over the characteristic diffusion timescale, and in some cases, estimate the time elapsed since the last accretion event \citep{hollands18}. In the case of the hydrogen abundance in helium-atmospheres, one can also infer about the full past history of accretion since hydrogen does not diffuse out of the convective layers \citep{gentile17}. Time variable abundances in the photosphere, although so far not observed \citep{farihi18,debes08}, could also be used to understand the accretion process. All of those inferred quantities need to be re-evaluated in light of 3D convection. We have determined that the mixed mass, or the total accreted mass over a characteristic diffusion timescale, is 1--2.5 orders of magnitude larger than the crude 1D estimate and we dedicate this section to an overview of the other quantities derived from the accretion-diffusion model.

One starting point is the impact of convective overshoot on the settling times, \tdiff, of trace elements in polluted white dwarfs. Accretion calculations are usually performed under the assumption that a steady state has been reached between accretion at the white dwarf and diffusion out of the base of the convection zone, for at least a few diffusion timescales. This implies that the abundance in the mixed region is homogeneous and in this scenario the accretion rate is given by 
\begin{equation}
 \dot{M}_i = \frac{M_{\mathrm{cvz}} X_i}{t_{\mathrm{diff}, i}} = 4\pi r^2 \rho v_{\mathrm{diff}}
 \label{eq:mdot}
\end{equation}
where $M_{\mathrm{cvz}}$ is the mixed mass and, for an element $i$, $\dot{M}_i$ is the accretion rate,  $t_{\mathrm{diff}, i}$ the diffusion timescale, $X_i$ the elemental atmospheric abundance, $\rho$ the density at the base of the mixed region and $r$ the stellar radius.

\begin{figure}
	\centering
	\subfloat{\includegraphics[width=\columnwidth]{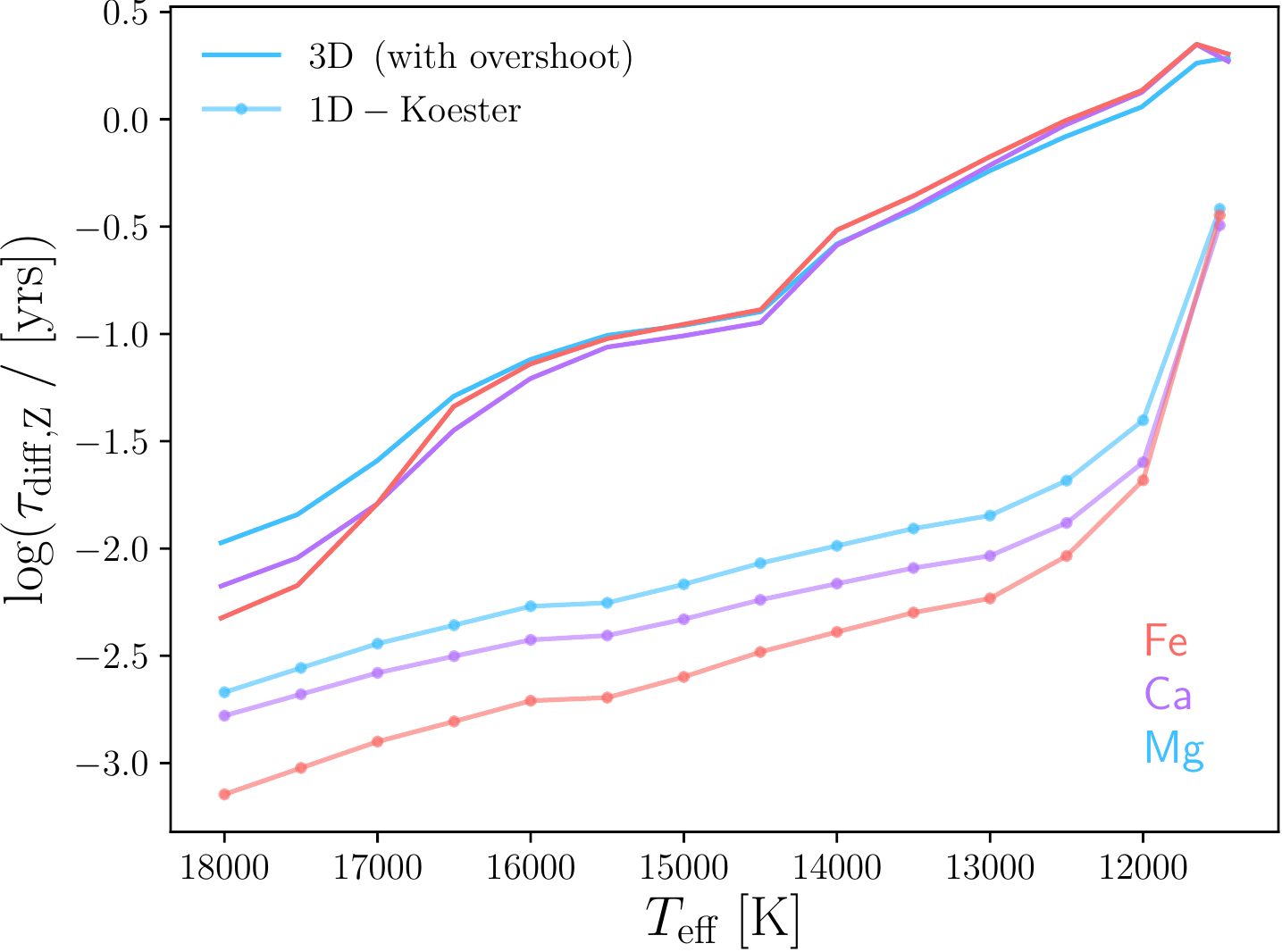}} \\
	\subfloat{\includegraphics[width=\columnwidth]{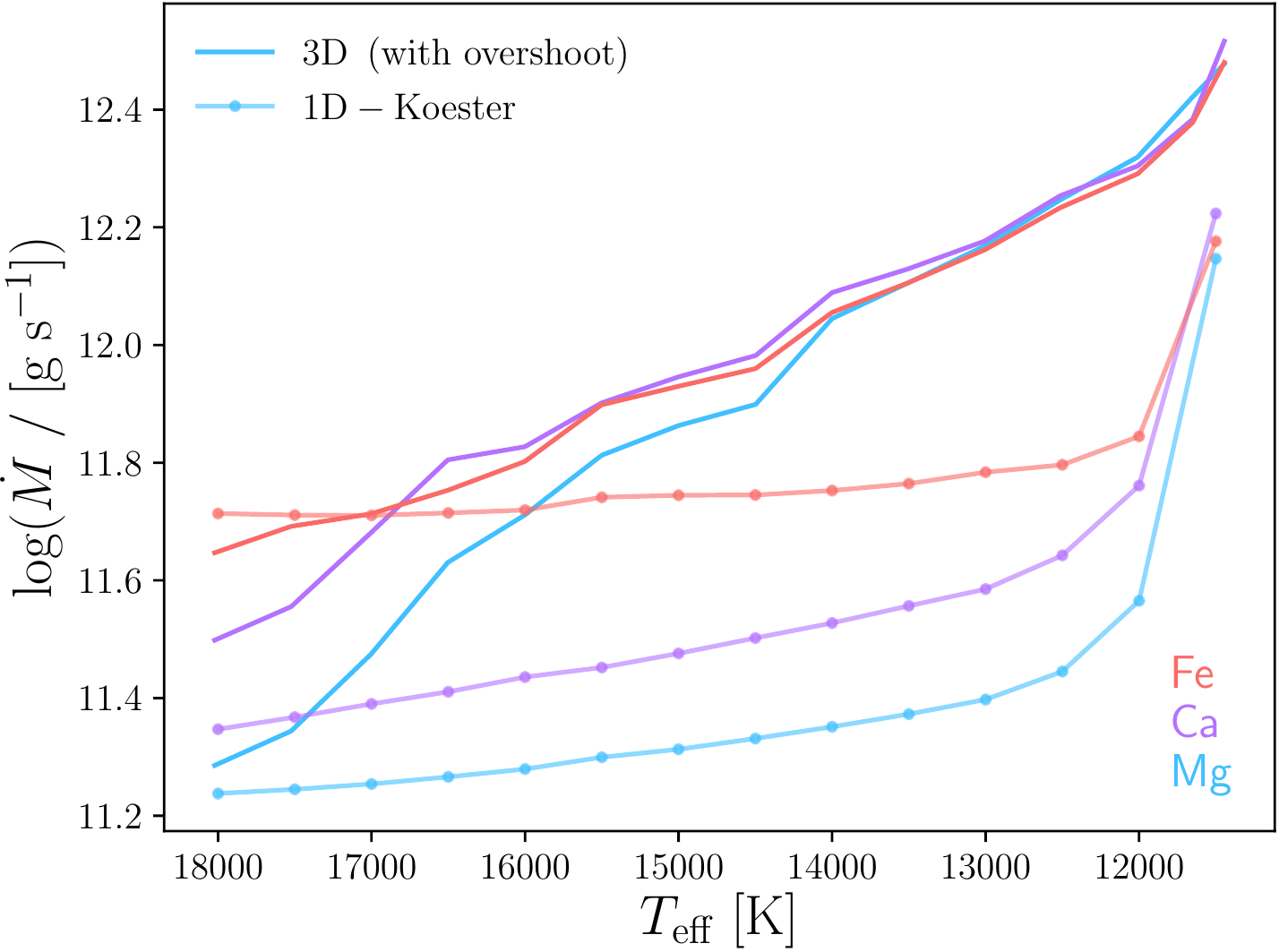}}
	\caption{Diffusion timescale (top panel) and inferred accretion rates (lower panel) when using the mixed masses, shown in Fig.~\ref{fg:full_1d_vs_3d}, borne from diffusion coefficients derived from our 3D grid using \vsq~(solid). For simplicity the abundance, $X_i$, in Eq.~\eqref{eq:mdot} used to infer an accretion rate is assumed to be unity. Also shown are 1D structures from Koester (solid-circles). All quantities are provided for Mg (blue), Ca (magenta) and Fe (red). The strong upward inflexion in the 1D accretion rates for models with $T_{\mathrm{eff}} < 12\,000$~K is a symptom of the strong downward inflexion seen in the mixed masses for the same models (see Fig.~\ref{fg:full_1d_vs_3d}).}
	 \label{fg:corrections}
\end{figure}

We use the 1D diffusion velocities presented in \citet{koester09} to calculate the change in settling timescale when one redefines the base of the mixed region. Fig.~\ref{fg:corrections} shows 3D corrections to diffusion timescales (top) and accretion rates (bottom). Convective overshoot implies that microscopic diffusion takes over as the dominant process of mass transport in deeper, denser layers where diffusion timescales are longer. For an inference of accretion rates we have the competing effects of larger masses and larger diffusion timescales that partially cancel each other out as the base of the mixed layer goes deeper. The net result from this competition is an increase in accretion rate of up to an order of magnitude (see bottom panel of Fig. \ref{fg:corrections}), which is not as large as the increase in mixed masses.

\begin{table}
	\centering
	\caption{Input parameters for accretion-diffusion calculations derived for our full grid of deep 3D simulations (see Table~\ref{tb:deep}). The mixed mass in units of enclosed mass, $M_{\mathrm{cvz}}$, diffusion velocity, $v_{\mathrm{diff}}$ and diffusion coefficient, $D$, are all evaluated at base of the 3D mixed region. Values are shown for mixed regions defined by diffusion coefficients with $D = v^2 \cdot t_{\mathrm{char}}$ (see Fig.~\ref{fg:full_1d_vs_3d}). For brevity, all values are expressed as a mean across those computed for magnesium, calcium and iron.}
	\label{tb:3D-mass}
	\begin{tabular}{|||||c|||||||||||||||||ccc||||||}
		\hline
		\teff  & $\log \langle M_{\mathrm{cvz}} \rangle$ & $\log \langle v_{\mathrm{diff}} \rangle$ & $\log \langle D \rangle $ \\
		$[\mathrm{K}]$ &  &  &  \\
		\hline
11445  &  $-$12.82  &   $-$7.62  &   $-$10.28  \\
11651  &  $-$12.89  &   $-$7.60  &   $-$9.97  \\
12009  &  $-$13.19  &   $-$7.43  &   $-$9.90  \\
12500  &  $-$13.41  &   $-$7.30  &   $-$9.64  \\
13005  &  $-$13.64  &   $-$7.17  &   $-$8.88  \\
13503  &  $-$13.89  &   $-$7.03  &   $-$8.76  \\
14000  &  $-$14.09  &   $-$6.91  &   $-$8.68  \\
14498  &  $-$14.58  &   $-$6.63  &   $-$8.51  \\
15000  &  $-$14.49  &   $-$6.68  &   $-$8.52  \\
15501  &  $-$14.60  &   $-$6.61  &   $-$8.47  \\
16002  &  $-$15.01  &   $-$6.42  &   $-$8.39  \\
16503  &  $-$15.24  &   $-$6.29  &   $-$8.31  \\
17004  &  $-$15.73  &   $-$6.02  &   $-$8.19  \\
17524  &  $-$16.12  &   $-$5.83  &   $-$8.11  \\
18022  &  $-$16.33  &   $-$5.70  &   $-$8.04  \\
		\hline
	\end{tabular}\\
\end{table}

To demonstrate the impact of 3D corrections derived from our numerical study we recalculate accretion rates for a sample of 48 polluted DAZ white dwarfs \citep{bergfors14}. As a function of effective temperature the change in accretion rate is given by
\begin{equation}
 \Delta \dot{M}(T_{\mathrm{eff}}) = \frac{1}{3}\sum_{\mathrm{Mg,Ca,Fe}} (\dot{M}^{\mathrm{3D}}_{el} - \dot{M}^{\mathrm{1D}}_{el})
 \label{eq:mdot-mean}
\end{equation}
where, for simplicity, the individual change in accretion rate from each element, $el$, considered - Mg, Ca and Fe - is weighted equally in the mean, corresponding to an equal abundance of all three elements which is reasonably close to solar abundance ratios. 

\begin{figure}
	\centering
	\includegraphics[width=\columnwidth]{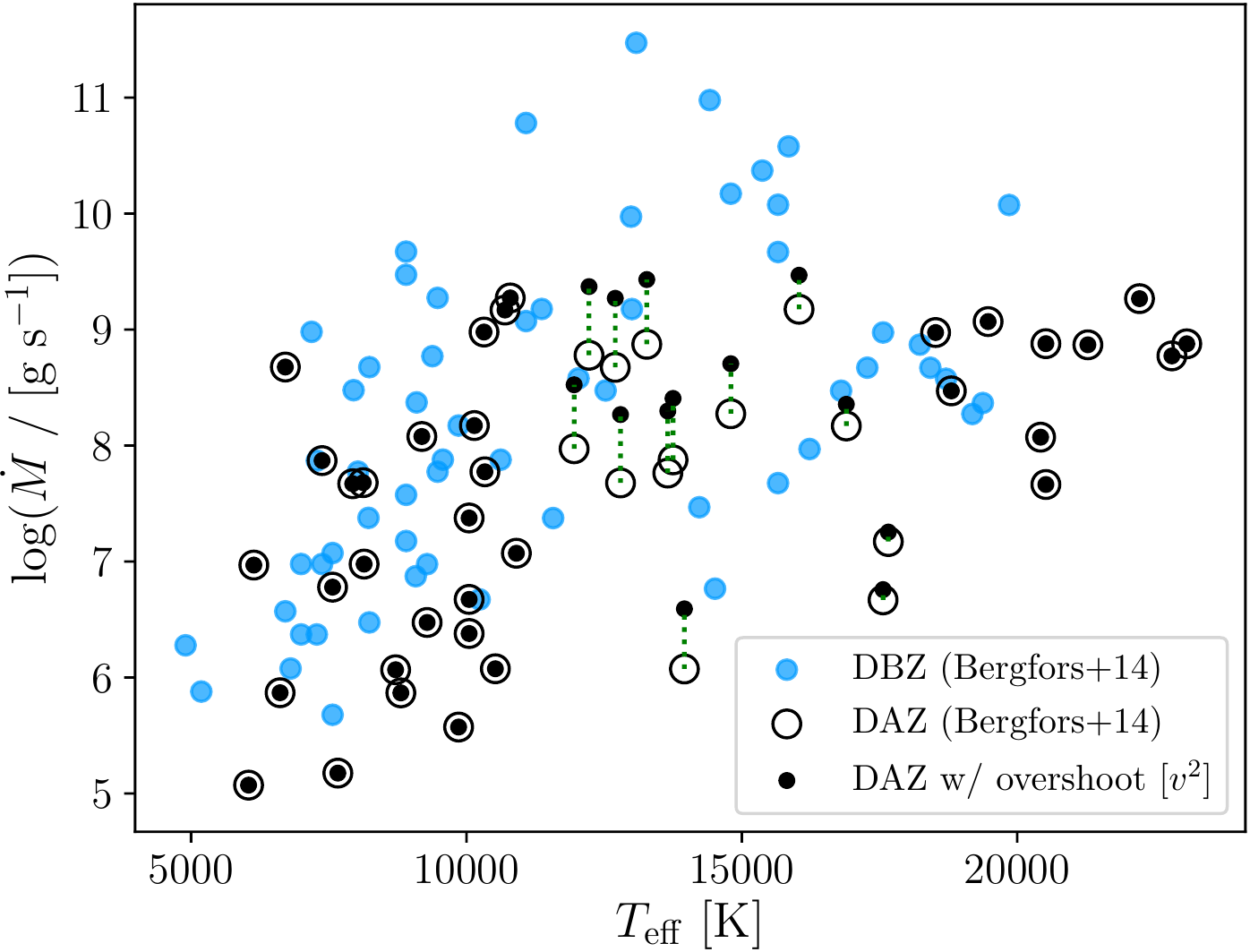}
	\caption{Sample of polluted white dwarfs comprising DAZs (black, open) and DBZs (blue) from \citet{bergfors14}. The DAZ sample is also shown after the inclusion of convective overshoot (black, filled).}
	\label{fg:bergfors-correction}
\end{figure}

\begin{figure}
	\centering
	\includegraphics[width=\columnwidth]{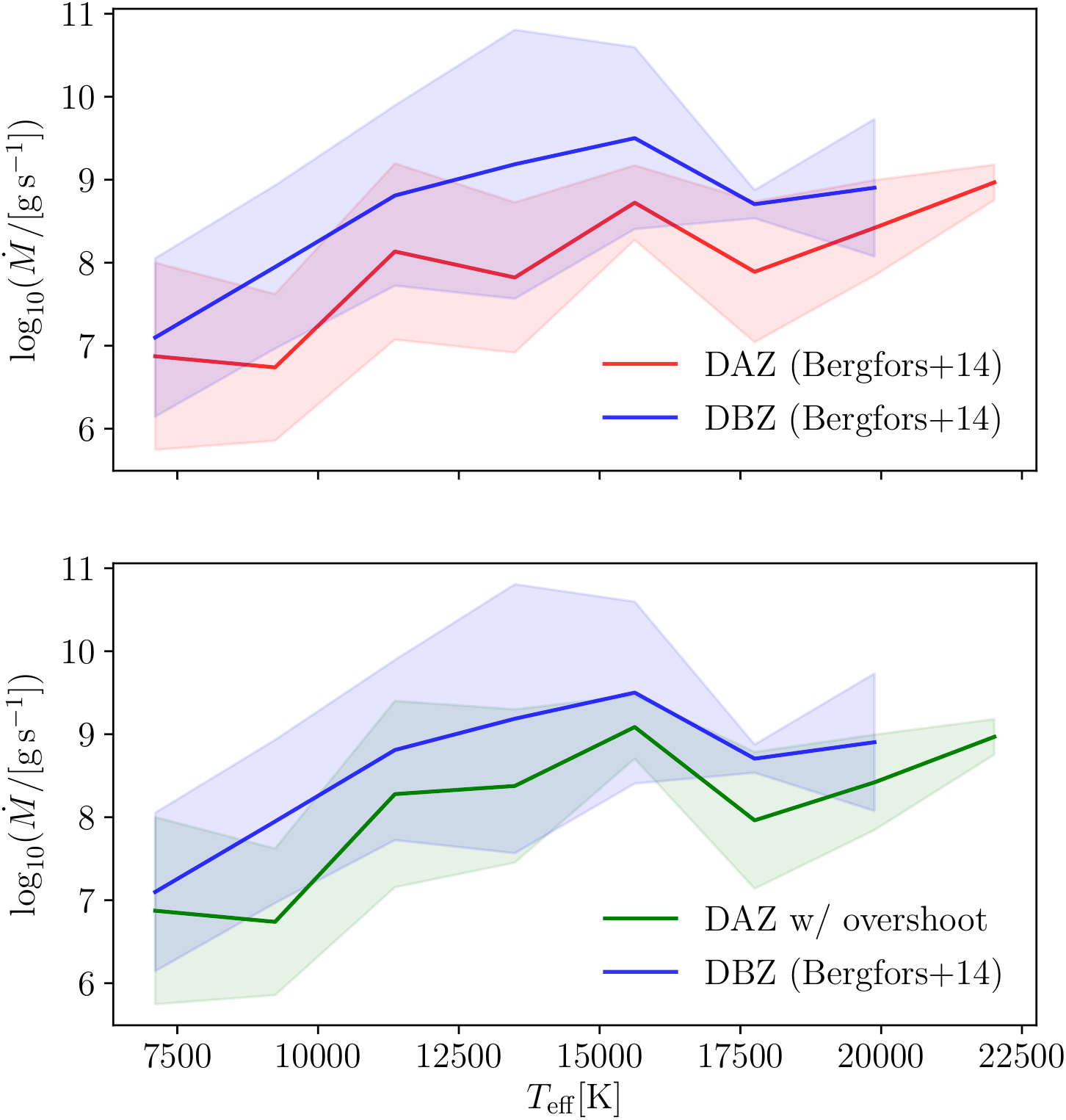}
	\caption{Accretion rates from \citet{bergfors14} and Fig. \ref{fg:bergfors-correction} mean-averaged within eight effective temperature bins. Shaded areas correspond to one standard deviation for the overshoot-corrected DAZ  (green), original DAZ (red) and DBZ (blue) samples.}
	\label{fg:histomean}
\end{figure}

Fig.~\ref{fg:bergfors-correction} shows the DAZ and DBZ samples taken from \citet{bergfors14} (open circles and blue filled circles, respectively) along with the 3D corrected DAZ sample (black, filled circles). It can be seen that within the effective temperature range considered in this paper (11\,400 - 18\,000 K), 3D corrections bring the mean accretion rates between the DAZ and DBZ samples into closer agreement. Furthermore, for the same region we find that the standard deviation of the corrected sample is contained within the standard deviation of the DBZ sample. This is shown in Fig.~\ref{fg:histomean}, where the mean (solid line) and one standard deviation (filled error regions) are plotted for the original sample from \citet{bergfors14} in the top panel and with the inclusion of macroscopic mixing due to overshoot shown in the lower panel. The lack of a correction below 11\,000 K is due to the computational limitations of simulating lower temperatures (due to their deep convection zones) and above 18\,000\,K we predict no correction due to overshoot, owing to the non-convective, radiative atmospheres of these warmer objects. A small correction due to overshoot is expected for the DBZ sample, but to a lesser extent than for the DAZs at comparable temperatures.

\section{Conclusions}
\label{conclusions}

Using the RHD code CO$^5$BOLD, we have presented the first macroscopic diffusion coefficients for DA white dwarfs in the range 12\,000~K $\leq T_{\mathrm{eff}} \leq 13\,500$~K derived directly from 3D simulations. We have shown that the phenomenon of convective overshoot is capable of mixing material over a much greater extent than the MLT approximation, as suggested before \citep{freytag96,tremblay15,kupka18}. These direct simulations have demonstrated that macroscopic diffusion decays with the vertical component of the velocities immediately beneath the unstable layers and with the same quantity squared for at least $\approx$ 2.5 -- 3.5 pressure scale heights beneath the unstable layers. For deeper layers, the 3D diffusion experiments were inconclusive because of the presence of waves amplified by boundary conditions, though the path integration method was less susceptible to artificial mixing in the wave-dominated regions.

We have developed an approach to acquire a lower limit on the mixed mass in the convection zone which consists of using wave-filtered convective velocities as a proxy for macroscopic diffusion. We identify the points at which the diffusion coefficients derived from the filtered velocity squared profiles intersect the microscopic diffusion coefficients for magnesium, calcium and iron, providing a conservative and elemental dependent estimate of the increase in mixed mass.

We have found evidence to suggest that settling times for pure-hydrogen atmosphere white dwarfs with $11\,400 \leq T_{\mathrm{eff}} / [\mathrm{K}] \leq 18\,000$ are likely to increase by 1.5 -- 3 orders of magnitude (see Fig.~\ref{fg:corrections}; top). This has implications for predicting the period over which we are unlikely to observe pollution variability in white dwarfs, providing the possibility for constraining this model. The often discussed instantaneity of accretion at warm DAZ white dwarfs is still valid compared to the Myr timescales of accretion at DBZ white dwarfs \citep{wyatt14}, but we have presented evidence that it may be slower than typically thought. Despite this large increase in settling time, we predict that derived accretion rates may increase by only up to an order of magnitude. In the immediate future we intend to create a semi-analytic model for DA white dwarfs cooler than the 3D grid presented here extends to, as well as for DB stars which should experience similar effects from convective overshoot. 

Our treatment of tracers as passive scalars contains the implicit approximation that the particles have no mass or charge. For future work we consider that this approximation could be lifted to provide the opportunity to investigate the role of other instabilities which may arise from, for example, the presence of a chemical gradient. 

Ultimately the intention behind this research is to provide robust constraints on accretion processes by improving the characterisation of the mass reservoir in white dwarf convection zones, and should serve as an important step to better understand the evolution of planetesimals in old planetary systems.

\section*{Acknowledgements}
The research leading to these results has received funding from the European Research Council under the European Union's Horizon 2020 research and innovation programme n. 677706 (WD3D). HGL acknowledges financial support by the Sonderforschungsbereich SFB\,881 ``The Milky Way System'' (subprojects A4) of the German Research Foundation (DFG). This research was supported in part by the National Science Foundation under Grant No. NSF PHY-1748958. We are grateful to Boris G\"{a}nsicke, Evan Bauer, Lars Bildsten and Christopher Manser for helpful discussions in regard to the implications for planetary accretion and stellar evolution.

%%%%%%%%%%%%%%%%%%%%%%%%%%%%%%%%%%%%%%%%%%%%%%%%%%



\bibliographystyle{mnras}
\bibliography{mybib} % if your bibtex file is called example.bib

%%%%%%%%%%%%%%%%%%%%%%%%%%%%%%%%%%%%%%%%%%%%%%%%%%

%%%%%%%%%%%%%%%%% APPENDICES %%%%%%%%%%%%%%%%%%%%%

\appendix

%%%%%%%%%%%%%%%%%%%%%%%%%%%%%%%%%%%%%%%%%%%%%%%%%%

% Don't change these lines
\bsp	% typesetting comment
\label{lastpage}
\end{document}